\documentclass{article}

\usepackage{arxiv}

\usepackage{amsmath}
\usepackage{amssymb}
\usepackage{xcolor}
\usepackage{algorithmic}
\usepackage{array}
\usepackage{todonotes}
\usepackage{xspace}
\usepackage{booktabs}
\usepackage{graphicx}
\usepackage{subcaption}
\usepackage{caption}
\usepackage{rotating}
\usepackage{algorithmic} 
\usepackage{algorithm}
\usepackage{hyperref}
\usepackage[table,dvipsnames]{xcolor}

\colorlet{mylinkcolor}{YellowOrange}
\colorlet{mycitecolor}{Aquamarine}
\colorlet{myurlcolor}{violet}
\hypersetup{ 
        colorlinks=true, 
        linkcolor=mylinkcolor, 
        citecolor=mycitecolor, 
        urlcolor=myurlcolor, 
        filecolor=myurlcolor }

\usepackage{makecell}

\usepackage{cleveref}
\usepackage{url}
\usepackage{listings}
\usepackage{multirow}
\usepackage[acronym,toc]{glossaries}
\usepackage{tikz}
\usetikzlibrary{positioning,arrows.meta}

\newcommand{\name}{\textsc{DeepWeave}\xspace}
\newcommand{\ourparagraph}[1]{\noindent\textbf{#1}}

\definecolor{bordeaux}{RGB}{128, 0, 32}

\usepackage[most]{tcolorbox}
\newtcolorbox{notations}{
    colback=gray!5,
    colframe=black,
    title=Notation
}

\definecolor{dplbackground}{RGB}{247,248,250}
\definecolor{dplframe}{RGB}{205,210,218}
\definecolor{dplkeyword}{RGB}{38,90,160}
\definecolor{dplcomment}{RGB}{95,110,105}
\definecolor{dplstring}{RGB}{145,65,120}

\lstdefinelanguage{DeepProbLog}{
    morekeywords={
        nn,
        true,
        false,
        not
    },
    morecomment=[l]{\%},
    morestring=[b]',
    sensitive=true
}

\lstdefinestyle{deepweave-logic}{
    language=DeepProbLog,
    basicstyle=\ttfamily\small,
    keywordstyle=\color{dplkeyword}\bfseries,
    commentstyle=\color{dplcomment}\itshape,
    stringstyle=\color{dplstring},
    backgroundcolor=\color{dplbackground},
    frame=single,
    rulecolor=\color{dplframe},
    framesep=4pt,
    xleftmargin=0.5em,
    xrightmargin=0.5em,
    aboveskip=0.6\baselineskip,
    belowskip=0.6\baselineskip,
    columns=fullflexible,
    keepspaces=true,
    showstringspaces=false,
    breaklines=true,
    breakatwhitespace=true,
    tabsize=2,
    numbers=none,
    captionpos=b
}

\lstnewenvironment{dplcode}[1][]{
    \lstset{
        style=deepweave-logic,
        #1
    }
}{}

\makeatletter
\newcommand{\crefnames}[3]{%
  \@for\next:=#1\do{%
    \expandafter\crefname\expandafter{\next}{#2}{#3}%
  }%
}
\makeatother

\crefnames{part,chapter,section}{\S}{\S}
\crefnames{paragraph,subparagraph}{\P}{\P}

\usepackage[inline]{enumitem}
\newenvironment{inlinelist}{\begin{enumerate*}[label=\emph{(\roman{*})}]}{\end{enumerate*}}

\usepackage{etoolbox}

\newcommand{\tablefont}{\fontsize{8}{9.5}\selectfont}
\makeatletter
\g@addto@macro\@floatboxreset{%
  \def\@tempa{table}%
  \ifx\@captype\@tempa\tablefont\setlength{\tabcolsep}{4pt}\fi}
\makeatother
\newcolumntype{L}[1]{>{\raggedright\arraybackslash}p{#1}}


\title{\name{}: Weaving Neural and Symbolic Evidence for ML-based Network Intrusion Detection System}

\author{
  Raoul Saad \quad Enkeleda Bardhi \quad Andrea Agiollo \quad Fernando Kuipers \\[1.5ex]
  Delft University of Technology \\
  Van Mourik Broekmanweg 6, Delft, The Netherlands, 2628 XE \\
  \texttt{raoul.e.saad@gmail.com}, \texttt{\{E.Bardhi-1, A.Agiollo-1, F.A.Kuipers\}@tudelft.nl}
}

\date{}

\renewcommand{\shorttitle}{\textit{DeepWeave}}

\hypersetup{
pdftitle={DeepWeave: Weaving Neural and Symbolic Evidence for ML-based Network Intrusion Detection System},
pdfauthor={Raoul Saad, Enkeleda Bardhi, Andrea Agiollo, Fernando Kuipers},
}

\begin{document}
\maketitle

\begin{abstract}
Machine learning-based network intrusion detection systems can recognize complex traffic patterns, but often require substantial labelled attack data, degrade under deployment shift, and provide limited evidence for analysts. 
We present \name{}, a probabilistic neuro-symbolic Network Intrusion Detection System (NIDS) that makes logic an integral part of detection. 
\name{} automatically constructs and validates interpretable rules, represents neural predictions as probabilistic predicates, and reconciles both evidence sources through explicit coupling policies over packet-, flow-, and hybrid-views. 
The resulting proofs provide intrinsic explanations for rule-supported alerts. 
Across multiple datasets, \name{} improves low-data detection, false-positive control, transferability, and robustness. 
With only \(50\) labelled packet observations per attack class, it improves macro-F1 from \(0.632\) to \(0.924\) and reduces the attack false-positive rate from \(0.075\) to \(0.008\) over the matched neural baseline. 
Its proofs are also more stable than post-hoc neural attributions, while batched inference keeps probabilistic reasoning practical.
These results demonstrate label-efficient, adaptable, and inspectable neuro-symbolic intrusion detection.
\end{abstract}

% keywords can be removed

\newacronym{nids}{NIDS}{Network Intrusion Detection System}
\section{Introduction}
\label{sec:introduction}

Network Intrusion Detection Systems (NIDSs) are a fundamental component of defense architectures.
By continuously inspecting packets, flows, or aggregated network events, a NIDS seeks to distinguish benign activity from traffic associated with malicious behaviour.
Malicious events are rare, benign traffic is heterogeneous, while both traffic patterns and adversaries evolve.
Moreover, sustained volumes of low-value alerts contribute to costly triage and alert fatigue~\cite{ban2021alertfatigue,tariq2025alertfatigue}.
A practical NIDS must not only be accurate on a static benchmark, but rather control false alarms, remain useful under traffic shifts, and provide evidence that an analyst can inspect~\cite{wang2025captain,sommer2010outside,wei2023xnids}.

Existing NIDSs largely follow symbolic or machine learning (ML) paradigms~\cite{roesch1999snort,mirsky2018kitsune,guo2026knowml,dong2023horuseye,rieger2023argus,barradas2021flowlens}.
Signature-based engines such as Snort and Suricata match explicit conditions over payloads, protocol fields, and flow state~\cite{roesch1999snort,oisf_suricata,di2026beyond}.
Their rules are auditable and dependable for known behaviors, but authoring and maintaining them requires substantial expertise, and environment-specific assumptions can produce false positives or miss unseen attack variants~\cite{sacher2020falsepositives,landauer2024alertdataset}.
%

% ML approaches instead infer decision boundaries from packet bytes, flow statistics, or temporal behavior and can capture patterns that are cumbersome to specify manually~\cite{buczak2016survey,khraisat2019survey}.\agtodo{rather than surveys add citations from top tiers}
% %
% Yet the resulting models can fail outside the conditions represented during training.
% %
% Intrusion-detection research has long warned that closed-world evaluation poorly reflects deployment, where traffic distributions and attacker behavior change~\cite{sommer2010outside}.
% %
% Dataset defects and collection artifacts can further inflate performance: analyses of CICIDS2017 identify errors in flow construction, labeling, and capture that materially affect evaluation~\cite{engelen2021troubleshooting,lanvin2022errors}.
% %
% More broadly, cross-dataset experiments show that models performing nearly perfectly within one dataset can approach chance-level accuracy on traffic collected in another network~\cite{cantone2024crossdataset}.
% %
% ML-based NIDSs are also exposed to feasible evasions that manipulate mutable traffic characteristics while preserving malicious functionality~\cite{he2023adversarial,wang2023robustness}.
% %
% These limitations motivate detection mechanisms whose decisions need not rest on one learned boundary.
% %%

ML approaches instead infer decision boundaries from packet bytes, flow statistics, or temporal behavior and can capture patterns that are cumbersome to specify manually~\cite{agiollo2023gnn4ifa,holland2021nprint,wei2023xnids,mirsky2018kitsune,bardhi2025mine}.
Yet, supervised models require labelled traffic that is costly and time-consuming to collect and annotate, particularly at the scale of operational networks~\cite{sommer2010outside,xu2024applying}.
Moreover, learned models may rely on dataset-specific correlations and fail under distribution shift, changing benign traffic, or previously unseen attacks~\cite{sommer2010outside,cantone2024crossdataset}.
Their limited transparency further complicates operational use since they provide little actionable evidence and post-hoc explanations fail to capture a meaningful rationale for the decision of the security analyst~\cite{wei2023xnids,subasi2024critical}. 
Finally, because an adversary can manipulate observable traffic characteristics while preserving malicious functionality, the learned decision boundary itself becomes a target for evasion~\cite{he2023adversarial}. 
Thus, although ML provides flexibility and powerful pattern-recognition capabilities, its limited generalization, explainability, and adversarial robustness prevent ML-based NIDS from constituting a complete solution. 

To overcome such limitations, recent studies have begun to explore neuro-symbolic intrusion detection~\cite{bizzarri2024synergistic,tran2025neurosymbolic}.
%
% Neuro-symbolic AI combines learned representations with explicit reasoning~\cite{marra2024survey}.
%
% DeepProbLog, in particular, embeds neural predicates in probabilistic logic programs, allowing neural outputs and symbolic rules to contribute to the same probabilistic inference~\cite{manhaeve2018deepproblog}.
% %
Neuro-symbolic approaches broadly integrate prior symbolic knowledge into neural learning by encoding logical constraints into the learning objective, compiling symbolic structure into neural architectures or coupling neural perception with discrete or probabilistic symbolic reasoning~\cite{marra2024survey,gibaut2023taxonomy}.
Existing work provides an initial exploration of neuro-symbolic NIDS, but remains limited in the scope of the symbolic knowledge and reasoning considered, offering neither a systematic procedure for constructing and validating expressive security rules nor a principled mechanism for resolving conflicts between neural and symbolic evidence.
%
% Thus, important challenges nevertheless remain in determining the origin and validity of symbolic knowledge, aggregating the evidence provided by multiple rules, and resolving conflicts between neural and symbolic predictions.
%

% This paper addresses these challenges through \name{}, a probabilistic neuro-symbolic NIDS built using DeepProbLog \cite{MANHAEVE2021103504}.
%
% DeepProbLog extends probabilistic logic programming with neural predicates, allowing the outputs of neural networks to participate as probabilistic facts in logical inference.
%
% Thus, \name{} treats the neural and symbolic components as distinct, potentially conflicting attack evidences.
%
This paper addresses these challenges through \name{}, a probabilistic neuro-symbolic NIDS which treats the neural and symbolic components as distinct, potentially conflicting attack evidences.
The neural component learns from packet-, flow-, or hybrid-level traffic and defines a neural predicate over benign and attack classes.
In parallel, the symbolic component evaluates interpretable predicates over traffic features.
\name{} constructs candidate rules automatically, identifying class-relevant features on held-out traffic data, deriving candidate predicates, and calibrating their behavior on traffic data.
% %
% \name{} assigns each rule an evidential role, depending on whether it supports or rejects evidence of an attack.
%
Lastly, \name{} combines neural and symbolic evidence through a \emph{coupling policy} that may enforce strict agreement between neural and logic or favor either component.

Relying on the neural symbolic weave, \name{} requires fewer data to optimize, as the knowledge encoded in the symbolic rules is excluded from the neural training process.
Additionally, \name{} generalizes well even under network or -- possibly adversarial -- traffic changes, as the symbolic guidance enables \name{} to ignore spurious signals coming from noisy data during learning.
Lastly, making its coupling policy explicit and relying on intelligible symbolic rules, \name{} exposes the supporting proof path for its predictions.
These proofs identify the satisfied predicates and rules supporting an alert, providing an explanation that is intrinsic to the model’s inference rather than generated by a separate post-hoc explainer.

This paper makes the following contributions:
\begin{itemize}
    \item We introduce \name{}, a Neuro-Symbolic NIDS combining neural attack evidence with symbolic reasoning over packet, flow, and hybrid traffic views.
    \item We design and develop a data-driven rule-construction pipeline that discovers and validate candidate rules without requiring any manual signature base.
    \item We formalize coupling policies that govern agreement and conflict between neural and symbolic evidence and generate proof-based explanations.
    \item We extensively evaluate \name{} through predictive performance, false alerts, cross-distribution generalization, explainability, and robustness to adversarial traffic.
\end{itemize}

The remainder of this paper is organized as follows. \Cref{sec:motivation_background} presents the paper motivation and the necessary background. \Cref{sec:deepweave} describes \name{} architecture,
while \Cref{sec:implementation} details its implementation and \Cref{sec:experiments} elaborates on its evaluation. Lastly, \Cref{sec:discussions_limitations} discuss limitations and \Cref{sec:related_work} the related work, while \Cref{sec:conclusions}
concludes the paper.

\section{Motivation and Background}
\label{sec:motivation_background}

\subsection{Motivating Rationale}
\label{sec:motivating_example}

\begin{figure}[h] 
\centering 
\begin{subfigure}[t]{0.49\columnwidth} 
\centering 
\includegraphics[ width=0.75\linewidth ]{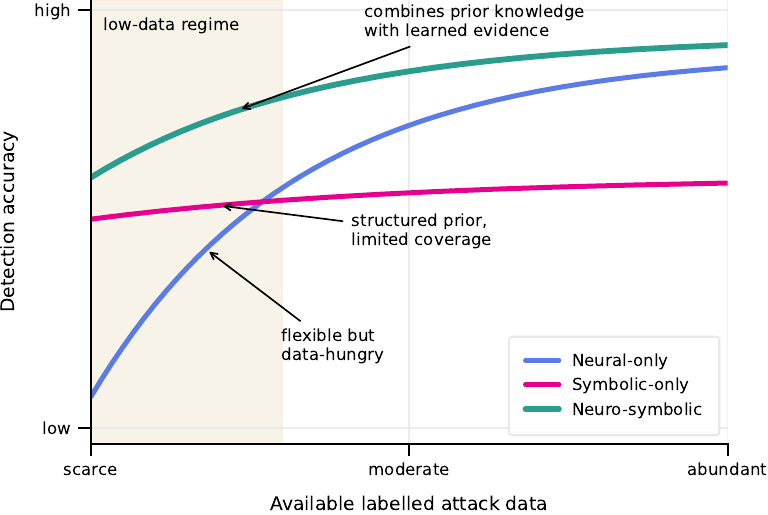} \caption{Detection accuracy as the amount of labelled data increases.} \label{fig:motivation_low_data} \end{subfigure} 
\hfill 
\begin{subfigure}[t]{0.49\columnwidth} 
\centering \includegraphics[ width=0.75\linewidth ]{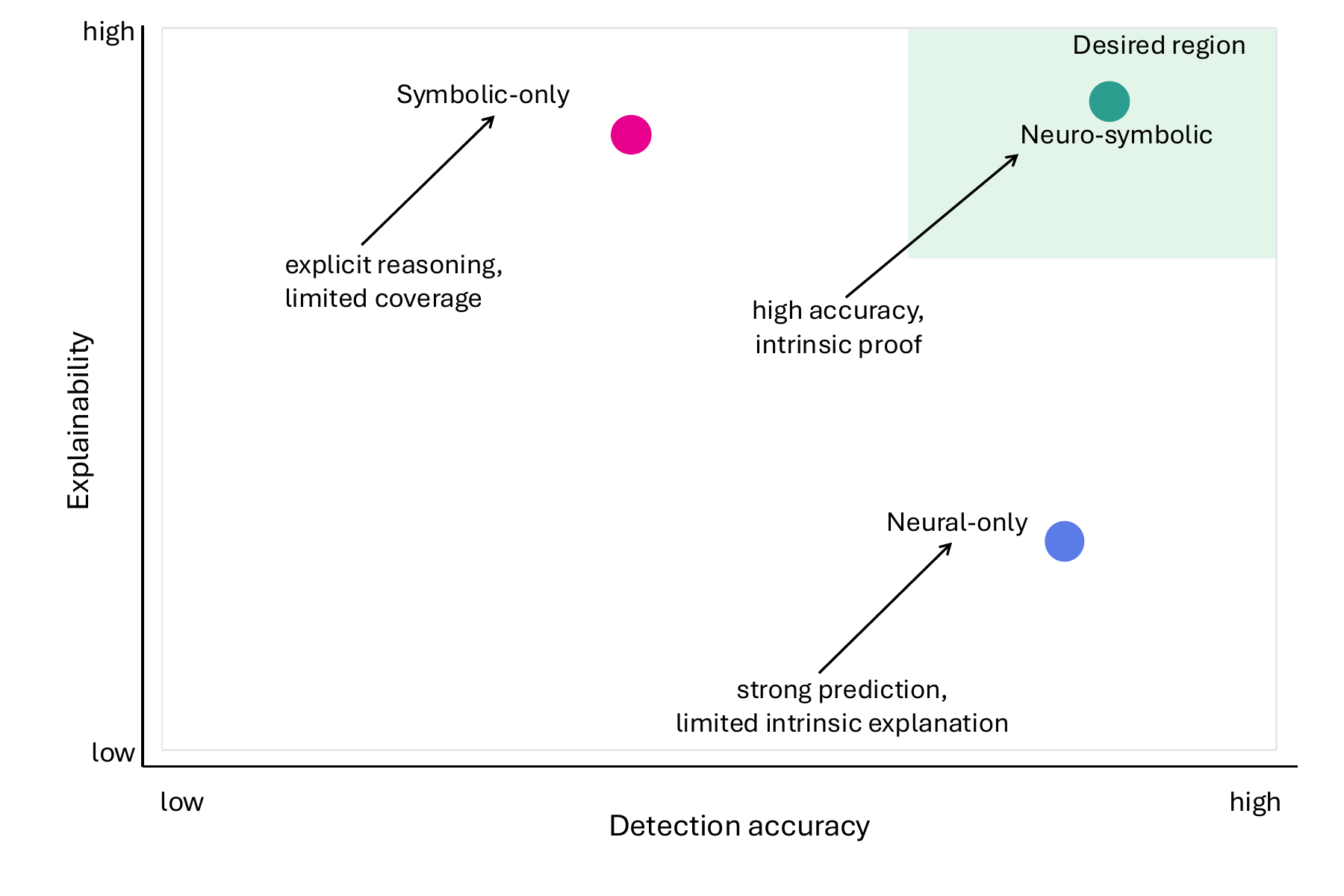} \caption{Trade-off between detection accuracy and explainability.} \label{fig:motivation_explainability} 
\end{subfigure} 
\caption{NeSy vs. ML-based and Symbolic NIDS.} 
%\caption{Motivation for neuro-symbolic intrusion detection.}
\label{fig:motivation_neurosymbolic} 
\end{figure}

\ourparagraph{Learning with scarce attack labels.} 
Operational NIDSs often deal with limited labelled attack traffic.
Although benign observations may be abundant, representative labels for each attack family are expensive to obtain and attacks are rare.
A practical NIDS must therefore learn from scarce attack labels.
Fitting dataset-specific correlations, ML-based detectors may fail to learn less frequent attack manifestations and their performance improves only when additional labels become available.
Anomaly-based ML detectors such as Kitsune~\cite{mirsky2018kitsune} reduce dependence on labelled attack data but still require the collection of large amount of benign traffic data to be optimized.
Meanwhile, symbolic detectors provide a complementary starting point as explicit conditions encoding attack-relevant properties require little data to be defined.
However, their accuracy plateaus as more data become available, since they require substantial effort to be maintained.
As Figure~\ref{fig:motivation_low_data} illustrates, a neuro-symbolic detector can instead use symbolic knowledge as a structured prior in the low-data regime, while allowing its neural component to learn additional patterns as labelled evidence increases.

\ourparagraph{Accuracy and explainability.} 
\Cref{fig:motivation_explainability} illustrates a second limitation of state-of-the-art NIDSs.
ML-based systems may achieve strong detection accuracy, but fail to provide a security rationale for their prediction.
Post-hoc explainability methods can associate neural decisions with influential features~\cite{wei2023xnids}, yet the explanation is generated after inference and may not correspond to an explicit reasoning process used to produce the decision. 
Conversely, symbolic systems expose the exact conditions that triggered an alert, but their detection accuracy is bounded by rule coverage.
These limitations become more consequential when a detector moves from training to production.
Changes in topology, services, benign traffic, measurement processes, or attacker behavior can invalidate correlations learned by the neural model and signatures assumed by symbolic rules.
A neuro-symbolic detector can achieve high detection accuracy and intrinsic transparency, since the neural component provides flexible pattern recognition, while validated rules provide inspectable evidences of their decision.

\subsection{Background}
\label{sec:background}

\ourparagraph{Network Intrusion Detection Systems.}
A NIDS monitors traffic at network observation points to detect malicious
activity or security-policy violations
~\cite{KhraisatCybersecurity2019,liao2013intrusion}.
Signature-based NIDSs match traffic against explicit descriptions of known
attacks, providing inspectable evidence but requiring continued rule maintenance
and sufficient signature coverage
~\cite{hubballi2014false,kumar2012signature}.
Anomaly-based NIDSs instead learn expected traffic behavior and identify
deviations, potentially detecting previously unseen activity but also flagging
benign changes and providing limited explanations
~\cite{liao2013intrusion,aldweesh2020deep}.
Both paradigms may operate on individual packets, aggregated flows, or
packet--flow pairs.
Packet representations preserve local protocol and header evidence, while flow
representations summarize communication-level properties such as duration,
volume, directionality, timing, and transport flags. Instead, hybrid representations
combine the current packet with its evolving flow context.

\paragraph{Detection task}
We formulate intrusion detection as classification over a label set
\(\mathcal{Y}=\{y_1,\ldots,y_K\}\) containing one benign class and one or more
attack classes.
For a packet, flow, or packet--flow representation \(x\in\mathcal{X}\), a NIDS
parameterized by \(\theta\) estimates \(p_\theta(Y=y\mid x)\) and predicts
%
% \begin{equation}
    $\hat{y}_{\theta}(x)
    =
    \arg\max_{y\in\mathcal{Y}} p_{\theta}(Y=y\mid x).$
% \end{equation}
%
An alert is raised when \(\hat{y}_{\theta}(x)\) belongs to an attack class.

\ourparagraph{Neuro-Symbolic AI.}
Neuro-symbolic AI (NeSy) encompasses a broad family of architectures that
combine the representation-learning capabilities of neural networks with
explicit knowledge representation and symbolic inference.
These architectures differ in how knowledge, learning, and reasoning interact
~\cite{marra2024survey}.
Symbolic knowledge may be translated into differentiable constraints that
regularize neural learning
~\cite{xu2018semanticloss,marra2019lyrics}, compiled into neural architectures
that approximate logical operations
~\cite{riegel2020logicalneuralnetworks}, or represented in an explicit reasoning
system whose predicates are partly evaluated by neural models
~\cite{manhaeve2018deepproblog,winters2021deepstochlog,yang2023neurasp}.
\name{} follows the last paradigm.
Neural models map unstructured inputs to distributions over symbolic values,
which subsequently participate in logical inference.
For example, a neural component may recognize a digit in an image, while
symbolic arithmetic rules determine how that digit contributes to the result of
a computation.

\paragraph{DeepProbLog}
We instantiate \name{} using DeepProbLog
~\cite{manhaeve2018deepproblog}, a neuro-symbolic language that extends the
probabilistic logic programming language ProbLog
~\cite{deRaedt2007problog} with neural predicates.
ProbLog represents uncertainty through probabilistic facts and annotated
disjunctions and computes query probabilities over the possible worlds induced
by these choices.
DeepProbLog preserves this semantics while allowing selected probabilities to
be produced by neural networks.

Given an input \(x\), a neural predicate produces a distribution over a finite
set of symbolic outputs \(z_1,\ldots,z_K\).
These outputs represent concepts that participate in inference together with
the program's facts and rules.
For example, the annotated disjunction
%
% \begin{equation}
$\mathtt{nn}(N_{\theta},x,\mathcal{Y})
\mathbin{::}
\mathtt{traffic\_class}(x,Y)$
% \end{equation}
%
uses \(N_\theta\) to assign probabilities to the possible traffic classes of
\(x\).
To answer a query, DeepProbLog grounds the relevant program, evaluates its
neural predicates, and compiles the resulting proof structure into an
arithmetic circuit
~\cite{manhaeve2018deepproblog}.
The query probability aggregates all valid proofs, including those containing
neural predictions and satisfied symbolic rules.
During end-to-end training, gradients of the query-level loss propagate through
probabilistic inference to the neural parameters, enabling learning without
direct supervision of the intermediate concepts \(z_1,\ldots,z_K\).

\section{\name{}}\label{sec:deepweave}

\subsection{Overview}
\label{sec:system_overview}
\begin{figure*}[!ht]
    \centering
    \includegraphics[width=0.85\linewidth]{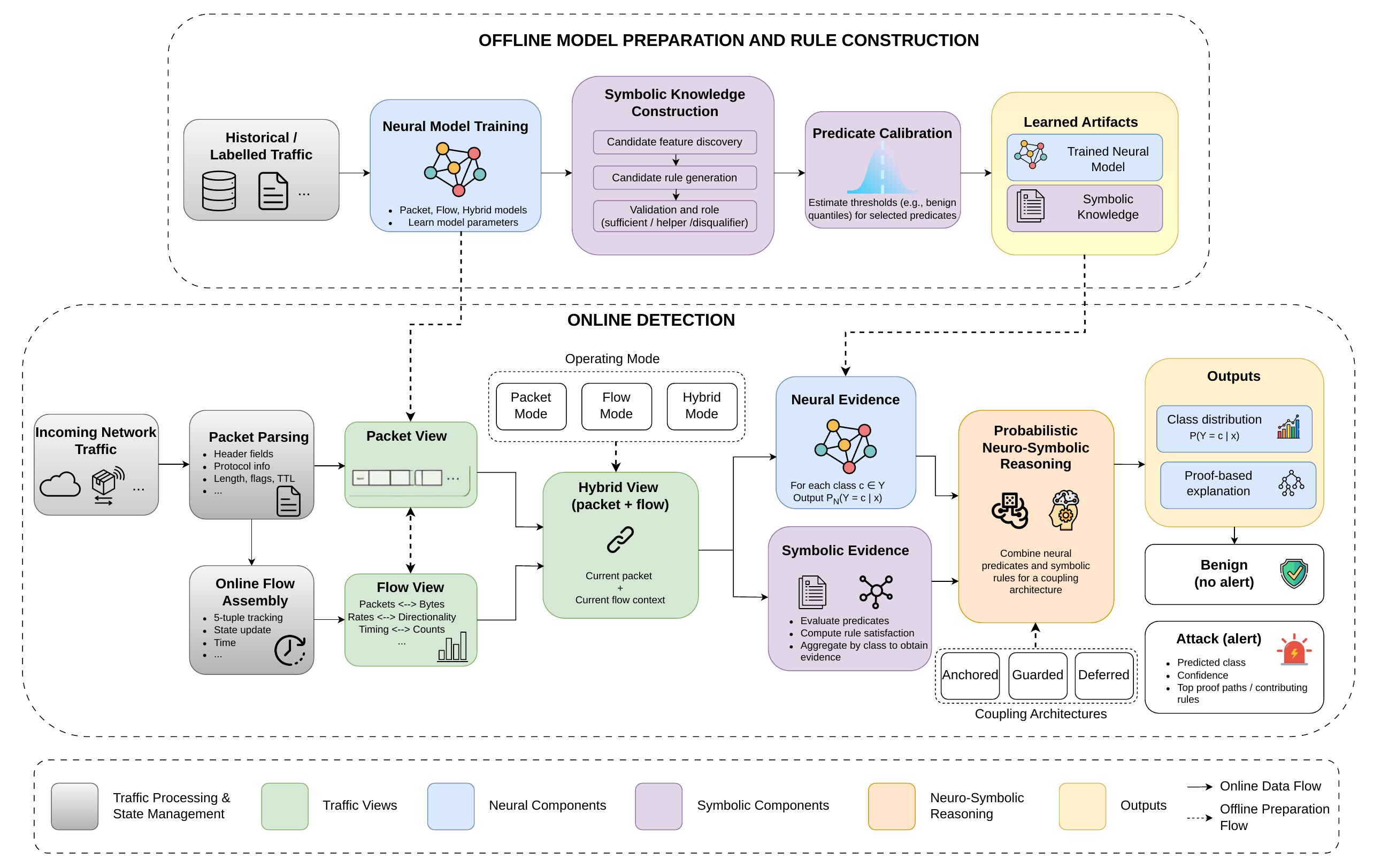}
    \caption{\name{} system overview. Incoming traffic is represented at the
    packet level, flow level, or jointly. Neural and symbolic paths derive
    complementary evidence from the selected representation. A probabilistic
    reasoning layer integrates this evidence to produce a class prediction and
    an accompanying symbolic explanation. Dashed arrows denote offline model
    preparation and rule construction; solid arrows denote online detection.}
    \label{fig:system_overview}
\end{figure*}

\name{} architecture is depicted in \Cref{fig:system_overview}.
\name{} operates in an offline preparation phase and an online detection phase.
During \emph{offline preparation}, the system constructs candidate symbolic rules, calibrates their predicates using reference traffic, and trains the neural and coupling architecture parameters.
The result is a detector containing a learned traffic model, a symbolic knowledge base, and the parameters governing their coupling.
%
% When the detector is deployed to a new network, environment-dependent predicates can be recalibrated using benign target traffic without reconstructing the underlying attack rules.
%
During \emph{online detection}, \name{} parses incoming traffic and transforms it into traffic view observations.
For each observation, the system constructs two complementary sources of evidence:
\begin{inlinelist}
    \item a neural path which learns statistical patterns from the traffic representation, and
    \item a symbolic path which evaluates attack-oriented predicates and rules.
\end{inlinelist}
A probabilistic neuro-symbolic reasoning layer combines these evidences to produce the final decision. 
Lastly, \name{} returns both an alert and its symbolic proof path to support its decision whenever it detects a malicious.

\subsubsection{Traffic views} \label{ssec:traffic_views} 

\name{} supports packet, flow, and hybrid traffic views.
Views vary in the available traffic context and symbolic knowledge, but share the same neuro-symbolic inference pipeline. 
The \emph{packet view} enables immediate prediction, but may miss behaviors that emerge only across multiple packets or connections. 
The \emph{flow view} aggregates packets from the same communication to capture temporal, directional, and rate-based behavior.
The \emph{hybrid view} jointly reasons over packets and their evolving flow context, allowing cross-view relationships to participate directly in inference rather than combining independent predictions.
We provide the complete formulation of the traffic views representation in~\Cref{sec:traffic_representation}.

\subsubsection{Neuro-Symbolic Detection}
\label{ssec:neurosymbolic_detection}

For each traffic observation, \name{} derives neural and symbolic evidence in parallel. 
The neural component learns traffic features patterns to output the candidate class probability, while the symbolic component evaluates rule-based attack oriented conditions. 
These conditions -- specified using rules -- relate observable traffic properties to higher-level attack behavior.
\name{} defines its symbolic conditions automatically during offline model preparation, where it constructs candidate rules, removes uninformative or redundant ones, and calibrates them (see \Cref{sec:symbolic_knowledge_representation,sec:symbolic_knowledge_construction}). 
%
% \name{} separates general attack behaviors from environment-specific thresholds.
%
The rules are structured to capture behavioral relationships associated with an attack, whereas they are calibrated to reflect the normal operating conditions of a particular environment. 
For example, a SYN-flood rule may detect repeated connection attempts that do not complete the TCP handshake, while the number of attempts required to trigger an alert must be calibrated to the network normal connection rate.
Thus, \name{} can apply the same rules across different networks and adapt them as network conditions change.

Rather than treating the neural and symbolic components as independent detectors, \name{} combines them inside a shared probabilistic reasoning model, which triggered the name ``Weave''.
This combination is defined by a coupling architecture (see~\Cref{sec:coupling_architectures}) which determines how neural and symbolic evidence interact during inference. 
Following the previous example, the neural network may predict the probability that an observed packet is a TCP SYN request, while additional symbolic predicates describe whether its flow contains a high rate of SYN packets and many incomplete handshakes. 
The coupling architecture then relates these uncertain observations to the SYN-flood class.
A strict coupling may require agreement between neural predictions and symbolic conditions, whereas a more selective coupling may invoke symbolic reasoning only to refine uncertain neural predictions.
These alternatives expose different trade-offs among detection coverage, false alarms, and robustness which we explore experimentally in~\Cref{sec:experiments}.
Lastly, when the selected coupling architecture produces an attack class prediction, \name{} raises an alert. 
In addition to the predicted class and its probability, \name{} exposes its reasoning process via the symbolic proof paths that contributed to the prediction. 
These proofs identify the satisfied predicates and rules supporting an alert, providing an explanation that is intrinsic to the model's inference rather than generated by a separate post-hoc explainer (see \Cref{sec:proof_explanations}).
%
% The proof path for the running example on SYN-flood is elaborated in~\Cref{secapp:running_example}.

\subsection{Design}\label{sec:design_details}
We first define the common representation through which packet, flow, and hybrid observations are supplied to the neural and symbolic components.
We then formalize the structure of the symbolic rules and describe the procedure used to construct them automatically.
Finally, we present the neural evidence, its integration into \name{}, and the coupling architectures that combine neural and symbolic evidence.

\subsubsection{Traffic Representation}
\label{sec:traffic_representation}
Let $p_t$ denote a packet observed at time $t$, and let $F_t$ denote the state of its associated flow.
The packet, flow, and hybrid observations are represented as
\begin{equation}
    x_t^P = \psi_P(p_t),
    \quad
    x_t^F = \psi_F(F_t),
    \quad
    x_t^H = \left(x_t^P,x_t^F,z_t^F\right),
    \label{eq:traffic_views}
\end{equation}
where $\psi_P$ and $\psi_F$ are packet-level and flow-level feature extractors.
The variable $z_t^F$ indicates whether the flow context required by the hybrid representation is available.
For the remainder of this section, $x^v$ denotes an observation under view $v\in\{P,F,H\}$.

\ourparagraph{Packet representation and flow assembly.}
The packet representation $x_t^P$ contains features such as network- and transport-layer protocol fields, packet length, flags, and other header-derived properties.
Packets are assigned to flows using a communication key derived from their network and transport endpoints and protocol.
For TCP and UDP traffic, this key corresponds to the conventional five-tuple: source address, destination address, source port, destination port, and transport protocol.
When bidirectional tracking is enabled, the endpoint ordering is normalized so that packets exchanged in either direction update the same flow.
Lastly, rather than recomputing the flow representation from all preceding packets, \name{} incrementally maintains the statistics required by $\psi_F$.
For instance, these statistics include packet and byte counts, communication direction, inter-arrival times, and packet-length distributions (for the full list see~\Cref{secapp:list_features}).
To allow hybrid detection to operate before the flow completion, neural or symbolic operations that depends on unavailable flow context -- i.e., $z_t^F=0$ -- are not evaluated.

\ourparagraph{Neural and symbolic feature interfaces.}
The neural and symbolic components access the same underlying observation through different feature interfaces.
The neural component receives an encoded representation $\eta_v(x^v)$, in which numeric features are standardized while categorical features are transformed into numeric encodings.
The symbolic component receives the semantically preserved feature values $v_f(x^v)$, where $v_f$ returns the original value of feature $f$ for input $x^v$.
This separation prevents neural preprocessing from altering the meaning of a symbolic condition.
For example, a rule referring to a packet-length threshold is evaluated in the original unit of the feature rather than against its standardized neural encoding.

\subsubsection{Symbolic Knowledge Representation}
\label{sec:symbolic_knowledge_representation}
The symbolic component operates over a vocabulary of predicates and a set of class-specific rules with their assigned roles.
The design does not require the rules to originate from a particular source.
They may be written by experts, imported from an external knowledge source, or constructed automatically.
In the evaluated system, \name{} uses the automated construction procedure described in \Cref{sec:symbolic_knowledge_construction}.
We use automatic construction because expert knowledge is not generally available in the probabilistic rule representation required by \name{}~\cite{wei2023xnids,wang2025captain}, while data-driven construction derives and calibrates rules directly from the features observable to the detector.
We then manually remove rules tied to capture-specific identifiers or known benchmark artifacts, reducing the risk of learning non-transferable correlations~\cite{sommer2010outside,cantone2024crossdataset}.

\paragraph{Symbolic Predicates.}
A predicate evaluates an interpretable condition over one feature or a small group of related features.
For a numeric feature $f$, \name{} supports upper-tail and lower-tail conditions:
\begin{align}
    \phi^{\uparrow}_{f,\tau}(x^v)
    &=
    \mathbb{I}\left[v_f(x^v)\geq\tau\right],
    \label{eq:upper_predicate}
    \\
    \phi^{\downarrow}_{f,\tau}(x^v)
    &=
    \mathbb{I}\left[v_f(x^v)\leq\tau\right],
    \label{eq:lower_predicate}
\end{align}
where $\tau$ is a threshold, $\mathbb{I}[\cdot]$ is the indicator function, and the symbols $\uparrow$ and $\downarrow$ identify whether unusually high or low values satisfy the predicate.
To further enhance the expressiveness of \name{} symbolic predicates and avoid the noisy character of pure thresholding, we also define their differentiable extension.
Rather than changing abruptly at a threshold, differentiable predicates provide a smooth transition near the predicate boundary leveraging a logistic sigmoid function.
The steepness of the logistic sigmoid is optimized through the neuro-symbolic training objective (see \Cref{sec:learning}).
Thus, we extend \Cref{eq:upper_predicate,eq:lower_predicate} to 
\begin{align}
    \phi^{\uparrow}_{f,\tau}(x^v)
    &=
    \sigma\left(
        \kappa_i(v_f(x^v)-\tau)
    \right),
    \label{eq:upper_predicate_sigmoid}
    \\
    \phi^{\downarrow}_{f,\tau}(x^v)
    &=
    \sigma\left(
        \kappa_i(\tau-v_f(x^v))
    \right),
    \label{eq:lower_predicate_sigmoid}
\end{align}
where $\sigma$ is the logistic sigmoid and $\kappa_i>0$ is the learnable parameter that controls the steepness.
For categorical features, predicates have the form
\begin{equation}
    \phi_{f,\mathcal{C}}(x^v)
    =
    \mathbb{I}\left[v_f(x^v)\in\mathcal{C}\right],
    \label{eq:categorical_predicate}
\end{equation}
where $\mathcal{C}$ is a set of values that the categorical feature can take.
Lastly, \name{} assigns a value $p_{\theta} \in [0,1]$ expressing the probability to which the observation satisfies the predicates $\phi^{\uparrow}_{f,\tau}$, $\phi^{\downarrow}_{f,\tau}$, $\phi_{f,\mathcal{C}}$.
$p_{\theta}$ is optimized during end-to-end training of the DeepProbLog program as it depends on the available traffic data (see \Cref{sec:learning}).

\paragraph{Symbolic Rules.}
A symbolic rule combines predicates into a class-specific condition.
A rule $r_j^c$ referring to class $c$ has the general form
\begin{equation}
    r_j^c(x^v)
    =
    \bigwedge_{\phi_i\in\Phi_j}
    \phi_i(x^v),
    \label{eq:symbolic_rule}
\end{equation}
where $\Phi_j$ contains the predicates used by the rule.
A rule therefore specifies three elements:
\begin{inlinelist}
    \item the class $c$ to which the rule refers;
    \item the predicates that form its antecedent; and
    \item the evidential role assigned to the rule.
\end{inlinelist}
The class-specific interpretation is important since a condition that supports one attack class does not necessarily provide evidence against every other class.

\paragraph{Rule Roles.}
Each rule $r_j^c$ receives one of three roles: 
% \ebtodo{make sure that these rule roles names are the same in all plots/tables}
%
\begin{description}
    \item[Sufficient:]
    a high-precision rule with a sufficiently low false-positive rate on benign traffic data, which identifies its associated attack class independently of the neural prediction.
    These resemble the type of detection rules that a signature-based NIDS would use in practice.
    \item[Helper:]
    identifies behavior associated with an attack, while not being sufficiently selective to determine the class independently.
    It may reinforce compatible neural evidence or challenge an uncertain benign prediction.
    \item[Disqualifier:]
    captures a condition that is inconsistent with a candidate class and blocks the assignment of prediction to the class they refer to.
\end{description}
The role assigned to a rule is distinct from its semantic, i.e., predicates and thresholds.
The semantic condition determines when the rule is satisfied, while its role determines how a coupling architecture is permitted to use it.
This separation allows the same rule to be used under different coupling architectures without modifying its conditions.

\subsubsection{Symbolic Knowledge Construction}
\label{sec:symbolic_knowledge_construction}
The automated construction procedure transforms labelled traffic into a set of symbolic rules following the pseudo-algorithm detailed in~\Cref{secapp:rule_algorithm}.
For each attack class $A$, it performs the following steps:
\begin{inlinelist}
    \item identify features that distinguish $A$ from benign traffic;
    \item instantiate predicates from the selected features;
    \item combine compatible predicates into candidate rules;
    \item validate the candidate rules on held-out construction data; and
    \item assign a role to each retained rule.
\end{inlinelist}
The automated procedure can optionally be followed by expert review to further remove statistically retained rules that are semantically implausible, operationally unsafe, or likely to encode artifacts of the traffic collection process.
The rule proposal and validation steps use disjoint traffic.
A construction partition is used for feature discovery, predicate instantiation, and candidate generation, while a separate validation partition is used for rule acceptance and role assignment.
In our experiments this traffic data belong to neither training nor testing set to avoid information leakage and performance overestimation.

\ourparagraph{Attack-Specific Feature Discovery.}
For each attack class $A$, the construction procedure compares the distribution of every candidate feature with its corresponding distribution in benign traffic.
This comparison identifies features whose values differ systematically between the attack and benign observations.
Additionally, for numeric features, \name{} records the directionality of such difference---i.e., if the attack class $A$ distribution contains values that are higher or lower than the benign ones. 

Both numeric and categorical features are eventually ranked.
For the former \name{} uses complementary measures of distributional separation, effect magnitude, ordering, shape, and univariate discrimination.
Instead, the latter are ranked using distributional distance, statistical association, and smoothed category-specific attack rates.
The evaluated measures include the Kolmogorov-Smirnov statistic~\cite{hodges1958smirnov}, Cliff's delta~\cite{cliff1993dominance}, Hellinger distance~\cite{gibbs2002probability}, and the univariate ROC-AUC for class discrimination~\cite{fawcett2006roc}.

Let $\mathcal{M}(f)$ be the ranking measures applicable to feature $f$.
After normalizing the individual measures, their values are aggregated into the attack-specific feature score
\begin{equation}
    S(f,A)
    =
    \frac{
        \sum_{m\in\mathcal{M}(f)}
        w_m\widehat{m}(f,A)
    }{
        \sum_{m\in\mathcal{M}(f)}w_m
    },
    \label{eq:feature_ranking}
\end{equation}
where $\widehat{m}(f,A)$ is the normalized value of measure $m$ for feature $f$ and attack class $A$, and $w_m$ is the configured weight of that measure.

\ourparagraph{Predicate Instantiation.}
\name{} selects the top-K features to transform them into candidate predicates of the forms defined in \Cref{eq:upper_predicate_sigmoid,eq:lower_predicate_sigmoid,eq:categorical_predicate}.
For a numeric feature, the directionality established during feature discovery determines whether \name{} constructs upper-tail ($\phi^{\uparrow}_{f,\tau}$) or lower-tail ($\phi^{\downarrow}_{f,\tau}$) predicates.
Candidate thresholds are obtained from selected quantiles of the benign feature distribution.
More in detail, let
\begin{equation}
    \tau_{f,q}
    =
    Q_q\left(f\mid Y=\mathit{benign}\right)
    \label{eq:benign_quantile}
\end{equation}
be the $q$-th quantile of numeric feature $f$ in benign reference traffic, where $q$ belongs to a user-specific set of quantiles.
\name{} then defines a candidate upper-tail (lower-tail) predicate following \Cref{eq:upper_predicate_sigmoid} (\Cref{eq:lower_predicate_sigmoid}) using $\tau_{f,q}$ as the threshold.
These directional predicate instantiation procedure allows \name{} to maintain flexibility.
The feature $f$ and directionality encode the behavioral relationship associated with an attack, while the threshold $\tau_{f,q}$ determines what constitutes an unusual value in a particular environment.
Therefore, \name{} can preserve a predicate such as ``unusually high flow rate'' while recalibrating its threshold from benign traffic whenever changes in the network occur.

For each categorical feature \(f\), \name{} constructs predicates of the form of \Cref{eq:categorical_predicate}, by enumerating all the values that \(f\) takes in the construction data.
Values with insufficient target-class support or excessive prevalence in non-target traffic are discarded. 

\ourparagraph{Candidate Rule Generation.} 
For each attack class $A$, \name{} combines its most promising atomic predicates into logical conjunctions of the form shown in \Cref{eq:symbolic_rule}. The maximal size of these conjunctions are user-defined. 
The search prioritizes predicates that provide sufficient support and discriminative performance, and avoids redundant combinations. 
Lastly, candidate conjunctions that lack sufficient support are discarded.
Restricting rules to a small number of predicates controls the search space, preserves interpretability, and reduces the risk of capturing accidental relationships specific to the construction data.
The search bounds and support requirements are configurable and their evaluated settings are reported in~\Cref{secapp:candidate_generation_config}. 
% \ebtodo{this seems to be duplicated wrt hyperparameters in the same section}

\ourparagraph{Rule Validation and Role Assignment.}
Each candidate rule is evaluated on the construction and validation partitions.
For a rule associated with attack class $A$, \name{} measures its precision and benign false-positive rate on both partitions, as well as its attack recall on the validation partition. 
The comparison between the two partitions acts as a stability check.
A rule is rejected when its precision deteriorates substantially on validation data or when its benign false-positive behavior does not remain within the permitted tolerance. 
The remaining rules are assigned a role according to their validated behavior. 
A rule with high attack precision and a very low benign false-positive rate is retained as a \emph{sufficient} rule. 
A rule that covers a sufficiently large portion of $A$ while maintaining an acceptable benign false-positive rate is retained as a \emph{helper} rule. 
Finally, a condition that identifies a class different from $A$ with high precision and low benign false-positive rate is retained as a \emph{disqualifier} for class $A$.
Candidates that satisfy neither the stability requirement nor the role-specific validation criteria are discarded. 
The exact hyperparameters of validation criteria together with candidate quantiles and role-assignment thresholds are reported in~\Cref{secapp:hyperparameter_settings_for_rule_construction}.

\subsubsection{Neural Evidence Interface}
\label{sec:neural_evidence}
The neural component maps the observation $\eta_v(x^v)$ -- encoded under view $v$ -- to a categorical distribution over the set of classes $\mathcal{Y}$:
\begin{equation}
    \mathbf{n}^v(x^v)
    =
    h_v\left(\eta_v(x^v);\theta_v\right),
    \label{eq:neural_evidence}
\end{equation}
where $h_v$ is a trainable classifier with parameters $\theta_v$.
The output expresses the neural evidence -- i.e., probability -- for class $c\in\mathcal{Y}$ to be true.
In this paper, we consider $\mathcal{Y}$ to be the set of attack classes $A_1, \ldots, A_k$ and the benign class.
Nevertheless, \name{} design is general and supports any output class that can be defined to enable the identification of attacks---e.g., the neural component can predict the probability that an observed packet is a TCP SYN request.
Lastly, the neural outputs are to be interpreted as probabilistic inputs to the reasoning program through the coupling architectures.
They are not the final \name{} prediction.

\name{} design does not depend on a particular neural architecture.
It requires only that the model produces a probability for each value in the class vocabulary and that these probabilities can be exposed through a neural predicate.
The concrete architecture and training configuration used in the evaluation are reported in \Cref{secapp:optimization_hyperparameters}.

\subsubsection{Coupling Architectures}
\label{sec:coupling_architectures}
A coupling architecture specifies which evidence combinations constitute successful proofs of an attack class.
The architectures share the same neural predicates and symbolic rules but differ in the authority assigned to the two evidence sources.
\name{} characterizes a coupling architecture according to two properties:
\begin{inlinelist}
    \item whether symbolic evidence can independently support a class or must accompany neural evidence, and
    \item whether symbolic reasoning is applied to every observation or only when a condition on the neural prediction is satisfied.
\end{inlinelist}
Considering these two properties, \name{} instantiates the following coupling architectures.

\ourparagraph{Guarded.}
requires class-matched agreement between the neural and symbolic components for every attack prediction. 
Let \(N^c\), \(S^c\), and \(H^c\) denote the probabilistic outcomes of the neural prediction, sufficient rules, and helper rules for attack class \(c\), respectively. 
The attack-class support is then defined as \(C_c=N^c\land(S^c\lor H^c)\), whereas benign support is \(C_b=N^b\land\neg\bigvee_{j\neq b}S^j\).
Thus, sufficient and helper rules may corroborate a neural attack prediction, but neither can independently assign an attack class. 
Only the higher-precision sufficient rules may challenge a neural benign prediction by suppressing it. 
Guarded coupling therefore filters unsupported neural attacks and false benign predictions while avoiding independent symbolic classification.

\ourparagraph{Anchored.}
generates classification support in one of two ways. 
First, a satisfied rule $S^c$ with sufficient role can independently support the class $c$. 
Alternatively, classification support can arise from the conjunction of a neural prediction \(N^c\) and the aggregate satisfaction of the class-specific rules with helper roles \(\bigwedge_j H^c_j\). 
In both cases, the resulting support is accepted as valid only when no disqualifier rule is valid. 
Consequently, sufficient rules can recover cases missed by the neural model, whereas less decisive helper rules must be corroborated by neural evidence, and all candidate classifications remain subject to symbolic rejection.

\ourparagraph{Deferred.}
relies mainly on the neural component and invokes symbolic rules only when the neural model is insufficiently decisive.
Let $\Delta(x^v)$ be the difference between the largest and second-largest class probabilities predicted by the neural component.
When $\Delta(x^v)\geq\tau_{\Delta}$, the Deferred coupling leverages the neural prediction without symbolic intervention, thus being equivalent to pure neural inference.
Meanwhile, when $\Delta(x^v)<\tau_{\Delta}$, the Deferred coupling reverts to predicting the attack class while leveraging the neural component together with sufficient, helper and disqualifier rules.
The intuition behind this coupling is that the margin $\Delta(x^v)$ is a measure of relative neural decisiveness which identifies when neural inference alone is sufficient to predict an attack instance.
Lastly, the threshold $\tau_{\Delta}$ determines the fraction of observations for which symbolic intervention is activated and thus is a hyperparameter that allows to balance the neural-symbolic trade-off.

\ourparagraph{Hard Gate.}
predicts a class $c$ only when both the neural and symbolic -- either sufficient or helper rules -- components provide evidence for $c$.
Thus, this coupling policy represents a logic \textsc{AND} gate between the neural and symbolic components and the stricter policy for attack detection.

\ourparagraph{Same-Class Boost.}
relaxes the Hard Gate requirements, adding the probability of sufficient symbolic evidence to the neural evidence, without requiring explicit agreement to support the same class. 
Therefore, the role of the symbolic component is to reinforce uncertain neural predictions when the defined rules are consistent with the information provided by the input.

\ourparagraph{Safety Net.}
predicts an attack class $c$ when either the neural or symbolic -- either sufficient or helper rules -- components provide evidence for $c$.
Therefore, this coupling represents the logical complement of the Hard Gate policy, giving the symbolic component the greatest independence.

\ourparagraph{Product of Experts.}
applies specifically for hybrid traffic view.
It predicts an attack class $c$ when neural agreement is achieved between both neural components -- i.e., packet- and flow-view -- or when the symbolic evidence coming from sufficient rules $S^c$ is satisfied.
Thus, the Product of Experts coupling favors classes jointly supported by both components, while not being as strict as Hard Gate.

\subsubsection{Learning and Inference}
\label{sec:learning}

For every class $c\in\mathcal{Y}$, \name{} evaluates the probability of class $c$ being true using the DeepProbLog program $\mathcal{P}_{\Gamma}$ which implements the coupling architecture $\Gamma$, the symbolic rules $r_j^c$ and the neural component $h_v$.
Within DeepProbLog, the class probability is obtained identifying a successful proof of the query $\operatorname{classify}(x^v,c)$, which asks what combination of neural and symbolic evidence permitted by $\Gamma$ leads to class $c$ being true and the associated probability.
Several proof paths may support the same class and DeepProbLog grounds the dependencies among probabilistic predicates before aggregating the probabilities of the successful proofs to obtain the final probability of class $c$.

For end-to-end training, \name{} minimizes the negative log-probability of the ground-truth class query $\mathcal{L}_{NS}(x^v,y) = -\log P_{\mathcal{P}_{\Gamma}} \left(\operatorname{classify}(x^v,y)\right)$.
Gradients propagate through the probabilistic program to the neural, predicate, and any learnable coupling parameters.
The logical rule structure, selected features, and semantic directionality of each predicate remain fixed after symbolic knowledge construction.
Meanwhile, environment-dependent thresholds are calibrated separately from the learned model parameters.
For example, for a threshold represented by benign quantile $q$, recalibration consists of estimating \Cref{eq:benign_quantile} from benign reference traffic in the target environment.
This changes the numeric boundary without changing the feature, directionality, associated class, or logical structure of the rule.

\subsubsection{Proof-Based Explanations}
\label{sec:proof_explanations}
For a predicted class, \name{} retains the successful proofs contributing to its class-query probability.
Therefore, a proof records:
\begin{inlinelist}
    \item the neural predicates used by the proof,
    \item the symbolic predicates satisfied by the observation,
    \item the feature values and thresholds involved in those predicates,
    \item the rules constructed from the predicates,
    \item the role of each contributing rule, and
    \item the coupling through which the evidence supported the class.
\end{inlinelist}
A proof can therefore distinguish, for example, between a sufficient rule that independently supported an attack, a helper rule that reinforced a neural prediction, and an attack signal that prevented acceptance of a benign prediction.
Meanwhile, the proof probability determines the contribution of a path to the class query.
When several proofs support the predicted class, \name{} ranks them according to their contribution and reports the most influential paths.
A visual example of one such explanation is provided in \Cref{sec:explanation_results}.

\section{Implementation}
\label{sec:implementation}
We implement \name{} in Python using PyTorch~\cite{paszke2019pytorch} and
DeepProbLog~\cite{MANHAEVE2021103504}.
%
% We release the full open source \name{} implementation at \href{https://anonymous.4open.science/r/DeepWeave-AAB2}{https://anonymous.4open.science/r/DeepWeave-AAB2}.

%while experiment-specific details such as training budgets, stopping criteria, random seeds, and evaluation hardware are reported in~\Cref{secapp:optimization_hyperparameters}.
%

\ourparagraph{Neural models.}
The neural components of the packet- and flow-view \name{} models are implemented as feed-forward Multi-Layer perceptrons (MLP) with identical hidden-layer architectures and view-specific input dimensions (see~\Cref{secapp:optimization_hyperparameters}).
In our experiments, we define a neural-only baseline which uses the same classifier as the corresponding neuro-symbolic model to isolate the effect of neuro-symbolic integration on attack detection performance.
In hybrid-view operation, \name{} instantiates two feed-forward MLPs as the packet and flow classifiers which remain separately accessible to the probabilistic program, allowing symbolic rules to distinguish packet- and flow-level evidence.

\ourparagraph{DeepProbLog Optimizations.} 
Each query carries a tensor handle that identifies the traffic view and observation, allowing neural and symbolic predicates to access their respective encodings through a shared reference. 
During training, \name{} evaluates the ground-truth class query and minimizes its negative log-probability.
Instead, during inference, it evaluates all candidate classes and selects the most probable one.
A naive implementation would ground and compile a separate arithmetic circuit for every observation and class query. 
Here, an arithmetic circuit is the compiled computation graph of a probabilistic query.
Its leaves contain neural or symbolic probabilities, while its internal operators combine them according to the program's logical structure.
Instead, we reformulate queries so that observations share the same symbolic structure and differ only through tensor substitutions. 
The resulting circuits are compiled once using placeholder handles and reused across observations. 
A batched graph semiring then supplies observation-dependent neural and symbolic probabilities as leaf vectors and evaluates each shared circuit elementwise over the batch.
We further reduce substitution overhead through direct placeholder lookup and reuse of input templates across neural predicates.
Together, these optimizations enable batch-level inference and produce a \(26.8\times\) speedup in our development benchmark without changing predicted labels.
Their impact on end-to-end latency is reported in ~\Cref{secapp:deepweave_batch_optimization}. 
\section{Experiments and Results}
\label{sec:experiments}
We evaluate \name{} via detection performance and efficiency under low-data regimes (\Cref{sec:detection_results}), generalization capability (\Cref{sec:generalization}), robustness to adversarial perturbations (\Cref{sec:adversarial_robustness}), explainability (\Cref{sec:explanation_results}), and computational efficiency (\Cref{sec:computational_efficiency}).
Unless stated otherwise, all methods rely on the same task definition, data preprocessing and partitions.
To avoid information leakage across \name{} optimization and testing, the test set is not used for symbolic knowledge construction, threshold calibration, hyperparameter selection, or model selection.

\subsection{Experimental Settings}\label{sec:experimental_settings}

% \subsection{Experimental Settings}
% \label{sec:experimental_settings}
% %
% This subsection defines the protocol shared across the detection,
% generalization, robustness, explanation, and ablation experiments.
% %
% Unless stated otherwise, all methods compared within an experiment use the same
% task definition, preprocessing state, training subsets, validation data, and
% held-out test data.
% %
% The test set is not used for preprocessing, symbolic knowledge construction,
% threshold calibration, hyperparameter selection, early stopping, or model
% selection.
% %

\ourparagraph{Datasets and tasks.}
The CIC-IDS2017~\cite{cic2017} dataset is used for \name{} flow-view testing, as it represents the state-of-the-art benchmark for NIDS.
Instead, for the packet- and hybrid-view testing we run experiments on the NF-CSE-CIC-IDS2018-v3 dataset~\cite{sarhan2020netflow} since it enables the reconstruction of flow features from packet captures using NetFlow-v3~\cite{claise2004cisco}.
Furthermore, we test \name{}'s generalizability capabilities training it on one of the above mentioned datasets and testing it on the NF-ToN-IoT-v3 dataset~\cite{luay2025nftoniotv3}.
This dataset is selected as it provides a distinct Internet-of-Things (IoT) environment which is radically different from the enterprise networks that characterize the other datasets.
Moreover, it shares the same NetFlow-v3 schema of NF-CSE-CIC-IDS2018-v3, ensuring that transferred rules retain the same feature semantics.
Lastly, in low-data experiments, we exclude classes with insufficient samples for reliable training and testing, mitigating instability caused by extreme class imbalance in the selected benchmarks~\cite{sommer2010outside,arp2022and}.

\ourparagraph{Data partitioning and preprocessing.}
Each dataset is divided into 60\%
training, 20\% validation, and 20\% test partitions using stratified sampling to maintain the natural classes distributions.
The validation and test partitions remain fixed across model variants and training-set sizes to ensure fairness.
For packet-view and hybrid-view experiments, we partition the data grouping them by flow so that all packets from the same communication remain in one partition and thus avoiding information leakage of packets coming from the same flow but belonging to train and test splits.
Where packet labels are derived by matching packets to labelled flow records, matching is completed before partitioning and unmatched or ambiguous packets are excluded.
Following established evaluation practice, we remove label-derived fields and capture-specific identifiers to prevent models from exploiting dataset-construction artifacts rather than attack behavior~\cite{arp2022and,engelen2021troubleshooting,sommer2010outside}.
IP addresses and timestamps are retained only when required for flow assembly or packet-flow association, while they are removed from the neural inputs.
Lastly, we standardize the input features only for the neural component, while symbolic predicates access feature values in their original format.
%

% The fixed prototype parameters are reported in
% Table~\ref{tab:implementation_config}.
%

\ourparagraph{Baselines.}
\label{sec:baselines}
We compare \name{} with Decision Tree, Logistic Regression, Random Forests, Histogram-based Gradient Boosting Classification Tree (HGB), an MLP having the same architecture of the \name{}'s neural component, a symbolic baseline containing only the symbolic rules in the \name{} model, and state-of-the-art compatible NIDSs, such as CNN-WDLSTM~\cite{hassan2020hybrid}, and SPHBI~\cite{elsherif2025sphbi} for packet-view, Kitsune~\cite{mirsky2018kitsune} and KnowML~\cite{guo2026knowml} for flow-view, and DST-IDS~\cite{QIU2022} for hybrid-view.
Kitsune is implemented both using its original AfterImage flow constructor and relying on the same flow features used by \name{}.
Meanwhile, KnowML is implemented in combination with the KitNet model used in the original paper.
KnowML and Kitsune are implemented in their original anomaly detection mode.
%
% We additionally compare with Suricata rules fetched from Emerging Threats Open Ruleset~\cite{proofpoint_etopen} and published repositories specific to each dataset. \ebtodo{check this with Raoul.}
% %
% Portability and robustness experiments focus on the matched MLP,
% symbolic-only model, KnowML, and alternative \name{} couplings, with PANTS~\cite{jin2025robustifying} informing the constraints used for adversarial traffic generation.\agtodo{revise afterwards}
% %
% Explanation experiments compare \name{}'s proof paths with neural
% attributions and, where compatible, xNIDS~\cite{wei2023xnids}.
%

\ourparagraph{Metrics.}
%
% \eb{Notes after the meeting: add ASR for the robustness experiment}
We measure detection performance using F1 scores and attack false positive rate (FPR).
For multi-class classification baselines, we consider Macro-F1 which assigns equal weight to every retained class.
Meanwhile, for the anomaly detection baselines, such a score represents the binary-F1.
Attack FPR is the fraction of benign observations classified as any attack.
Where relevant, we additionally report per-class precision and recall, PR-AUC and ROC-AUC.
For adversarial robustness experiments, we use metrics related to attack success rate (ASR) and attacked F1 that represents the F1 score on perturbed data.

%%%%%%%%%%%%%%%%%%%%%%%%%%%%%%%%%%%%%%%%%%%%%%%%%%%%%%%%%%%
\subsection{Detection Performance and Efficiency}
\label{sec:detection_results}

To measure \name{} detection performance and efficiency, we follow a low-data protocol where training subsets are parametrized by \(k\in\{50,500,1000,2500,5000,10000\}\). 
Each retained attack class contributes \(k\) labelled observations.
Benign observations are sampled to match the total attack count,
maintaining a controlled balance between benign and aggregate attack traffic.
For every value of \(k\), we draw five independent training subsets and
re-initialize each model from scratch.
Neural-only and neuro-symbolic models use matched subsets and random seeds.
Rather than reconstructing symbolic rules for each \(k\) batch, we derive them once from a larger historical knowledge pool.
The resulting rule base is fixed and shared across experiments, reducing construction cost and preventing batch-specific rule variation.

\ourparagraph{Packet-view.}
Results for the packet-view detection are presented in \Cref{tab:packet_sota_sweep_reduced} for $k\in\{50,1000,5000\}$, while the extended results are shown in \Cref{tab:packet_sota_sweep} of \Cref{secapp:all_results}.
When compared to the simple neural baseline, \name{} improves the F1 rate of up to 34.5\% at $k=50$ using the Anchored coupling architecture.
Similarly, \name{} reduces false-positive rate from 0.075 to 0.001.
Although CNN-WDLSTM~\cite{hassan2020hybrid} and SPHBI~\cite{elsherif2025sphbi} systems improve over the simple neural baselines, they underperformed compared to \name{} in low-data regime ($k=50$).
The \name{} Anchored coupling model achieves an improvement of $8.1\%$ in terms of F1, while reducing the false positive rate by 0.011 points.

As the available data grow in size, all data-driven baselines -- including the symbolic baseline that learn symbolic rules thresholds from data -- improve in performance with some of them achieving results similar to \name{}.
%
% Meanwhile, \name{} retains its detection performance virtually unmodified, showing the effectiveness of the neuro-symbolic coupling for extracting attack information when labeled data are scarse.
%
Across different values of $k$, \name{} shows high performance stability, highlighting the effectiveness of neuro-symbolic integration.
Lastly, \name{} shows substantial detection improvements over Suricata rules which become effective only when considering  custom rules defined using the rate and behaviour of the attack derived from the documented CIC-IDS2017 attack profiles---i.e., ET Open + published.
However, even custom made Suricata rules result in a high number of false positives compared to \name{}, showing the limitations of purely signature-based NIDS and confirming our intuition of \Cref{sec:motivating_example}.

\begin{table}[h]
\centering
\scriptsize
\setlength{\tabcolsep}{4pt}
\caption{Packet-view comparison on CIC-IDS2018, across per-class training sizes of $k{=}50$, $k{=}1000$ and $k{=}5000$. F1 indicates the macro-F1 for multi-class classification models and binary-F1 for Suricata. Both F1 and FPR are reported as average $\pm$ standard deviation. Best per column in bold.}
\label{tab:packet_sota_sweep_reduced}
\resizebox{\columnwidth}{!}{%
\begin{tabular}{l rr rr rr}
\toprule
& \multicolumn{2}{c}{${k{=}50}$}
& \multicolumn{2}{c}{${k{=}1000}$}
& \multicolumn{2}{c}{${k{=}5000}$}\\
\cmidrule(lr){2-3} \cmidrule(lr){4-5} \cmidrule(lr){6-7}
System & F1 & FPR & F1 & FPR & F1 & FPR\\
\midrule
Neural baseline
& $0.632{\scriptstyle\pm.044}$ & $0.075{\scriptstyle\pm.029}$
& $0.843{\scriptstyle\pm.024}$ & $0.020{\scriptstyle\pm.010}$
& $0.863{\scriptstyle\pm.019}$ & $0.013{\scriptstyle\pm.006}$\\
Symbolic baseline
& $0.725{\scriptstyle\pm.005}$ & $0.046{\scriptstyle\pm.000}$
& $0.834{\scriptstyle\pm.013}$ & $0.018{\scriptstyle\pm.013}$
& $0.857{\scriptstyle\pm.006}$ & $0.007{\scriptstyle\pm.000}$\\
Decision tree
& $0.688{\scriptstyle\pm.035}$ & $0.064{\scriptstyle\pm.019}$
& $0.776{\scriptstyle\pm.059}$ & $0.082{\scriptstyle\pm.127}$
& $0.881{\scriptstyle\pm.048}$ & $0.026{\scriptstyle\pm.017}$\\
Logistic regression
& $0.542{\scriptstyle\pm.016}$ & $0.318{\scriptstyle\pm.031}$
& $0.493{\scriptstyle\pm.016}$ & $0.362{\scriptstyle\pm.017}$
& $0.499{\scriptstyle\pm.017}$ & $0.366{\scriptstyle\pm.023}$\\
Random forest
& $0.884{\scriptstyle\pm.052}$ & $0.017{\scriptstyle\pm.010}$
& $0.924{\scriptstyle\pm.061}$ & $0.012{\scriptstyle\pm.021}$
& $\mathbf{0.995{\scriptstyle\pm.003}}$ & $\mathbf{0.000{\scriptstyle\pm.000}}$\\
HGB
& $0.811{\scriptstyle\pm.045}$ & $0.021{\scriptstyle\pm.013}$
& $0.857{\scriptstyle\pm.128}$ & $0.070{\scriptstyle\pm.123}$
& $0.943{\scriptstyle\pm.031}$ & $0.001{\scriptstyle\pm.001}$\\
\midrule
\name{} (Anchored)
& $\mathbf{0.977{\scriptstyle\pm.013}}$ & $\mathbf{0.001{\scriptstyle\pm.001}}$
& $0.969{\scriptstyle\pm.023}$ & $\mathbf{0.000{\scriptstyle\pm.000}}$
& $0.985{\scriptstyle\pm.007}$ & $\mathbf{0.000{\scriptstyle\pm.000}}$\\
\name{} (Guarded)
& $0.924{\scriptstyle\pm.009}$ & $0.008{\scriptstyle\pm.002}$
& $0.936{\scriptstyle\pm.026}$ & $0.001{\scriptstyle\pm.000}$
& $0.916{\scriptstyle\pm.084}$ & $0.001{\scriptstyle\pm.000}$\\
\name{} (Hard Gate)
& $0.823{\scriptstyle\pm.023}$ & $0.005{\scriptstyle\pm.002}$
& $0.931{\scriptstyle\pm.012}$ & $0.001{\scriptstyle\pm.000}$
& $0.927{\scriptstyle\pm.012}$ & $\mathbf{0.000{\scriptstyle\pm.000}}$\\
\name{} (Safety Net)
& $0.876{\scriptstyle\pm.018}$ & $0.030{\scriptstyle\pm.008}$
& $0.867{\scriptstyle\pm.045}$ & $0.018{\scriptstyle\pm.008}$
& $0.878{\scriptstyle\pm.046}$ & $0.012{\scriptstyle\pm.006}$\\
\name{} (Same-Class Boost)
& $0.644{\scriptstyle\pm.030}$ & $0.071{\scriptstyle\pm.017}$
& $0.838{\scriptstyle\pm.017}$ & $0.021{\scriptstyle\pm.009}$
& $0.822{\scriptstyle\pm.094}$ & $0.016{\scriptstyle\pm.015}$\\
\name{} (Deferred)
& $0.951{\scriptstyle\pm.019}$ & $0.007{\scriptstyle\pm.003}$
& $0.977{\scriptstyle\pm.003}$ & $0.001{\scriptstyle\pm.000}$
& $0.974{\scriptstyle\pm.002}$ & $\mathbf{0.000{\scriptstyle\pm.000}}$\\
\midrule
SPHBI~\cite{elsherif2025sphbi}
& $0.896{\scriptstyle\pm.019}$ & $0.012{\scriptstyle\pm.004}$
& $\mathbf{0.992{\scriptstyle\pm.005}}$ & $0.001{\scriptstyle\pm.000}$
& $0.992{\scriptstyle\pm.010}$ & $\mathbf{0.000{\scriptstyle\pm.000}}$\\
CNN-WDLSTM~\cite{hassan2020hybrid}
& $0.742{\scriptstyle\pm.030}$ & $0.034{\scriptstyle\pm.014}$
& $0.845{\scriptstyle\pm.017}$ & $0.008{\scriptstyle\pm.003}$
& $0.884{\scriptstyle\pm.016}$ & $0.007{\scriptstyle\pm.001}$\\
\midrule
Suricata (ET Open)$^\dagger$
& $0.449$ & $0.279$
& $0.449$ & $0.279$
& $0.449$ & $0.279$\\
Suricata (ET Open + published)$^\dagger$
& $0.624$ & $0.280$
& $0.624$ & $0.280$
& $0.624$ & $0.280$\\
\bottomrule
\end{tabular}}

\end{table}

\ourparagraph{Flow-view.}
Results for the flow-view detection are presented in \Cref{tab:flow_sota_sweep_reduced} for $k\in\{50,1000,5000\}$, while the extended results are shown in \Cref{tab:flow_sota_sweep} of \Cref{secapp:all_results}.
We implement Kitsune using both the same flow features used by \name{} and the original AfterImage pipeline (noted with $^\dagger$ in the table).
Such a pipeline is unsupervised and trains once on the first 3 million packets of the benign Monday capture, so its performance does not change across $k$s.
Its testing runs over the raw Tue-Fri captures rather than the held-out 20\% split and we report its most favourable run with threshold $q{=}0.995$ found experimentally.

Across all $k$ values, \name{} represents the best performing model.
While at $k=50$, KnowML achieves a higher F1 score, its score is based on a binary classification task -- i.e., predicting either attack or benign --, while \name{} is a multi-class classifier.
When considering binary F1, \name{} achieves a much higher performance around 0.95 (see \Cref{tab:flow_sota_sweep}).
Moreover, \name{} achieves a much lower FPR with a 0.124 points reduction when comparing the Anchored coupling architecture with KnowML.

Similarly to the packet-view results, as the value $k$ increases most baselines improve. 
However, they are not capable of reaching the same level of performance of \name{}, as shown by the $9.6\%$ F1 gap between the Guarded coupling and the best performing baseline (HGB).
Anomaly-based approaches in Kitsune and KnowML, however, do not improve their performance with larger $k$.
We analyze this effect in App \ref{secapp:per_class_performance} and find that it depends on these models being effective for detecting volumetric attacks that distort traffic statistics -- e.g., DoS, DDoS, PortScan --, while collapsing on low-footprint attacks, where the anomaly score is uninformative.
This confirms the findings of \cite{guo2026knowml}, showing how both Kitsune and KnowML perform particularly well mainly for DoS or DoS-related attacks.

Regarding the coupling architectures, the results show the importance of the symbolic component in detecting the attack at flow-view.
Coupling architectures that rely mostly on the neural component, such as Deferred and Safety Net, show limited performance at low $k$.
Overall, these results confirm our intuition on low-data regime operation depicted in \Cref{sec:motivating_example}. 

\begin{table}[h]
\centering
\scriptsize
\setlength{\tabcolsep}{3pt}
\caption{Flow-view comparison on CIC-IDS2017 across training
$k=50$, $k=1000$, and $k=5000$. F1 indicates the macro-F1 for multi-class classification models and binary-F1 for the anomaly-based baselines of Kitsune and KnowML. Both F1 and FPR are reported as average $\pm$ standard deviation. Best per column in bold.}
\label{tab:flow_sota_sweep_reduced}
\resizebox{\columnwidth}{!}{%
\begin{tabular}{l rr rr rr}
\toprule
& \multicolumn{2}{c}{$k{=}50$} & \multicolumn{2}{c}{$k{=}1000$} & \multicolumn{2}{c}{$k{=}5000$} \\
\cmidrule(lr){2-3}\cmidrule(lr){4-5}\cmidrule(lr){6-7}
System & F1 & FPR & F1 & FPR & F1 & FPR \\
\midrule
Neural baseline & $0.595{\scriptstyle\pm.016}$ & $0.071{\scriptstyle\pm.015}$ & $0.762{\scriptstyle\pm.024}$ & $0.021{\scriptstyle\pm.007}$ & $0.789{\scriptstyle\pm.004}$ & $0.013{\scriptstyle\pm.000}$ \\
Symbolic baseline & $0.584{\scriptstyle\pm.060}$ & $0.142{\scriptstyle\pm.000}$ & $0.626{\scriptstyle\pm.002}$ & $0.034{\scriptstyle\pm.000}$ & $0.603{\scriptstyle\pm.055}$ & $0.124{\scriptstyle\pm.009}$ \\
Decision tree & $0.452{\scriptstyle\pm.098}$ & $0.174{\scriptstyle\pm.072}$ & $0.673{\scriptstyle\pm.059}$ & $0.096{\scriptstyle\pm.026}$ & $0.726{\scriptstyle\pm.055}$ & $0.053{\scriptstyle\pm.014}$ \\
Logistic regression & $0.443{\scriptstyle\pm.014}$ & $0.275{\scriptstyle\pm.026}$ & $0.496{\scriptstyle\pm.009}$ & $0.231{\scriptstyle\pm.013}$ & $0.508{\scriptstyle\pm.004}$ & $0.221{\scriptstyle\pm.007}$ \\
Random forest & $0.679{\scriptstyle\pm.044}$ & $0.049{\scriptstyle\pm.021}$ & $0.782{\scriptstyle\pm.033}$ & $0.032{\scriptstyle\pm.011}$ & $0.771{\scriptstyle\pm.011}$ & $0.029{\scriptstyle\pm.003}$ \\
HGB & $0.574{\scriptstyle\pm.077}$ & $0.115{\scriptstyle\pm.074}$ & $0.778{\scriptstyle\pm.058}$ & $0.037{\scriptstyle\pm.019}$ & $0.825{\scriptstyle\pm.059}$ & $0.027{\scriptstyle\pm.018}$ \\
\midrule
\name{} (Anchored) & $0.833{\scriptstyle\pm.016}$ & $\mathbf{0.008{\scriptstyle\pm.001}}$ & $\mathbf{0.913{\scriptstyle\pm.005}}$ & $0.009{\scriptstyle\pm.001}$ & $0.919{\scriptstyle\pm.006}$ & $0.008{\scriptstyle\pm.001}$ \\
\name{} (Same-Class Boost) & $0.672{\scriptstyle\pm.013}$ & $0.045{\scriptstyle\pm.010}$ & $0.765{\scriptstyle\pm.021}$ & $0.026{\scriptstyle\pm.007}$ & $0.810{\scriptstyle\pm.013}$ & $0.018{\scriptstyle\pm.007}$ \\
\name{} (Guarded) & $0.849{\scriptstyle\pm.016}$ & $0.011{\scriptstyle\pm.005}$ & $0.904{\scriptstyle\pm.009}$ & $\mathbf{0.008{\scriptstyle\pm.000}}$ & $\mathbf{0.921{\scriptstyle\pm.002}}$ & $\mathbf{0.007{\scriptstyle\pm.000}}$ \\
\name{} (Hard Gate) & $0.838{\scriptstyle\pm.018}$ & $0.010{\scriptstyle\pm.004}$ & $0.902{\scriptstyle\pm.022}$ & $0.010{\scriptstyle\pm.004}$ & $0.916{\scriptstyle\pm.006}$ & $0.009{\scriptstyle\pm.004}$ \\
\name{} (Safety Net) & $0.703{\scriptstyle\pm.029}$ & $0.016{\scriptstyle\pm.007}$ & $0.673{\scriptstyle\pm.022}$ & $0.069{\scriptstyle\pm.010}$ & $0.715{\scriptstyle\pm.042}$ & $0.061{\scriptstyle\pm.018}$ \\
\name{} (Deferred) & $0.593{\scriptstyle\pm.020}$ & $0.074{\scriptstyle\pm.020}$ & $0.728{\scriptstyle\pm.008}$ & $0.032{\scriptstyle\pm.001}$ & $0.789{\scriptstyle\pm.008}$ & $0.020{\scriptstyle\pm.007}$ \\
% \name{} (Protocol) & $\mathbf{0.883{\scriptstyle\pm.006}}$ & $\mathbf{0.005{\scriptstyle\pm.001}}$ & $\mathbf{0.933{\scriptstyle\pm.003}}$ & $\mathbf{0.007{\scriptstyle\pm.001}}$ & $0.887{\scriptstyle\pm.070}$ & $\mathbf{0.007{\scriptstyle\pm.000}}$ \\
\midrule
Kitsune~\cite{mirsky2018kitsune} + Flow features & $0.754{\scriptstyle\pm.022}$ & $0.066{\scriptstyle\pm.017}$ & $0.740{\scriptstyle\pm.005}$ & $0.028{\scriptstyle\pm.004}$ & $0.752{\scriptstyle\pm.004}$ & $0.018{\scriptstyle\pm.006}$ \\
Kitsune~\cite{mirsky2018kitsune} + AfterImage (faithful)$^\dagger$ & $0.734$ & $0.102$ & $0.734$ & $0.102$ & $0.734$ & $0.102$ \\
% Suricata (all ET rules)$^\dagger$ & $0.418$ & $0.360$ & $0.418$ & $0.360$ & $0.418$ & $0.360$ \\
% Suricata (ET operational + published)$^\dagger$ & $0.577$ & $0.017$ & $0.577$ & $0.017$ & $0.577$ & $0.017$ \\
% KnowML~\cite{guo2026knowml} + MLP (supervised) & $0.587{\scriptstyle\pm.176}$ & $0.239{\scriptstyle\pm.106}$ & $0.795{\scriptstyle\pm.147}$ & $0.096{\scriptstyle\pm.010}$ & $0.918{\scriptstyle\pm.068}$ & $0.035{\scriptstyle\pm.005}$ \\
KnowML~\cite{guo2026knowml} + KitNET (unsupervised) & $\mathbf{0.883{\scriptstyle\pm.019}}$ & $0.132{\scriptstyle\pm.025}$ & $0.851{\scriptstyle\pm.036}$ & $0.175{\scriptstyle\pm.051}$ & $0.792{\scriptstyle\pm.024}$ & $0.260{\scriptstyle\pm.035}$ \\
\bottomrule
\end{tabular}}
\end{table}

\ourparagraph{Hybrid-view.}
Results for the hybrid-view detection are presented in \Cref{tab:hybrid_sota_sweep_reduced} for $k\in\{50,1000,5000\}$, while the extended results are shown in \Cref{tab:hybrid_couplings} of \Cref{secapp:all_results}.
The neural baseline consists of a product embedding model composed of two neural networks that encode the packet and flow data and a product layer that combines the two neural embeddings before a final classification layer.
Additionally, since \name{} does not operate on payload features, we adapt the DST-IDS~\cite{QIU2022} fusion to the same packet and flow classifiers, rather than using the original payload-based mechanism for fair comparison.
The neural product and DST-IDS baselines are trained using early stopping over fifty epochs, while all \name{} models are trained for ten epochs.

\name{} represents the best performing model across all data sizes $k$.
Moreover, its performance improvement is more evident for smaller values of $k$, achieving a $7.5\%$ F1 improvement and a 0.027 FPR reduction at $k=50$.
For larger $k$, the neural and DST-IDS baselines improve their performance, but never catch up with \name{}, showcasing the effectiveness of neuro-symbolic integration when dealing with complex input data.
Overall, these results confirm the findings of packet- and flow-view on the data efficiency of \name{}.
Moreover, they also prove that the explicit cross-view rules constructed during the \name{}'s offline preparation are particularly useful when the packet and flow classifiers have insufficient data to learn an effective combination policy.

\begin{table}[h]
\centering
\scriptsize
\setlength{\tabcolsep}{3.5pt}
\caption{Hybrid-view comparison on CIC-IDS2018 across
per-class training sizes of $k=50$, $k=1000$, and $k=5000$.
Results report F1 and FPR as average $\pm$ standard deviation. Best per column in bold.}
\label{tab:hybrid_sota_sweep_reduced}
\resizebox{\columnwidth}{!}{%
\begin{tabular}{l *{3}{rr}}
\toprule
& \multicolumn{2}{c}{${k{=}50}$}
& \multicolumn{2}{c}{${k{=}1000}$}
& \multicolumn{2}{c}{${k{=}5000}$} \\
\cmidrule(lr){2-3} \cmidrule(lr){4-5} \cmidrule(lr){6-7}
System & F1 & FPR & F1 & FPR & F1 & FPR \\
\midrule
Symbolic-only baseline
& $0.841{\scriptstyle\pm.023}$ & $0.006{\scriptstyle\pm.000}$
& $0.857{\scriptstyle\pm.014}$ & $0.086{\scriptstyle\pm.018}$
& $0.849{\scriptstyle\pm.001}$ & $0.077{\scriptstyle\pm.007}$ \\
Neural product (baseline)
& $0.782{\scriptstyle\pm.064}$ & $0.029{\scriptstyle\pm.015}$
& $0.950{\scriptstyle\pm.022}$ & $\mathbf{0.003}{\scriptstyle\pm.001}$
& $0.970{\scriptstyle\pm.012}$ & $\mathbf{0.002}{\scriptstyle\pm.001}$ \\
DST-IDS fusion~\cite{QIU2022}
& $0.827{\scriptstyle\pm.065}$ & $0.024{\scriptstyle\pm.014}$
& $0.962{\scriptstyle\pm.014}$ & $\mathbf{0.003}{\scriptstyle\pm.001}$
& $0.970{\scriptstyle\pm.008}$ & $\mathbf{0.002}{\scriptstyle\pm.001}$ \\
\midrule
\name{} (Product of Experts)
& $0.867{\scriptstyle\pm.035}$ & $0.020{\scriptstyle\pm.016}$
& $0.961{\scriptstyle\pm.023}$ & $0.006{\scriptstyle\pm.005}$
& $\mathbf{0.985{\scriptstyle\pm.002}}$ & $0.005{\scriptstyle\pm.004}$ \\
\name{} (Guarded)
& $0.759{\scriptstyle\pm.036}$ & $0.008{\scriptstyle\pm.004}$
& $0.952{\scriptstyle\pm.015}$ & $0.004{\scriptstyle\pm.004}$
& $0.971{\scriptstyle\pm.015}$ & $\mathbf{0.002}{\scriptstyle\pm.001}$ \\
\name{} (Anchored)
& $0.805{\scriptstyle\pm.019}$ & $\mathbf{0.003}{\scriptstyle\pm.002}$
& $0.818{\scriptstyle\pm.032}$ & $0.091{\scriptstyle\pm.024}$
& $0.828{\scriptstyle\pm.044}$ & $0.096{\scriptstyle\pm.035}$ \\
\name{} (Same-Class Boost)
& $0.709{\scriptstyle\pm.040}$ & $0.049{\scriptstyle\pm.016}$
& $0.954{\scriptstyle\pm.007}$ & $0.005{\scriptstyle\pm.004}$
& $0.960{\scriptstyle\pm.011}$ & $0.003{\scriptstyle\pm.000}$ \\
\name{} (Deferred)
& $\mathbf{0.870{\scriptstyle\pm.037}}$ & $0.018{\scriptstyle\pm.016}$
& $0.961{\scriptstyle\pm.023}$ & $0.006{\scriptstyle\pm.005}$
& $\mathbf{0.985{\scriptstyle\pm.002}}$ & $0.005{\scriptstyle\pm.004}$ \\
\name{} (Hard Gate)
& $0.749{\scriptstyle\pm.041}$ & $0.016{\scriptstyle\pm.014}$
& $\mathbf{0.963{\scriptstyle\pm.014}}$ & $0.004{\scriptstyle\pm.001}$
& $0.969{\scriptstyle\pm.010}$ & $0.003{\scriptstyle\pm.000}$ \\
\name{} (Safety Net)
& $0.862{\scriptstyle\pm.043}$ & $0.017{\scriptstyle\pm.011}$
& $0.956{\scriptstyle\pm.038}$ & $0.020{\scriptstyle\pm.020}$
& $0.982{\scriptstyle\pm.006}$ & $0.008{\scriptstyle\pm.002}$ \\
\bottomrule
\end{tabular}}

\end{table}
%

%%%%%%%%%%%%%%%%%%%%%%%%%%%%%%%%%%%%%%%%%%%%%%%%%%%%%%%%%%
%
\subsection{Generalization}
\label{sec:generalization}
We evaluate whether \name{} generalizes well across datasets and network topologies operating a cross-environment transfer from the enterprise
CIC-IDS2018 dataset to the IoT NF-ToN-IoT-v3 dataset---i.e., training a model on the former and testing it on the latter.
Although these datasets differ in network devices,
traffic distributions, and attack families, both adopt NetFlow-v3 flow-views.
Thus, symbolic rules constructed in the source environment can be evaluated in the target without modifying \name{}'s logical structure.

Source models are trained using \(k=500\) labelled samples per attack class of CIC-IDS2018.
The fine-tuning to the IoT dataset then proceeds in two stages.
First, the learned feature extractor is frozen, and only the classifier head and environment-dependent symbolic thresholds are updated.
Second, the entire neural model is unfrozen and optimized with a reduced learning rate, while symbolic-threshold adaptation continues.
Rules transferred from the source to the target environment retain their logical structure but are recalibrated using only information from the target
environment.
The fine-tuning uses \(k\in\{0,10,50\}\) samples per attack class, with $k=0$ representing the most challenging zero-shot generalization setting.
We compare four models:
\begin{inlinelist}
    \item an \emph{oracle} model trained and evaluated directly on the IoT target data which represents the achievable performance upper bound,
    \item a purely \emph{neural} model trained only in the enterprise domain,
    \item a \emph{regularized neural} model which applies weight decay during fine-tuning -- to check whether \name{} improvement is attributable solely to regularization --, and
    \item a \name{} model where both the neural component and rule structures of the original \name{} model are used without modifications.
    % \item a partial \name{} transfer model where only the neural components is kept fixed and new rules are constructed for the target dataset.
\end{inlinelist}
Lastly, we focus on the flow traffic-view and use the Guarded coupling architecture for \name{} as it is the best performing coupling on average over the flow traffic-view.
%
% The \emph{oracle} is trained and evaluated directly in the target environment.
% %
% \emph{Neural transfer} evaluates a source-trained neural model either zero-shot
% or after limited target-domain fine-tuning.
% %
% \emph{Regularized neural transfer} additionally applies weight decay,
% controlling for the possibility that any neuro-symbolic improvement is
% attributable solely to regularization.
% %
% Finally, \emph{neuro-symbolic transfer} combines the transferred neural model
% with symbolic evidence.
% %
% We consider two variants: \emph{Transfer-NeSy}, which transfers both the neural
% model and broad rule structures, and \emph{NeSy local-rules}, which transfers
% only the neural model and constructs new rules for the target environment.
%
\begin{figure}[t]
\centering
\begin{subfigure}{0.32\columnwidth}
    \centering
    \includegraphics[width=0.8\linewidth]{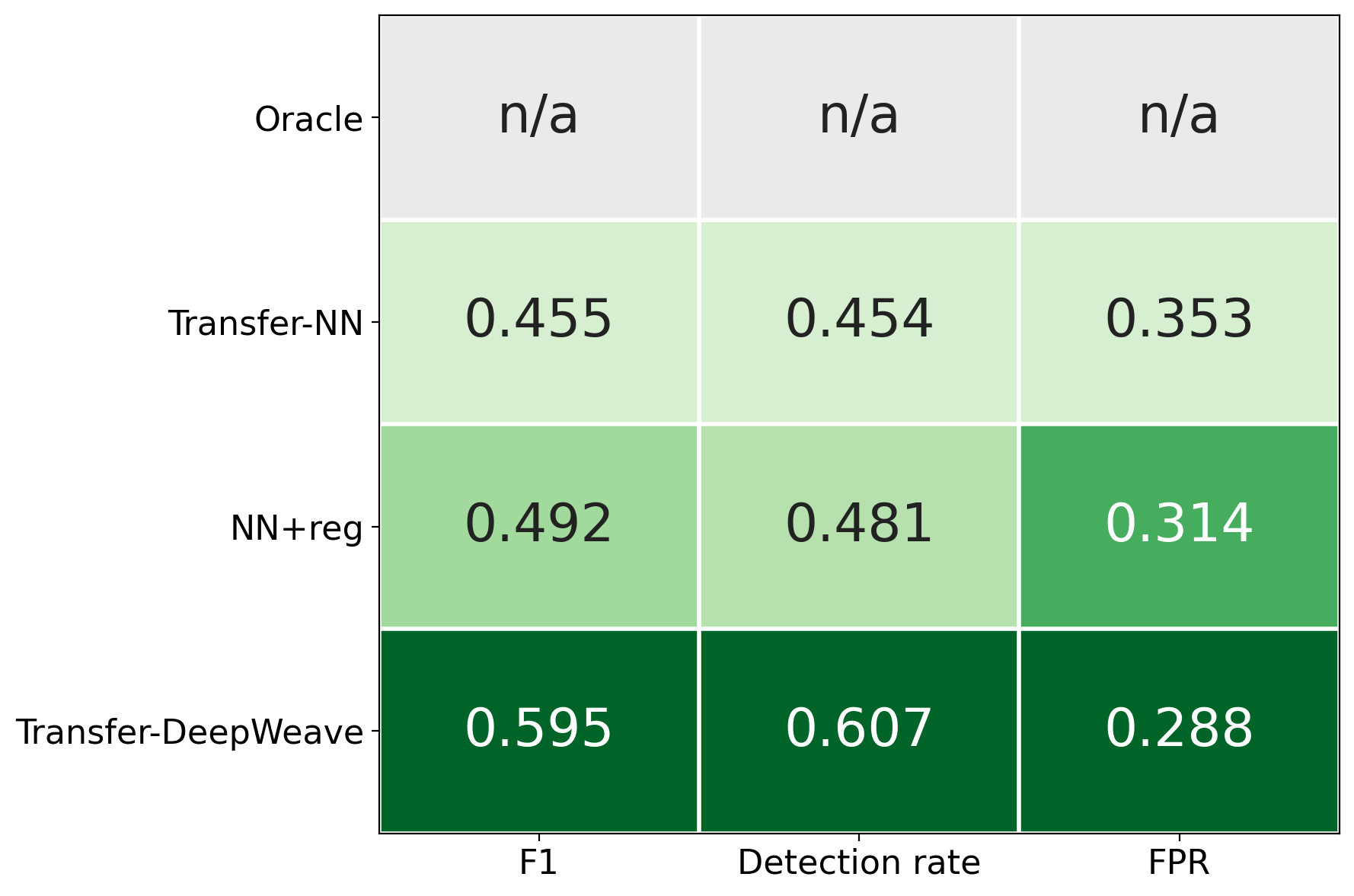}
    \caption{Zero-shot transfer}
    \label{fig:portability_zero_shot}
\end{subfigure}
\hfill
\begin{subfigure}{0.32\columnwidth}
    \centering
    \includegraphics[width=0.8\linewidth]{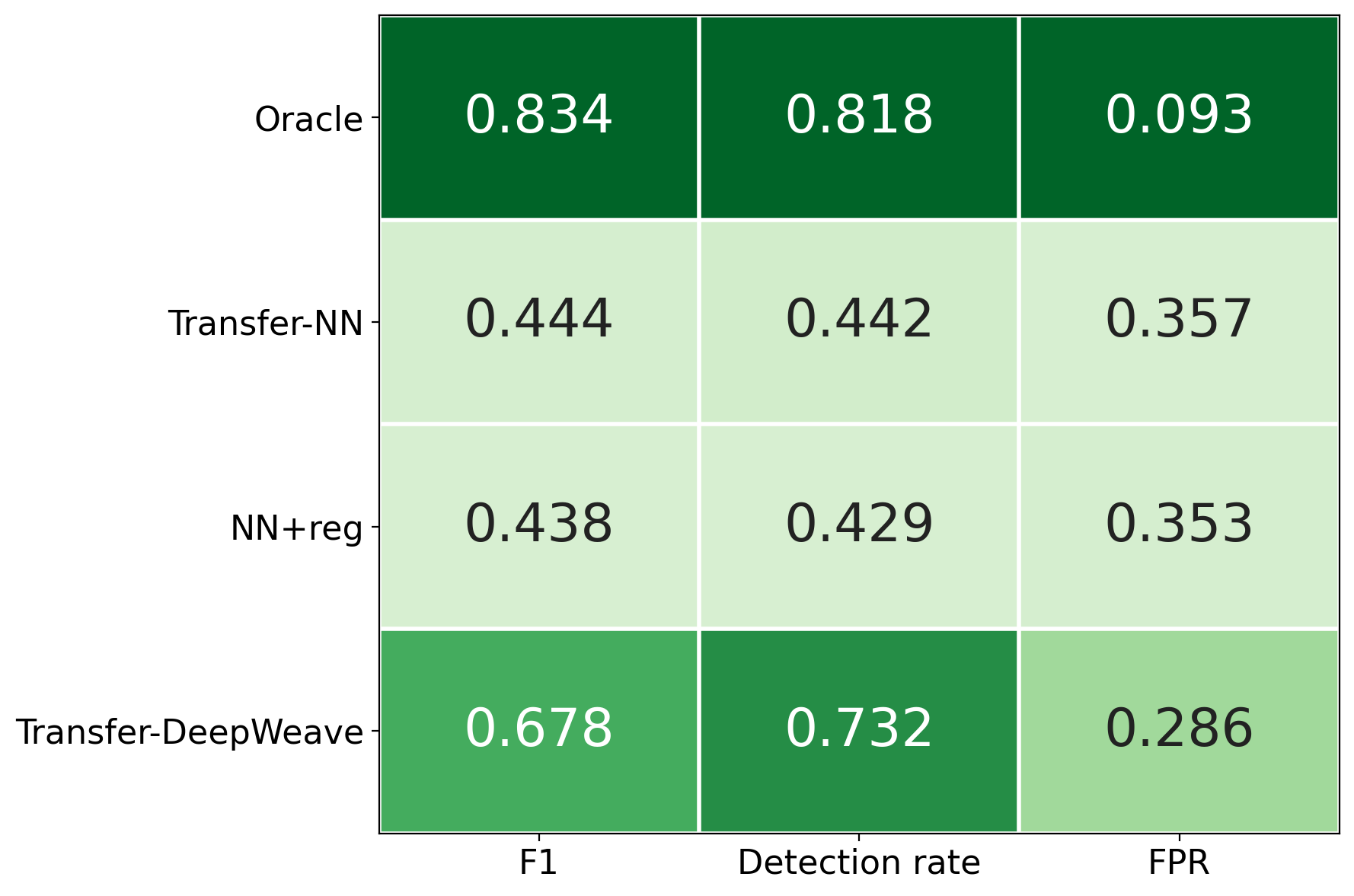}
    \caption{Fine-tuning k=10}
    \label{fig:portability_fine_tune_k10}
\end{subfigure}
\hfill
\begin{subfigure}{0.32\columnwidth}
    \centering
    \includegraphics[width=0.8\linewidth]{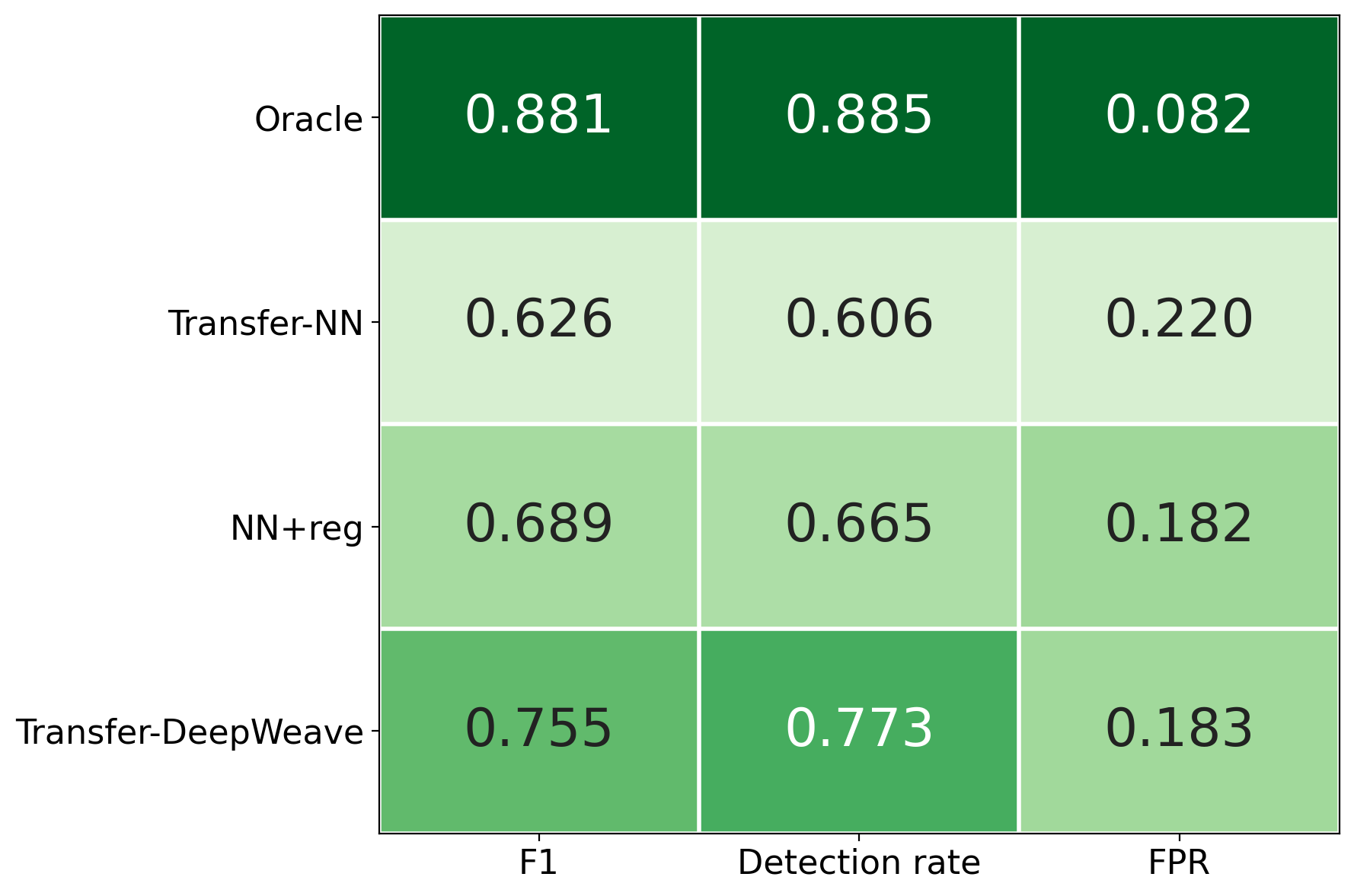}
    \caption{Fine-tuning k=50}
    \label{fig:portability_fine_tune_k50}
\end{subfigure}
\caption{Generalization testing from enterprise scenario (CIC-IDS2018) to IoT (NF-ToN-IoT-v3).}
\label{fig:rq2_portability}
\end{figure}
\Cref{fig:rq2_portability} reports the enterprise-to-IoT transfer
results.
As expected, the oracle model achieves the highest performance on the target dataset, being directly trained on it.
Meanwhile, \name{} shows effective generalization capabilities -- considering both F1 and FPR scores --, outperforming the purely neural counterparts -- both with and without regularization -- by 14\% in the zero-shot setting and by up to 23\% after fine-tuning.
As more fine-tuning data become available, the purely neural baselines adapt to the target dataset, closing the gap with \name{}.
However, they never reach the same level of performance of \name{} for the tested settings. 
%

%%%%%%%%%%%%%%%%%%%%%%%%%%%%%%%%%%%%%%%%%%%%%%%%%%%%%%%%%%
%
\subsection{Adversarial Robustness}
\label{sec:adversarial_robustness}

We evaluate the adversarial robustness of \name{} on the CIC-IDS2017 flow-view.
To model a realistic adversary, we adopt the attacker capabilities modeled in PANTS~\cite{jin2025robustifying}. 
For ease of implementation, we update the affected flow features directly rather than replicating their full packet-level synthesis.
Specifically, for a finite set of attacker admissible actions we modify the corresponding flow representations and evaluate the resulting attack flows without modifying the training data nor adversarially retraining the models.
The threat model combines two attacker positions.
An in-path attacker can delay packets in one direction, while a compromised end host can delay or pad its own forward packets, inject dummy forward packets, or both.
For each flow, we limit adversarial packet injection to 20\% of the original packet count, and bound the padding to the longest packet in the original flow.
Lastly, reverse traffic is held fixed, so the attacker can modify only its own traffic.

To isolate the effect of neuro-symbolic integration on adversarial robustness, we compare the different \name{} coupling architectures against a purely symbolic baseline -- composed of only symbolic rules -- and a purely neural baseline---the neural component.
Moreover, because the CIC-IDS2018 is unbalanced -- DoS and PortScan account for 68.2\% and 28.5\% of the malicious flows, respectively, -- measuring only the ASR would obscure failures on minority attack classes.
We therefore report weighted ASR along with the ASR for each attack class, and the F1 detection performance over the attacked flows.
%

% \begin{figure}[t]
% \centering
% \includegraphics[width=0.72\columnwidth]{img/robustness_adversarial.png}
% \caption{Performance under the proportional, PANTS-inspired feature-space
% attack on the natural-prior CIC-IDS2017 test split. Only malicious flows are
% perturbed, and no adversarial retraining is performed.}
% \label{fig:robust_adversarial}
% \end{figure}

\begin{figure}
  \centering
  \begin{subfigure}{0.49\columnwidth}
    \centering
    \includegraphics[width=0.65\linewidth]{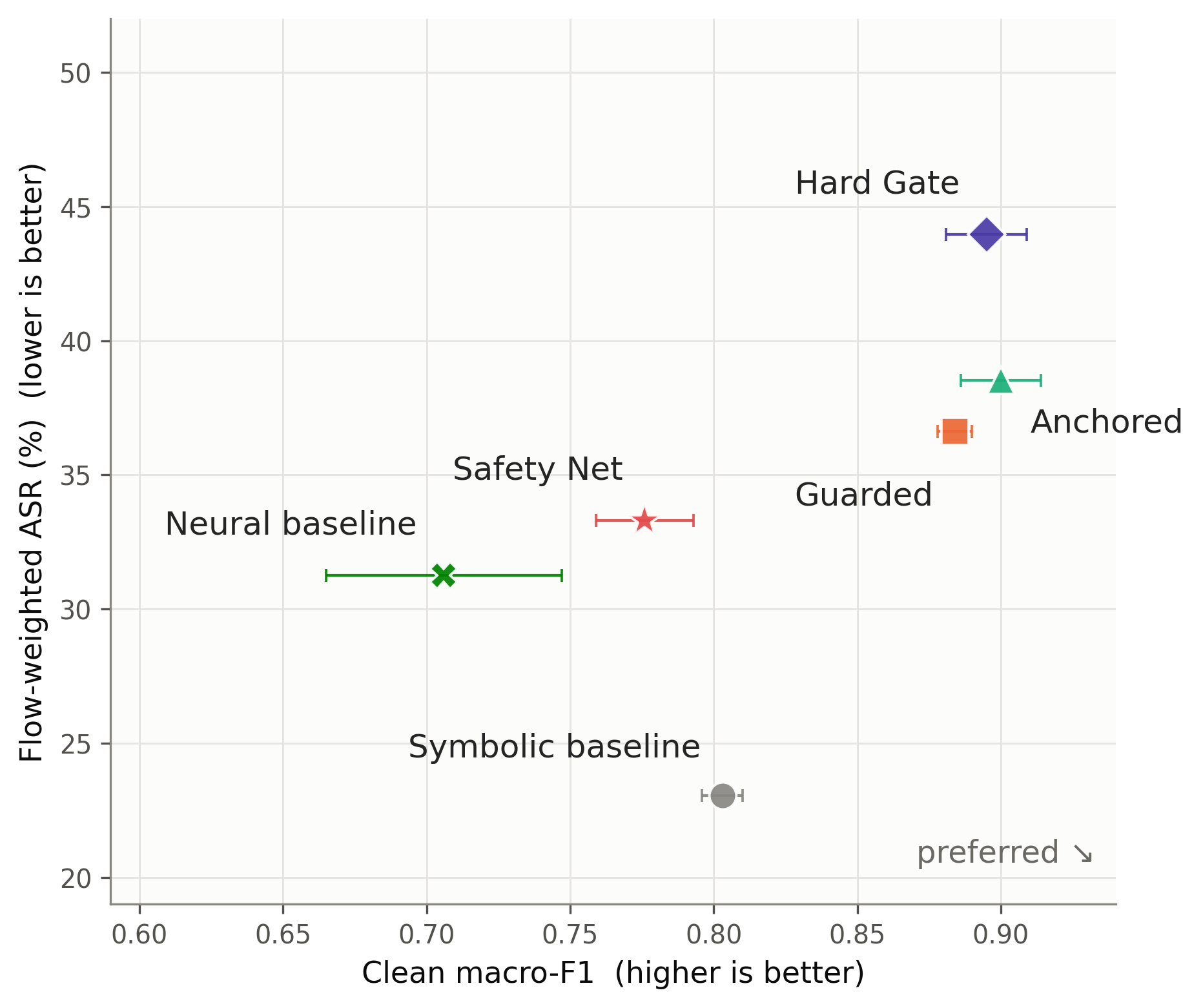}
    \caption{ASR vs.\ clean accuracy}
    \label{fig:robust_weighted_asr}
  \end{subfigure}
  \hfill
  \begin{subfigure}{0.49\columnwidth}
    \centering
    \includegraphics[width=0.65\linewidth]{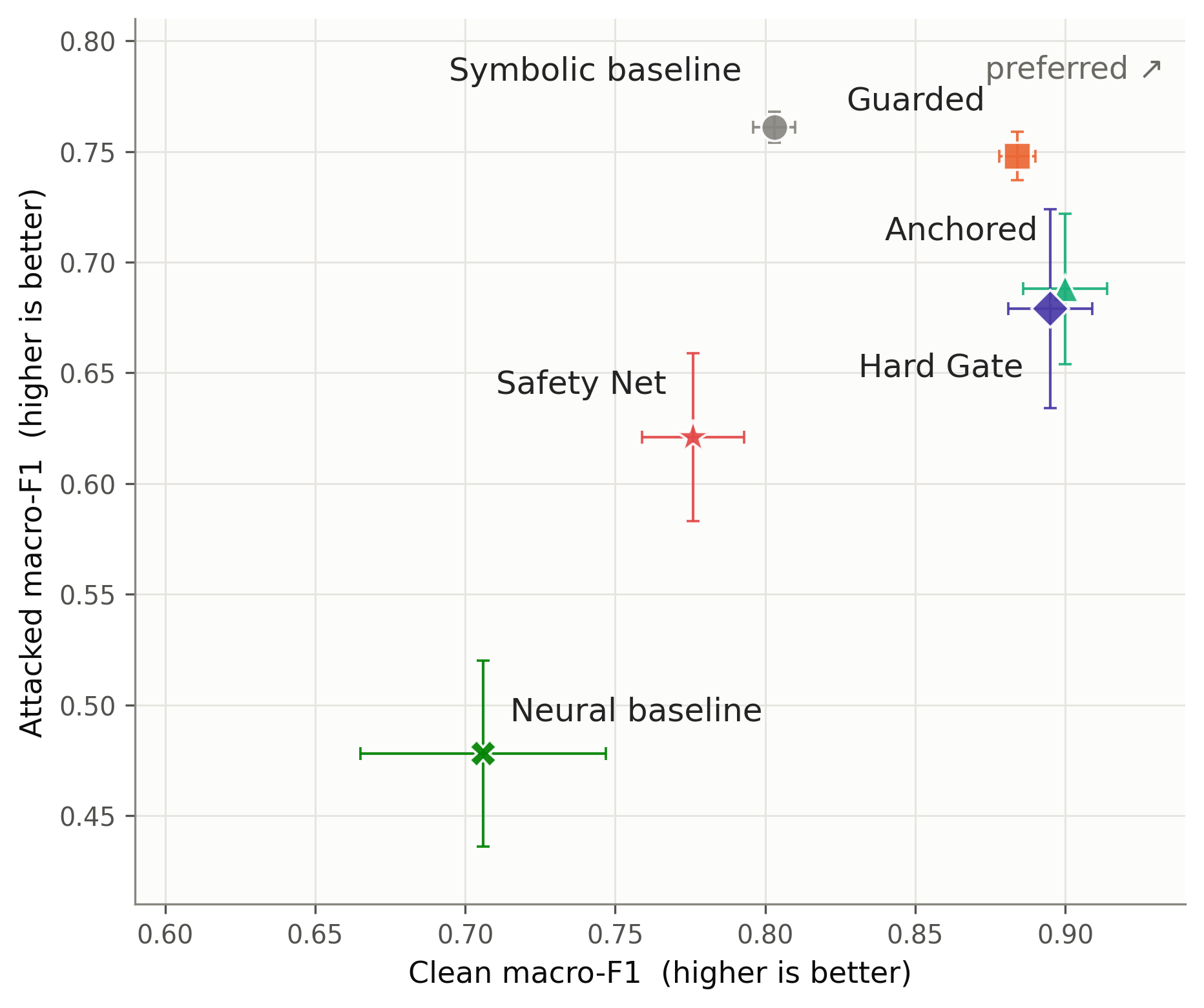}
    \caption{Performance under attack}
    \label{fig:robust_attacked_f1}
  \end{subfigure}
  \caption{Adversarial robustness analysis.}
  \label{fig:robust_adversarial_tradeoff}
\end{figure}

As shown in \Cref{fig:robust_adversarial_tradeoff}, the symbolic-only system is the most robust under adversarial attacks.
Its ASR is 23.1\%, while its F1 performance decreases from 0.803 to 0.761, corresponding to 94.8\% performance retention.
This robustness is consistent with the symbolic detection relying primarily on properties such as ports, flags, and categorical protocol conditions that a constrained attacker is not able to modify.
Meanwhile, the neural baseline achieves a weighted ASR of 31.3\%, showcasing surprisingly high robustness.
This is largely due to it being robust mainly against perturbations that affect the DoS class, where its achieved ASR is only 13.3\% (see~\Cref{tab:pants_relative_per_class} in~\Cref{secapp:robustness_extra}).
As DoS accounts for 68.2\% of the malicious test flows, being robust against such perturbations make the model implicitly more robust.
Nevertheless, the neural baseline does not represent a reliable model, since its F1 performance decreases to only 0.478 when considering adversarial flows.
Further showing that ASR alone is not sufficient to measure the usable robustness of a NIDS.

Guarded represents the strongest coupling architecture, achieving a class-averaged ASR of 25.3\% (see \Cref{tab:pants_relative_per_class}).
Thus, \name{} is more robust than its neural counterpart, although not being able to match the symbolic-only system.
Moreover, \name{} (Guarded) provides more consistent robustness across classes w.r.t. the neural baseline, thus being more reliable in practice (see \Cref{tab:pants_relative_per_class}).
\name{} robustness largely depends on the chosen coupling architecture as more neural-driven couplings result in weaker models such as Hard Gate and Safety Net (see \Cref{tab:pants_relative_per_class}).
Lastly, as show in~\ref{fig:robust_attacked_f1}, \name{} allows to define coupling architectures that remain effective also on adversarially perturbed traffic.
For example, Guarded achieves a post-attack macro-F1 performance of 0.748, thus achieving a 84.7\% performance retention.
\subsection{Explainability}
\label{sec:explanation_results}
\begin{figure}[h]
    \centering
    \begin{subfigure}[t]{0.49\columnwidth}
        \centering
        \includegraphics[width=0.65\linewidth]{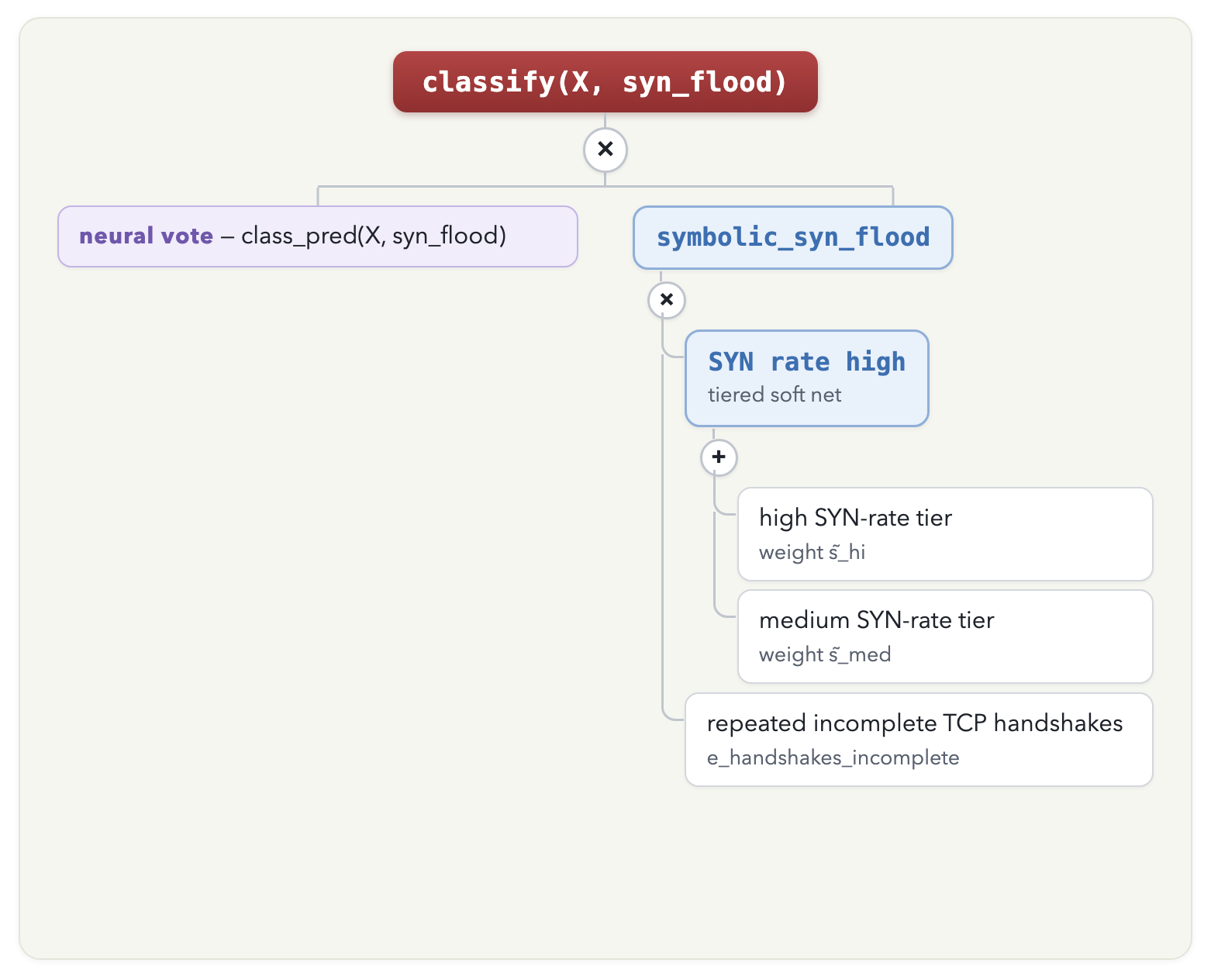}
        \caption{Global explanation.}
        \label{fig:global_explanation}
    \end{subfigure}
    \hfill
    \begin{subfigure}[t]{0.49\columnwidth}
        \centering
        \includegraphics[width=0.65\linewidth]{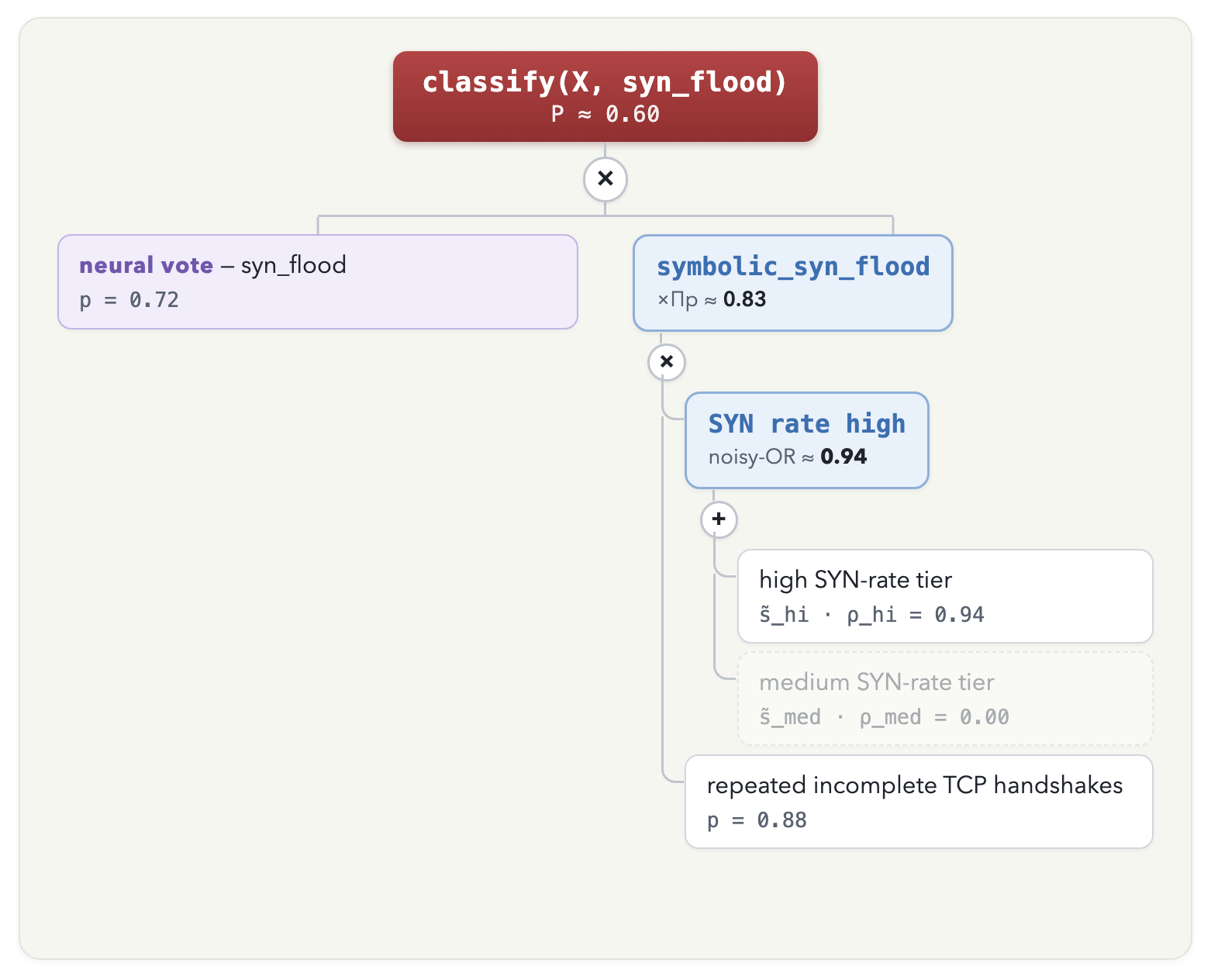}
        \caption{Local explanation.}
        \label{fig:local_explanation}
    \end{subfigure}
    \caption{Visualization of \name{}'s global and local explanations.}
    \label{fig:explanation_example}
\end{figure}

We evaluate whether \name{} produces explanations that reflect its predictive
reasoning and provide meaningful evidence for security analysis.
Unlike post-hoc attribution methods, \name{} explains predictions through the
proof paths used to compute their class probabilities, making its explanations
intrinsic to the model.
\Cref{fig:explanation_example} illustrates both a global explanation of the
model's decision structure (\Cref{fig:global_explanation}) and a local
explanation of one prediction (\Cref{fig:local_explanation}).
The example uses Hard Gate coupling and the flow-view model trained on
CIC-IDS2017 for SYN-flood detection, while $\times$ and $+$ denote logical AND and OR,
respectively.
The optimized rule checks for an unusually high SYN rate and repeated
incomplete TCP handshakes.
SYN-rate evidence is evaluated at two thresholds, whose steepness is optimized
during training (\Cref{sec:symbolic_knowledge_construction}), to distinguish
clearly anomalous high rates from stealthier medium-rate flooding.
The symbolic result is gated by a neural component that directly estimates
SYN-flood evidence.
Consequently, \name{} predicts SYN flood only when the neural traffic pattern
and the symbolic attack conditions agree.
For a particular alert, \Cref{fig:local_explanation} reports the activated
predicates and their probabilities, allowing an analyst to inspect detected
patterns rather than isolated features---e.g., a very high SYN rate with
probability $0.94$ and repeated incomplete handshakes with probability $0.88$.
Additional global and local examples appear in
\Cref{fig:additional_explanation_examples} of
\Cref{secapp:explainability}.

Following the xNIDS evaluation setting~\cite{wei2023xnids}, we compare \name{}
with SHAP~\cite{lundberg2017shap} and Integrated Gradients
(IG)~\cite{sundararajan2017integrated}, applied to the matched neural flow
classifier.
Across five CIC-IDS2017 training replicates, \name{} selects the same winning
rule for $96.4\%$ of observations, whereas IG and SHAP select the same top
feature for only $24.3\%$ and $18.8\%$, respectively.
\Cref{secapp:explainability} provides the complete protocol, representative
comparisons, and per-class stability results.

Lastly, we replicate xNIDS's deletion-based fidelity evaluation
~\cite{wei2023xnids}.
For each alert, we progressively nullify features selected by its explanation
and measure the probability retained by the original predicted class; lower
retention indicates that the removed evidence was more important to the
prediction.
\name{} selects the conditions of the clause responsible for the alert, whereas
IG and SHAP rank features by the absolute magnitude of their per-sample
attributions.
As shown in \Cref{tab:deletion_fidelity}, removing the complete \name{}
explanation -- only $2.4$ conditions on average -- reduces probability retention
to $0.11$.
By comparison, removing the highest-ranked IG or SHAP feature retains $0.90$
and $0.91$ of the original decision, respectively, while removing their top ten
features still retains between $0.65$ and $0.77$.
\name{} is also substantially sparser: its explanations contain $2.4$
conditions on average, compared with $28$ IG features and $16$ SHAP features
required to account for $90\%$ of total attribution.
Thus, \name{} identifies a compact set of conditions directly responsible for
each prediction, while the post-hoc methods produce less sparse explanations
whose highest-ranked features have a smaller effect when removed.

\begin{table}[tp]
    \centering
    \fontsize{8}{9.5}\selectfont   % not needed if you use the updated template.tex
    \caption{Deletion fidelity on CIC-IDS2017 flow-view traffic with Guarded
    coupling. Values report the probability retained by the originally
    predicted class; lower is more faithful.}
    \label{tab:deletion_fidelity}
    \begin{tabular}{lccccc}
        \toprule
        & \multicolumn{4}{c}{\makecell{Probability retention after \\
          deleting top-$m$ features/rules}}
        & \makecell{Explanation size \\
          (features/rules)} \\
        \cmidrule(lr){2-5}
        Explanation method & $m{=}1$ & $m{=}3$ & $m{=}5$ & $m{=}10$ & \\
        \midrule
        \name{}
        & $\mathbf{0.13}$ & $\mathbf{0.11}$ & $\mathbf{0.11}$
        & $\mathbf{0.11}$ & $\mathbf{2.4\pm0.7}$ \\
        Integrated Gradients (MLP)
        & $0.90$ & $0.85$ & $0.77$ & $0.65$ & $27.9\pm8.4$ \\
        SHAP (MLP)
        & $0.91$ & $0.86$ & $0.82$ & $0.77$ & $15.7\pm4.3$ \\
        \bottomrule
    \end{tabular}
\end{table}

\subsection{Computational Efficiency}
\label{sec:computational_efficiency}
\begin{figure}[t]
    \centering
    \begin{subfigure}[t]{0.49\columnwidth}
        \centering
        \includegraphics[width=0.65\linewidth]{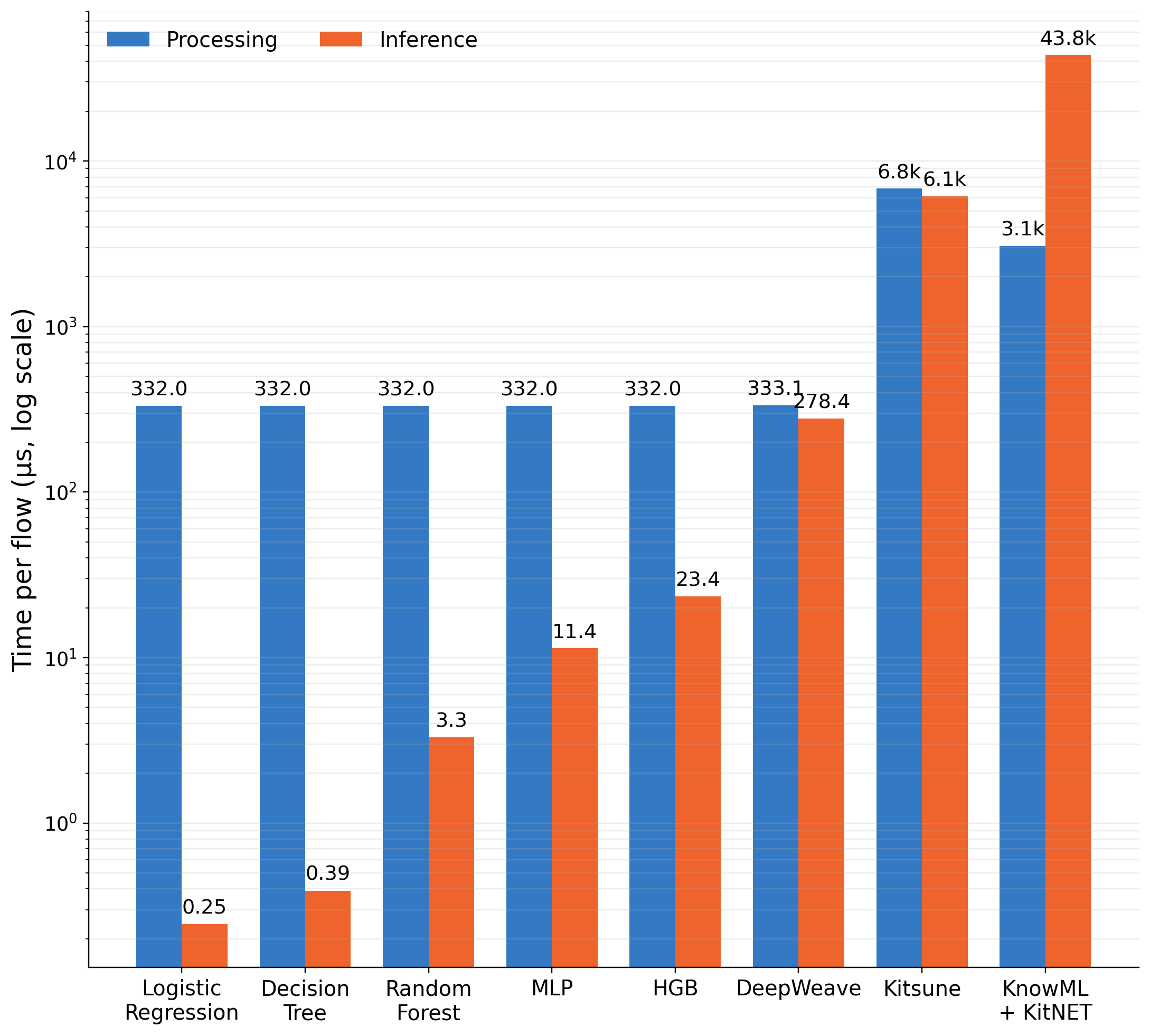}
        \caption{Effective cost per flow.}
        \label{fig:computational_efficiency_per_flow}
    \end{subfigure}
    \hfill
    \begin{subfigure}[t]{0.49\columnwidth}
        \centering
        \includegraphics[width=0.65\linewidth]{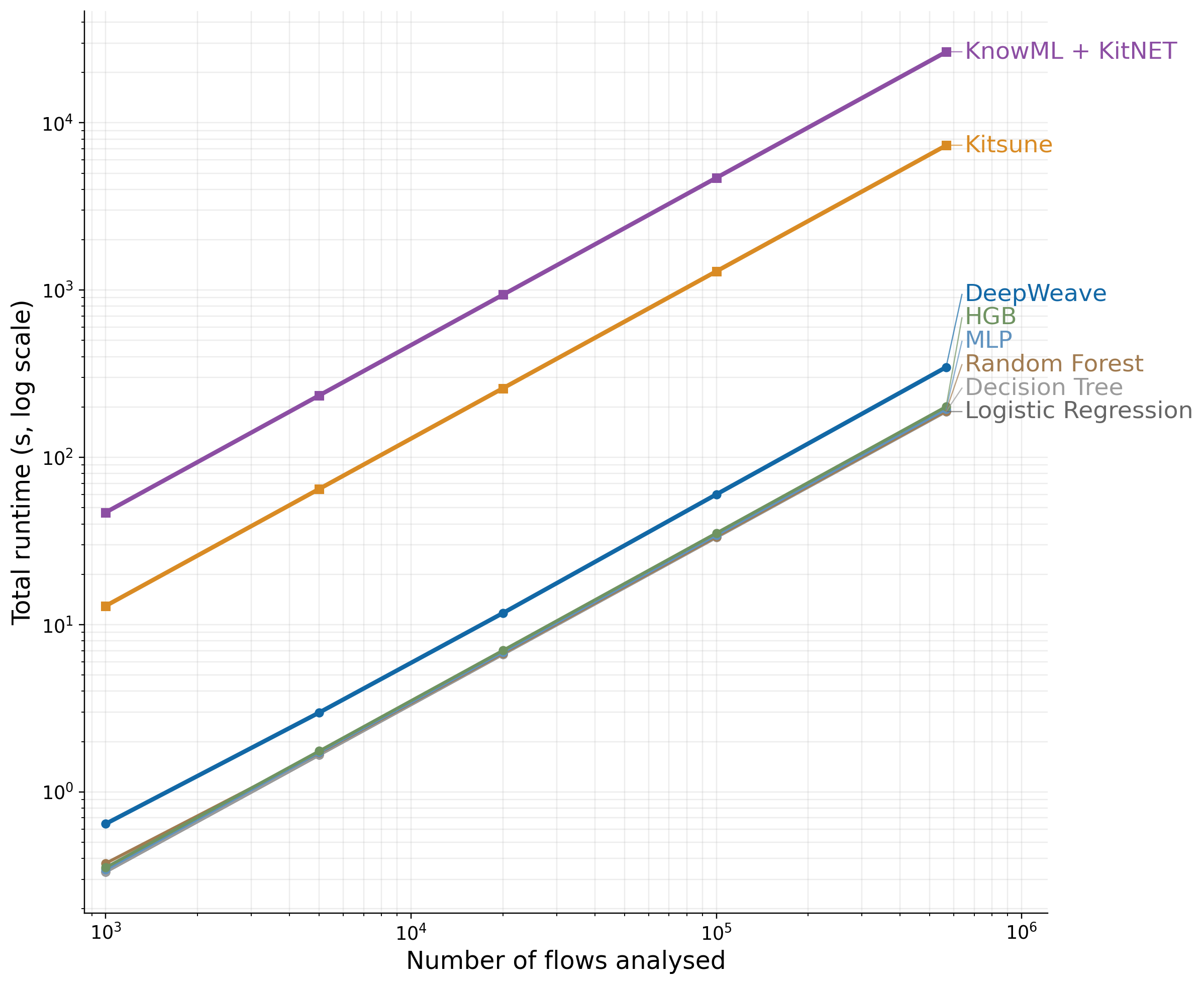}
        \caption{Processing and inference time.}
        \label{fig:computational_efficiency_total}
    \end{subfigure}
    \caption{Computational efficiency of \name{} and the baselines:
    effective per-flow processing and inference cost at the complete workload
    (left), and total runtime by workload (right).}
    \label{fig:computational_efficiency}
\end{figure}

We evaluate the cost of the complete detection pipeline as the number of
analyzed flows increases, including both the transformation of incoming traffic
into model-ready representations and prediction by an already trained detector.
We compare \name{} with the ML baselines in \Cref{sec:baselines},
Kitsune~\cite{mirsky2018kitsune}, and KnowML+KitNET~\cite{guo2026knowml}.
\Cref{fig:computational_efficiency} reports (a) the effective per-flow cost at
the complete workload and (b) total runtime for increasing workloads.
\emph{Processing} covers the operations required to construct a valid detector
representation from packets or flow records, whereas \emph{inference} covers
the operations required to produce a class or anomaly prediction from that
representation.

Packet-to-flow assembly dominates processing for all flow-input systems.
Before assembly, \name{} requires approximately \(1.34\,\mu\text{s}\) per flow
for feature scaling, symbolic-atom evaluation, quantile lookup, and fact
grounding, about \(5\times\) the \(0.27\,\mu\text{s}\) required by the
conventional ML baselines.
After adding the shared \(331.75\,\mu\text{s}\) assembly cost, however,
\name{}'s total processing time is only \(0.32\%\) higher.
Thus, the similar processing times do not imply that symbolic representation is
free, but rather that common packet ingestion and flow assembly dominate model-specific
preparation.

Inference shows a larger difference.
HGB processes the complete \(565{,}525\)-flow workload in \(13.21\) seconds,
or \(23.35\,\mu\text{s}\) per flow, whereas \name{} requires \(157.47\)
seconds, or \(278.45\,\mu\text{s}\) per flow.
Despite being slower than the flow-based ML classifiers, \name{} remains
substantially faster than the evaluated packet-native systems in their current
implementations.
Kitsune requires \(418.8\,\mu\text{s}\) per packet for AfterImage processing
and \(377.4\,\mu\text{s}\) for inference through its 19-autoencoder KitNET
ensemble.
KnowML processes features faster than Kitsune -- approximately \(3.08\) versus
\(6.81\) ms per resulting flow -- but its 148-autoencoder KitNET ensemble
requires approximately \(43.82\) ms of inference per flow, compared with
\(6.14\) ms for Kitsune.
Consequently, Kitsune's estimated complete-workload runtime is approximately
\(21\times\) \name{}'s measured runtime, while KnowML+KitNET is approximately
\(77\times\) slower.

Packet- and flow-level timings are not directly interchangeable, but they
contextualize probabilistic-reasoning cost relative to complete NIDS pipelines.
For deployment planning, the total-runtime curves are the most informative
comparison because they include batching, vectorization, query construction,
and framework overhead.
The isolated inference bars should therefore be interpreted as
workload-specific effective costs rather than workload-independent constants.
\Cref{secapp:additional_computational_efficiency} separately examines how
symbolic-program structure affects cost under a fixed workload.

\section{Discussion and Limitations}
\label{sec:discussions_limitations}

\ourparagraph{Value and limits of symbolic knowledge.}
Reusable rules provide structured evidence when labels or traffic context are
scarce, reducing what the neural component must learn and helping control false
alarms~\cite{MANHAEVE2021103504,wang2025captain}.
However, rules may capture environment-specific or collection-specific
correlations and therefore require provenance tracking, recalibration, and
periodic validation under deployment shift
~\cite{sommer2010outside,cantone2024crossdataset}.
Our label-efficiency results consequently assume limited current labels
supplemented by historical symbolic knowledge, rather than equal information
budgets for rule discovery and neural training.
DeepWeave's probabilities should likewise be interpreted as structured
classification evidence, not a complete generative model of network traffic.

\ourparagraph{Coupling is an operational policy.}
%
% The coupling policy determines the authority assigned to neural and symbolic
% evidence.
% %
% Agreement-based couplings reduce false alarms but fail when either required
% evidence source is disrupted, whereas permissive couplings preserve alternative
% proof paths under corruption or evasion at the cost of additional benign
% alerts.
% %
% There is therefore no universally optimal policy: precision-oriented
% deployments may prefer strict corroboration, while noisy or adversarial
% environments may require greater evidential independence.
% %
% DeepWeave makes this precision--robustness trade-off explicit rather than
% embedding it within an opaque fusion layer.
Coupling is an operational policy. It determines the authority assigned to neural and symbolic evidence, so no policy is universally optimal: precision-oriented deployments may prefer strict corroboration, whereas noisy or adversarial environments may require greater evidential independence. DeepWeave makes this precision–robustness trade-off explicit rather than embedding it within an opaque fusion layer.

\ourparagraph{Faithful proofs do not guarantee analyst utility.}
DeepWeave's explanations are intrinsic because the reported predicates and
clauses belong to proofs contributing to the prediction, avoiding the semantic
gap of post-hoc explanations~\cite{wei2023xnids}.
Our results establish proof coverage, clause reliability, and stability, but do
not show that proofs improve analyst decisions or reduce investigation effort.
Operational studies indicate that explanations are useful only when they fit
analyst workflows and support appropriate reliance
~\cite{nyreyu2022xaiops,roch2024autonomy,
kersten2025alertinvestigation}.
Analyst studies are therefore needed to measure comprehension, triage time,
decision quality, and automation bias.

\ourparagraph{Computational and external validity.}
DeepWeave is slower than optimized ML classifiers because inference includes
grounding, predicate evaluation, symbolic solving, and answer decoding.
Batching makes this overhead practical for the evaluated workloads, although
line-rate packet deployment may require parallelization or lightweight
screening, as explored by in-network detection systems
~\cite{barradas2021flowlens,dong2023horuseye,
mirnajafizadeh2024isdc}.
Scalability also depends on rule interactions, since dense logical dependencies
may increase compiled-circuit size.
Finally, public benchmarks do not capture long-term drift, sensor failures,
evolving applications, or analyst feedback.
Our results therefore establish DeepWeave's behavior in the evaluated
environments, not universal superiority in operational networks.

\section{Related Work}
\label{sec:related_work}

\ourparagraph{ML-based NIDS.}
Learning-based NIDSs reduce dependence on manually maintained signatures by
learning from packet, flow, or contextual observations.
Kitsune models benign traffic online with an autoencoder ensemble
\cite{mirsky2018kitsune}, while HorusEye combines in-switch anomaly screening
with deeper flow analysis \cite{dong2023horuseye}.
ARGUS models contextual IoT-device behavior \cite{rieger2023argus}, and
5G-SPECTOR applies protocol-aware analysis to cellular attacks
\cite{wen2024spector}.
These systems demonstrate flexible or context-aware detection, but their attack
knowledge is either implicit in learned models or specialized to one protocol
environment; learned detectors can also degrade across network distributions
\cite{sommer2010outside,cantone2024crossdataset}.
Deployment-oriented systems instead optimize data collection and inference:
FlowLens constructs compact flow markers for programmable switches
\cite{barradas2021flowlens}, and ISDC distributes flow selection and aggregation
across the network \cite{mirnajafizadeh2024isdc}.
\name{} complements this work by integrating packet-, flow-, and hybrid-level
learning with explicit probabilistic reasoning and proof-based explanations.

\ourparagraph{Explainable and Robust ML-based Intrusion Detection.}
xNIDS produces post-hoc explanations that account for temporal context and
feature dependencies and can generate defense rules \cite{wei2023xnids}.
However, its explanations are produced after prediction, whereas \name{} uses
symbolic clauses directly in inference and reports contributing proofs.
Robustness work emphasizes that adversarial network inputs must remain
realizable and preserve malicious functionality.
PANTS combines adversarial search with SMT constraints to generate valid
network inputs and robustify a target classifier \cite{jin2025robustifying}.
PANTS uses symbolic reasoning externally for adversarial generation; \name{}
embeds symbolic attack conditions within the detector and evaluates explicit
neural--symbolic coupling under shift, corruption, and constrained evasion.

\ourparagraph{Neuro-Symbolic NIDS.}
Bizzarri et al.\ guide neural training with logical constraints over benign and malicious traffic \cite{bizzarri_icccn}, while other recent pipelines combine neural representations with clustering, tree-based prediction, uncertainty, or transfer learning \cite{tran2025neurosymbolic}.
These studies establish the promise of symbolic supervision, but do not reason through a shared probabilistic logic program whose proofs determine predictions.
In \name{}, neural outputs instantiate probabilistic predicates and symbolic rules contribute directly to training, inference, and explanation.
The system additionally addresses how rules are constructed and validated, how multiple uncertain rules are aggregated, and how conflicts between neural and symbolic evidence are resolved through explicit coupling policies.

\section{Conclusions}
\label{sec:conclusions}

This paper presents \name{}, a probabilistic neuro-symbolic NIDS that combines learned traffic patterns with validated symbolic rules through explicit coupling policies. 
Across packet-, flow-, and hybrid-views, \name{} improves low-data detection and false-positive control, transfers useful knowledge across networks, and produces more stable explanations than post-hoc neural attributions. 
Despite the additional reasoning overhead, batching enables \name{} to process the complete flow workload at \(278.45\,\mu\text{s}\) per flow, remaining substantially faster than the evaluated packet-native pipelines. 
Nevertheless, accuracy and robustness depend on the chosen coupling policy as symbolic reasoning cannot recover absent information.
Overall, \name{} demonstrates that explicit integration of logic into ML-based NIDS provide data-efficient, adaptable, and inspectable detection at practical computational cost.
%
% Our results also show that accuracy and robustness depend on the coupling policy and available features, while enforcing that symbolic reasoning cannot recover absent distinctions and is not inherently robust. 
% %
% Overall, \name{} demonstrates that making logic a first-class component of ML-based NIDS can provide data-efficient, adaptable, and inspectable detection at practical computational cost.

\bibliographystyle{plain}
\bibliography{references}  %%% Uncomment this line and comment out the ``thebibliography'' section below to use the external .bib file (using bibtex) .





% \end{thebibliography}
\clearpage   % start the appendix on a fresh page so its floats stay after the references
\appendix
\section{Use of Generative AI}
\label{secapp:use_genai}
Generative AI tools -- mainly GPT 5.6 -- were used to improve the English language, clarity, and readability of the manuscript. Their use was limited primarily to grammar correction, sentence restructuring, and stylistic refinement. The authors developed the research methodology, implemented the system, conducted and interpreted the experiments, and wrote the substantive content of the manuscript. All AI-assisted text was reviewed, revised, and approved by the authors, who take full responsibility for the final content.

% \section{Dataset Details}
% \label{secapp:dataset_details}

\section{Additional \name{} Details}
\subsection{Detailed list of used features}
\label{secapp:list_features}
% \eb{asked Raoul to fill this section}
\begin{table*}[h]
\centering
\scriptsize
\caption{Packet-level feature set of the standardized CSE-CIC-IDS2018 view (22 per-packet header fields). Identifier and label columns (\texttt{ip.src}, \texttt{ip.dst}, \texttt{frame.time\_epoch}, \texttt{Attack}) are excluded from the model input. Features marked (c) are treated as categorical.}
\label{tab:annex_features_stdpacket}
\begin{tabular}{@{}l p{0.25\textwidth} l p{0.25\textwidth}@{}}
\toprule
Feature & Description & Feature & Description \\
\midrule
\texttt{ip.proto} (c) & IP protocol number (6 TCP, 17 UDP, \ldots) & \texttt{ip.len} & IP total length field \\
\texttt{ip.ttl} & IP time-to-live value & \texttt{tcp.seq} & TCP sequence number (relative) \\
\texttt{ip.dsfield} (c) & IP differentiated-services field (DSCP/ECN) & \texttt{tcp.ack} & TCP acknowledgement number (relative) \\
\texttt{frame.len} & Total captured frame length in bytes & \texttt{tcp.len} & TCP payload length in bytes \\
\texttt{srcport} & Transport-layer source port & \texttt{icmp.type} (c) & ICMP message type \\
\texttt{dstport} (c) & Transport-layer destination port & \texttt{icmp.code} (c) & ICMP message code \\
\texttt{tcp.flags} (c) & TCP flags byte of the segment & \texttt{dns.id} & DNS transaction identifier \\
\texttt{tcp.window\_size\_value} & Advertised TCP receive window (unscaled) & \texttt{dns.qry.type} (c) & DNS query type (A, AAAA, MX, \ldots) \\
\texttt{tcp.hdr\_len} & TCP header length in bytes & \texttt{dns.resp.ttl} & TTL of the DNS answer record \\
\texttt{udp.length} & UDP datagram length (header + payload) & \texttt{ftp.response.code} (c) & FTP server response code \\
\texttt{frame.time\_delta} & Inter-arrival time since previous packet & \texttt{frame.protocols} (c) & Protocol stack string (e.g.\ eth:ip:tcp:http) \\
\bottomrule
\end{tabular}
\end{table*}

\begin{table*}[h]
\centering
\scriptsize
\caption{Flow-level feature set of the CIC-IDS2017 view (79 CICFlowMeter features). Durations and inter-arrival times (IAT) are in microseconds; Fwd denotes the direction of the flow's first packet, Bwd the reverse. Features marked (c) are treated as categorical.}
\label{tab:annex_features_flow2017a}
\begin{tabular}{@{}l p{0.255\textwidth} l p{0.255\textwidth}@{}}
\toprule
Feature & Description & Feature & Description \\
\midrule
\texttt{Source Port} (c) & Transport-layer source port & \texttt{Bwd IAT Total} & Sum of IATs between backward packets \\
\texttt{Destination Port} (c) & Transport-layer destination port & \texttt{Bwd IAT Mean} & Mean backward IAT \\
\texttt{Protocol} (c) & IP protocol number (6 TCP, 17 UDP, \ldots) & \texttt{Bwd IAT Std} & Std.\ deviation of backward IAT \\
\texttt{Flow Duration} & Time between first and last packet & \texttt{Bwd IAT Max} & Maximum backward IAT \\
\texttt{Total Fwd Packets} & Packet count, forward direction & \texttt{Bwd IAT Min} & Minimum backward IAT \\
\texttt{Total Backward Packets} & Packet count, backward direction & \texttt{Fwd PSH Flags} (c) & Forward packets with TCP PSH set \\
\texttt{Total Length of Fwd Packets} & Sum of payload sizes, forward & \texttt{Bwd PSH Flags} & Backward packets with TCP PSH set \\
\texttt{Total Length of Bwd Packets} & Sum of payload sizes, backward & \texttt{Fwd URG Flags} (c) & Forward packets with TCP URG set \\
\texttt{Fwd Packet Length Max} & Largest packet size, forward & \texttt{Bwd URG Flags} & Backward packets with TCP URG set \\
\texttt{Fwd Packet Length Min} & Smallest packet size, forward & \texttt{Fwd Header Length} & Total header bytes, forward \\
\texttt{Fwd Packet Length Mean} & Mean packet size, forward & \texttt{Bwd Header Length} & Total header bytes, backward \\
\texttt{Fwd Packet Length Std} & Std.\ deviation of packet size, forward & \texttt{Fwd Packets/s} & Forward packet rate \\
\texttt{Bwd Packet Length Max} & Largest packet size, backward & \texttt{Bwd Packets/s} & Backward packet rate \\
\texttt{Bwd Packet Length Min} & Smallest packet size, backward & \texttt{Min Packet Length} & Smallest packet, both directions \\
\texttt{Bwd Packet Length Mean} & Mean packet size, backward & \texttt{Max Packet Length} & Largest packet, both directions \\
\texttt{Bwd Packet Length Std} & Std.\ deviation of packet size, backward & \texttt{Packet Length Mean} & Mean packet size, whole flow \\
\texttt{Flow Bytes/s} & Flow byte rate, both directions & \texttt{Packet Length Std} & Std.\ deviation of packet size, whole flow \\
\texttt{Flow Packets/s} & Flow packet rate, both directions & \texttt{Packet Length Variance} & Variance of packet size, whole flow \\
\texttt{Flow IAT Mean} & Mean IAT between consecutive flow packets & \texttt{FIN Flag Count} (c) & Packets with FIN set \\
\texttt{Flow IAT Std} & Std.\ deviation of flow IAT & \texttt{SYN Flag Count} (c) & Packets with SYN set \\
\texttt{Flow IAT Max} & Maximum flow IAT & \texttt{RST Flag Count} (c) & Packets with RST set \\
\texttt{Flow IAT Min} & Minimum flow IAT & \texttt{PSH Flag Count} (c) & Packets with PSH set \\
\texttt{Fwd IAT Total} & Sum of IATs between forward packets & \texttt{ACK Flag Count} (c) & Packets with ACK set \\
\texttt{Fwd IAT Mean} & Mean forward IAT & \texttt{URG Flag Count} (c) & Packets with URG set \\
\texttt{Fwd IAT Std} & Std.\ deviation of forward IAT & \texttt{CWE Flag Count} (c) & Packets with CWR set \\
\texttt{Fwd IAT Max} & Maximum forward IAT & \texttt{ECE Flag Count} (c) & Packets with ECE set \\
\texttt{Fwd IAT Min} & Minimum forward IAT & \texttt{Down/Up Ratio} & Download-to-upload packet ratio \\
\texttt{Average Packet Size} & Mean size over all packets & \texttt{Subflow Bwd Bytes} & Mean byte count per backward subflow \\
\texttt{Avg Fwd Segment Size} & Mean payload (segment) size, forward & \texttt{Init\_Win\_bytes\_forward} & TCP initial window announced forward \\
\texttt{Avg Bwd Segment Size} & Mean payload (segment) size, backward & \texttt{Init\_Win\_bytes\_backward} & TCP initial window announced backward \\
\texttt{Fwd Avg Bytes/Bulk} & Mean bytes per bulk burst, forward & \texttt{act\_data\_pkt\_fwd} & Forward packets with $\geq$1 byte of TCP payload \\
\texttt{Fwd Avg Packets/Bulk} & Mean packets per bulk burst, forward & \texttt{min\_seg\_size\_forward} & Minimum segment (header) size, forward \\
\texttt{Fwd Avg Bulk Rate} & Mean bulk transfer rate, forward & \texttt{Active Mean} & Mean active time before going idle \\
\texttt{Bwd Avg Bytes/Bulk} & Mean bytes per bulk burst, backward & \texttt{Active Std} & Std.\ deviation of active time \\
\texttt{Bwd Avg Packets/Bulk} & Mean packets per bulk burst, backward & \texttt{Active Max} & Maximum active time \\
\texttt{Bwd Avg Bulk Rate} & Mean bulk transfer rate, backward & \texttt{Active Min} & Minimum active time \\
\texttt{Subflow Fwd Packets} & Mean packet count per forward subflow & \texttt{Idle Mean} & Mean idle time before becoming active \\
\texttt{Subflow Fwd Bytes} & Mean byte count per forward subflow & \texttt{Idle Std} & Std.\ deviation of idle time \\
\texttt{Subflow Bwd Packets} & Mean packet count per backward subflow & \texttt{Idle Max} & Maximum idle time \\
\bottomrule
\end{tabular}
\end{table*}
\begin{table*}[h]
\centering
\scriptsize
\caption{Flow-level feature set of the hybrid view on the standardized CSE-CIC-IDS2018 data (49 NetFlow-v3-schema features, computed online from the packet stream per bidirectional 5-tuple). In denotes the client-to-server direction, out the reverse; durations and inter-arrival times (IAT) are in milliseconds.}
\label{tab:annex_features_hybridflow}
\begin{tabular}{@{}l p{0.21\textwidth} l p{0.21\textwidth}@{}}
\toprule
Feature & Description & Feature & Description \\
\midrule
\texttt{L4\_SRC\_PORT} & Source port of the flow's first packet & \texttt{SRC\_TO\_DST\_AVG\_THROUGHPUT} & Average throughput in, bits per second \\
\texttt{L4\_DST\_PORT} & Destination port of the flow's first packet & \texttt{DST\_TO\_SRC\_AVG\_THROUGHPUT} & Average throughput out, bits per second \\
\texttt{PROTOCOL} & IP protocol number of the flow & \texttt{NUM\_PKTS\_UP\_TO\_128\_BYTES} & Packets with frame length $\leq$128 bytes \\
\texttt{L7\_PROTO} & Application protocol code (from protocol stack, well-known-port fallback) & \texttt{NUM\_PKTS\_128\_TO\_256\_BYTES} & Packets with frame length 129--256 bytes \\
\texttt{IN\_BYTES} & Total bytes, client to server & \texttt{NUM\_PKTS\_256\_TO\_512\_BYTES} & Packets with frame length 257--512 bytes \\
\texttt{IN\_PKTS} & Total packets, client to server & \texttt{NUM\_PKTS\_512\_TO\_1024\_BYTES} & Packets with frame length 513--1024 bytes \\
\texttt{OUT\_BYTES} & Total bytes, server to client & \texttt{NUM\_PKTS\_1024\_TO\_1514\_BYTES} & Packets with frame length 1025--1514 bytes \\
\texttt{OUT\_PKTS} & Total packets, server to client & \texttt{TCP\_WIN\_MAX\_IN} & Maximum TCP window advertised in \\
\texttt{TCP\_FLAGS} & Cumulative OR of TCP flags, whole flow & \texttt{TCP\_WIN\_MAX\_OUT} & Maximum TCP window advertised out \\
\texttt{CLIENT\_TCP\_FLAGS} & Cumulative OR of TCP flags, in & \texttt{ICMP\_TYPE} & ICMP type and code ($\mathrm{type} \times 256 + \mathrm{code}$) \\
\texttt{SERVER\_TCP\_FLAGS} & Cumulative OR of TCP flags, out & \texttt{ICMP\_IPV4\_TYPE} & ICMP type alone \\
\texttt{FLOW\_DURATION\_MILLISECONDS} & Time between first and last packet & \texttt{DNS\_QUERY\_ID} & DNS transaction identifier \\
\texttt{DURATION\_IN} & Duration of the in packet stream & \texttt{DNS\_QUERY\_TYPE} & DNS query type \\
\texttt{DURATION\_OUT} & Duration of the out packet stream & \texttt{DNS\_TTL\_ANSWER} & TTL of the DNS answer record \\
\texttt{MIN\_TTL} & Minimum IP TTL in the flow & \texttt{FTP\_COMMAND\_RET\_CODE} & FTP server response code in the flow \\
\texttt{MAX\_TTL} & Maximum IP TTL in the flow & \texttt{SRC\_TO\_DST\_IAT\_MIN} & Minimum packet IAT, in \\
\texttt{LONGEST\_FLOW\_PKT} & Largest frame length in the flow & \texttt{SRC\_TO\_DST\_IAT\_MAX} & Maximum packet IAT, in \\
\texttt{SHORTEST\_FLOW\_PKT} & Smallest frame length in the flow & \texttt{SRC\_TO\_DST\_IAT\_AVG} & Mean packet IAT, in \\
\texttt{MIN\_IP\_PKT\_LEN} & Smallest IP total length in the flow & \texttt{SRC\_TO\_DST\_IAT\_STDDEV} & Std.\ deviation of packet IAT, in \\
\texttt{MAX\_IP\_PKT\_LEN} & Largest IP total length in the flow & \texttt{DST\_TO\_SRC\_IAT\_MIN} & Minimum packet IAT, out \\
\texttt{SRC\_TO\_DST\_SECOND\_BYTES} & Byte rate in, over in-stream duration & \texttt{DST\_TO\_SRC\_IAT\_MAX} & Maximum packet IAT, out \\
\texttt{DST\_TO\_SRC\_SECOND\_BYTES} & Byte rate out & \texttt{DST\_TO\_SRC\_IAT\_AVG} & Mean packet IAT, out \\
\texttt{RETRANSMITTED\_IN\_BYTES} & Retransmitted bytes, in (repeated TCP sequence numbers) & \texttt{DST\_TO\_SRC\_IAT\_STDDEV} & Std.\ deviation of packet IAT, out \\
\texttt{RETRANSMITTED\_IN\_PKTS} & Retransmitted packets, in & & \\
\texttt{RETRANSMITTED\_OUT\_BYTES} & Retransmitted bytes, out & & \\
\texttt{RETRANSMITTED\_OUT\_PKTS} & Retransmitted packets, out & & \\
\bottomrule
\end{tabular}
\end{table*}

\paragraph{Packet view.} As shown in~\Cref{tab:annex_features_stdpacket}, the packet view is restricted to 22 header fields readable at parsing time, with no payload inspection, so the detector remains applicable to encrypted traffic and can emit a verdict per packet without waiting for flow termination; address and timestamp columns are excluded from the model input to prevent identity and temporal leakage. Furthermore, using the standardized NetFlow-v3 version of the CSE-CIC-IDS2018 allows for a complete construction of the flow features proposed by the NetFlow-v3 through these 22 packet header features.\\

\paragraph{Flow view.} As shown in~\cref{tab:annex_features_flow2017a}, the flow view uses all 79 CICFlowMeter features proposed in the CIC-IDS2017 dataset: volumetric counts, packet-size and inter-arrival-time statistics per direction, TCP flag counts, and activity/idle timing. No manual feature selection is applied, resulting in a fair comparison with the machine-learning baselines (every model sees the same input) as well as SOTA models, and because the rule-mining stage performs its own feature selection, extracting a small interpretable subset as symbolic predicates while the neural branch consumes the full vector.\\

\paragraph{Hybrid View.} As shown in~\Cref{tab:annex_features_hybridflow}, the hybrid view follows the 49 features of the NetFlow-v3 schema of the standardized datasets but are computed incrementally
from the packet stream itself, per bidirectional 5-tuple. This gives the hybrid model the aggregate context that single packets lack (directional volumes, durations, retransmission and inter-arrival statistics) while preserving streaming semantics.

\subsection{Algorithm for automatic rule construction}
\label{secapp:rule_algorithm}
Algorithm~\ref{alg:rule_pipeline} gives the full offline rule-mining pipeline, with the concrete thresholds used in the experiments. The four stages -- scoring, calibration, program assembly, and soft rewrite -- run entirely on the scoring reference pool.
\begin{algorithm}[ht]
\caption{Offline symbolic-rule mining}
\label{alg:rule_pipeline}
\begin{algorithmic}[1]
\REQUIRE training/scoring pool $S$, validation split $V$, attack classes $\mathcal{A}$, top count $K=12$, statistic weights $w$
\ENSURE DeepProbLog program $\Pi$
\STATE $\mathcal{R} \gets \emptyset$
\FORALL{attack $a \in \mathcal{A}$}
  \FORALL{candidate feature $f$ eligible for the symbolic vocabulary}
    \STATE compute KS, Cliff's $\delta$, Hellinger, folded AUC (numeric) or Cramer's V, TVD, categorical AUC (categorical) on $S$ for $a$ versus benign
    \STATE $\mathrm{score}(f,a) \gets$ weighted mean of the per-attack min-max normalised statistics, using $w$
  \ENDFOR
  \STATE $F_a \gets$ top-$K$ features of $a$ by $\mathrm{score}(\cdot,a)$, each carrying its predicate hint (quantile tier or category)
  \STATE $C_a \gets$ candidate rule bodies: each top feature alone, plus short conjunctions of their hint predicates
  \FORALL{candidate body $c \in C_a$}
    \STATE measure precision $p$, recall $r$, benign FPR $\phi$ of $c$ on $V$; measure $p_{\mathrm{tr}},\phi_{\mathrm{tr}}$ on $S$
    \STATE $\mathit{stable} \gets (p_{\mathrm{tr}}-p \le 0.20) \wedge (\phi_{\mathrm{tr}}-\phi \le 0.05)$ \COMMENT{drop rules selective only on mining data}
    \IF{$\mathit{stable} \wedge p \ge 0.90 \wedge \phi \le 0.015$}
      \STATE role$(c) \gets$ \textsc{Sufficient} \COMMENT{may support $a$ on its own}
    \ELSIF{$\mathit{stable} \wedge r \ge 0.35 \wedge \phi \le 0.050$}
      \STATE role$(c) \gets$ \textsc{Helper} \COMMENT{fires only behind the neural prediction}
    \ELSIF{$\mathit{stable} \wedge p \ge 0.90 \wedge \phi \le 0.020$ on a rival class}
      \STATE role$(c) \gets$ \textsc{disqualifier}
    \ELSE
      \STATE discard $c$
    \ENDIF
    \STATE $\mathcal{R} \gets \mathcal{R} \cup \{c\}$
  \ENDFOR
\ENDFOR
\STATE group each class's sufficient and helper rules into per-class pockets; wrap pockets in learnable predicates $t(p)\!::\!symbolic\_a\_sufficient(X)$ whose initial $p$ is the calibrated confidence
\STATE attach the query head $classify/2$ for the configured coupling; emit hard externals
\STATE rewrite each hard external to its soft module and split multi-threshold bodies into single-external helpers
\RETURN $\Pi$
\end{algorithmic}
\end{algorithm}

\ourparagraph{Candidate-generation configuration.} \label{secapp:candidate_generation_config} 
In our experiments, atomic predicates are ranked by validation performance, and the search is restricted to the \(40\) highest-ranked candidates for each attack class.
Predicates with fewer than \(10\) validation true positives are excluded.
The pipeline evaluates non-redundant conjunctions of up to three predicates and applies the same minimum-support requirement to each conjunction.
To bound construction cost, it emits at most \(5{,}000\) candidate conjunctions and terminates the search after \(120\) seconds.
The predicate budget, rule length, support threshold, candidate limit, and time limit are configurable parameters rather than fixed properties of \name{}.
Table~\ref{tab:rule-role-gates} features the criteria used for assigning a role for each candidate rule. A candidate that failed its role gate, stability requirement, schema check, or provenance check is \texttt{rejected}. For an accepted rule, the rule strength (i.e. its initial learnable probability weight) is set to its validation precision, clipped to the interval $[0.05, 0.99].$ 

\begin{table}[h]
    \centering
    \caption{Shared validation gates for assigning roles to generated rules.}
    \label{tab:rule-role-gates}
    \begin{tabular}{lc}
        \toprule
        \textbf{Proposed Role} &
        \textbf{Acceptance criteria on validation set} \\
        \midrule
        Sufficient    & Precision $\geq$ 0.90 and FPR $\leq$ 0.015 \\
        Helper    & Recall $\geq$ 0.35 and FPR $\leq$ 0.050 \\
        Disqualifier & Precision $\geq$ 0.90 and FPR $\leq$ 0.020 \\
        \bottomrule
    \end{tabular}
\end{table}

\ourparagraph{Hyperparameter settings for rule construction.} \label{secapp:hyperparameter_settings_for_rule_construction}The rule miner ranks features separately for each attack-versus-benign comparison and retains the 12 highest-ranked stable features. A feature is stable when its validation score is at least $0.50$, its construction-to-validation score gap is below $0.20$, and its rank shifts by at most five positions. The ranking combines the Kolmogorov-Smirnov statistic, Cliff's delta, Hellinger distance, AUC, Cram\'{e}r's~$V$, and total variation distance with weights $0.30$, $0.20$, $0.20$, $0.20$, $0.05$, and $0.05$, respectively; weights are renormalized when a statistic does not apply.

Numeric predicates use lower-tail quantiles $Q_L=\{0.001,0.003,0.005,0.03,0.10,0.30\}$ and upper-tail quantiles $Q_U=\{0.80,0.90,0.97,0.995,0.997,0.999\}$, including intervals between adjacent quantiles. A categorical value is retained when it occurs in the target class and its non-target occurrence rate is below $0.02$.

\subsection{Optimization Hyperparameters}
\label{secapp:optimization_hyperparameters}
Table~\ref{tab:implementation_config} reports the parameters used in the experiments. We selected them through a two-stage search on the CIC-IDS2017 validation split. The first stage fixed training at five epochs and evaluated the 24 combinations
\[
b\in\{8,16,32\}, \qquad
\eta_{\mathrm{NN}},\eta_{\mathrm{DPL}},\eta_{\mathrm{pred}}
\in\{10^{-3},3\times10^{-3}\},
\]
where $b$ is the batch size and the three learning rates control the neural network, probabilistic logic parameters, and soft-predicate networks. We ranked each configuration by
\[
S=F_{1,\mathrm{macro}}-0.5\,\mathrm{FPR}_{\mathrm{attack}}
-0.25\,\sigma(F_{1,\mathrm{macro}}).
\]

The selected configuration, $b=8$, $\eta_{\mathrm{NN}}=\eta_{\mathrm{DPL}}=3\times10^{-3}$, and $\eta_{\mathrm{pred}}=10^{-3}$, obtained $0.835$ macro-F1, $0.030$ attack FPR, and a selection score of $0.820$.

The second stage fixed this configuration and compared $E\in\{5,10,25,50\}$ epochs over five replicates. The corresponding macro-F1 scores were $0.906\pm0.004$, $0.895\pm0.025$, $0.904\pm0.018$, and $0.891\pm0.028$. None differed significantly from the ten-epoch configuration, which we retained as a conservative balance between convergence and cost.

\begin{table}[h]
\centering
\small
\caption{Fixed \name{} implementation parameters.}
\begin{tabular}{ll}
\toprule
\textbf{Component} & \textbf{Configuration} \\
\midrule
Neural model & MLP: 256--128--64 \\
Activation and dropout & ReLU, 0.3 \\
Optimizer & Adam \\
Neural learning rate & \(3\times10^{-3}\) \\
Rule-weight learning rate & \(3\times10^{-3}\) \\
Predicate learning rate & \(1\times10^{-3}\) \\
Training batch size & 8 \\
Inference batch size & 8192 \\
Initial predicate steepness & 50 \\
Temperature schedule & \(1.0\rightarrow0.5\) \\
\bottomrule
\end{tabular}
\label{tab:implementation_config}
\end{table}

\subsection{Effect of vectorized inference on \name{}}
\label{secapp:deepweave_batch_optimization}

The initial evaluator invoked the DeepProbLog solving path for each flow, repeating symbolic traversal and framework overhead even though all queries shared an arithmetic circuit. The optimized evaluator instead stacks the neural-leaf values of multiple queries and evaluates the shared circuit once per batch.

The diagnostic measured query construction, neural-predicate evaluation, symbolic solving, and answer decoding on $1{,}000$ held-out CIC-IDS2017 flows. Three matched executions used a batch size of one for the original evaluator and $1{,}024$ for the optimized evaluator. Training, backpropagation, feature preprocessing, and packet-to-flow assembly were excluded.

Table~\ref{tab:deepweave_speedup} reports a reduction in median runtime from $11.781$ to $0.440$ seconds. The effective cost falls from $11{,}781.0$ to $439.6\,\mu\mathrm{s}$ per flow, while throughput increases from $84.9$ to $2{,}274.9$ flows per second. This corresponds to a $26.8\times$ evaluator speedup and a $96.27\%$ runtime reduction for the tested workload.

The same checkpoint produced identical predictions for all $1{,}000$ flows. Vectorization therefore changes execution without changing the classifier output.

\begin{table}[tp]
    \centering
    \fontsize{8}{9.5}\selectfont   % not needed if you use the updated template.tex
    \caption{Matched-checkpoint comparison of the original per-flow \name{} evaluator and the optimized vectorized evaluator on 1,000 held-out CIC-IDS2017 flows.}
    \label{tab:deepweave_speedup}
    \begin{tabular}{lrrrrrr}
        \toprule
        Evaluator &
        Batch size &
        Time (s) [IQR] &
        $\mu$s/flow &
        Flows/s &
        Speedup &
        Agreement \\
        \midrule
        Original \name{}
            & 1
            & 11.781 [11.646, 11.797]
            & 11,781.0
            & 84.9
            & 1.0$\times$
            & 100\% \\
        Optimized \name{}
            & 1,024
            & 0.440 [0.405, 0.443]
            & 439.6
            & 2,274.9
            & 26.8$\times$
            & 100\% \\
        \bottomrule
    \end{tabular}
\end{table}

\section{Packet, Flow and Hybrid Extensive Results}
% \label{secapp:all_results}

\subsection{Detection Performance and Efficiency Extensive results}
\label{secapp:all_results}
This section reports the training-size sweep for the packet, flow, and hybrid views. We compare multiclass macro-F1 (MF1), attack false-positive rate (FPR), and detection rate (DR), and measure data efficiency by the labelled budget $k$ needed to reach a given MF1.

\subsubsection{Packet view}
Table~\ref{tab:packet_sota_sweep} shows the largest gains at small training sizes. At $k=50$, Anchored, Deferred, and Guarded reach $0.977$, $0.951$, and $0.924$ MF1, respectively, compared with $0.632$ for the neural baseline, $0.811$ for HGB, and $0.884$ for the random forest. These three couplings already exceed the neural baseline's best result of $0.869$ at $k=10000$. They therefore reach the same performance level with $50$ rather than $10{,}000$ labelled examples per attack class, corresponding to a $200\times$ reduction in required training data. The gap narrows with more labels: the random forest reaches $0.995$ MF1 at $k=5000$, while Anchored and Deferred reach $0.985$ and $0.974$.

The coupled models also maintain low FPR. Anchored records $0.001$ at $k=50$ and values that round to zero thereafter, while Deferred, Guarded, and Hard Gate remain at or below $0.008$. Most methods already attain high DR, so differences in FPR and attack-family assignment account for most of the MF1 gap. Suricata illustrates this distinction: dataset-specific rules raise its DR from $0.184$ to $0.868$, but its FPR remains $0.280$.

\begin{table*}[h]
\centering
\scriptsize
\setlength{\tabcolsep}{3pt}
\caption{Packet-view comparison on CIC-IDS2018, across the per-class training-size $k$. F1 indicates the macro-F1 for multi-class classification models and binary-F1 for binary classification baselines (Suricata). Both F1 and FPR are reported as average $\pm$ standard deviation. Best F1 (highest) and FPR (lowest) for each $k$ are shown in bold. Suricata is signature-based and training-size-independent, thus is not affected by $k$.}
\label{tab:packet_sota_sweep}
\resizebox{\textwidth}{!}{%
\begin{tabular}{l rr rr rr rr rr rr rr rr}
\toprule
& \multicolumn{2}{c}{${k{=}50}$} & \multicolumn{2}{c}{${k{=}100}$} & \multicolumn{2}{c}{${k{=}250}$} & \multicolumn{2}{c}{${k{=}500}$} & \multicolumn{2}{c}{${k{=}1000}$} & \multicolumn{2}{c}{${k{=}2500}$} & \multicolumn{2}{c}{${k{=}5000}$} & \multicolumn{2}{c}{${k{=}10000}$}\\
\cmidrule(lr){2-3} \cmidrule(lr){4-5} \cmidrule(lr){6-7} \cmidrule(lr){8-9} \cmidrule(lr){10-11} \cmidrule(lr){12-13} \cmidrule(lr){14-15} \cmidrule(lr){16-17}
System & F1 & FPR & F1 & FPR & F1 & FPR & F1 & FPR & F1 & FPR & F1 & FPR & F1 & FPR & F1 & FPR\\
\midrule
Neural baseline & $0.632{\scriptstyle\pm.044}$ & $0.075{\scriptstyle\pm.029}$ & $0.702{\scriptstyle\pm.029}$ & $0.063{\scriptstyle\pm.011}$ & $0.774{\scriptstyle\pm.052}$ & $0.033{\scriptstyle\pm.017}$ & $0.776{\scriptstyle\pm.026}$ & $0.031{\scriptstyle\pm.007}$ & $0.843{\scriptstyle\pm.024}$ & $0.020{\scriptstyle\pm.010}$ & $0.856{\scriptstyle\pm.017}$ & $0.011{\scriptstyle\pm.004}$ & $0.863{\scriptstyle\pm.019}$ & $0.013{\scriptstyle\pm.006}$ & $0.869{\scriptstyle\pm.011}$ & $0.012{\scriptstyle\pm.004}$\\
Symbolic baseline & $0.725{\scriptstyle\pm.005}$ & $0.046{\scriptstyle\pm.000}$ & $0.768{\scriptstyle\pm.064}$ & $0.033{\scriptstyle\pm.016}$ & $0.833{\scriptstyle\pm.016}$ & $0.020{\scriptstyle\pm.013}$ & $0.812{\scriptstyle\pm.016}$ & $0.026{\scriptstyle\pm.016}$ & $0.834{\scriptstyle\pm.013}$ & $0.018{\scriptstyle\pm.013}$ & $0.846{\scriptstyle\pm.005}$ & $0.011{\scriptstyle\pm.000}$ & $0.857{\scriptstyle\pm.006}$ & $0.007{\scriptstyle\pm.000}$ & $0.842{\scriptstyle\pm.017}$ & $0.007{\scriptstyle\pm.000}$\\
Decision tree & $0.688{\scriptstyle\pm.035}$ & $0.064{\scriptstyle\pm.019}$ & $0.757{\scriptstyle\pm.060}$ & $0.103{\scriptstyle\pm.110}$ & $0.778{\scriptstyle\pm.083}$ & $0.094{\scriptstyle\pm.096}$ & $0.801{\scriptstyle\pm.047}$ & $0.081{\scriptstyle\pm.102}$ & $0.776{\scriptstyle\pm.059}$ & $0.082{\scriptstyle\pm.127}$ & $0.834{\scriptstyle\pm.045}$ & $0.048{\scriptstyle\pm.043}$ & $0.881{\scriptstyle\pm.048}$ & $0.026{\scriptstyle\pm.017}$ & $0.928{\scriptstyle\pm.049}$ & $0.006{\scriptstyle\pm.009}$\\
Logistic regression & $0.542{\scriptstyle\pm.016}$ & $0.318{\scriptstyle\pm.031}$ & $0.493{\scriptstyle\pm.027}$ & $0.343{\scriptstyle\pm.021}$ & $0.490{\scriptstyle\pm.047}$ & $0.352{\scriptstyle\pm.028}$ & $0.553{\scriptstyle\pm.023}$ & $0.303{\scriptstyle\pm.007}$ & $0.493{\scriptstyle\pm.016}$ & $0.362{\scriptstyle\pm.017}$ & $0.489{\scriptstyle\pm.024}$ & $0.380{\scriptstyle\pm.011}$ & $0.499{\scriptstyle\pm.017}$ & $0.366{\scriptstyle\pm.023}$ & $0.500{\scriptstyle\pm.011}$ & $0.378{\scriptstyle\pm.015}$\\
Random forest & $0.884{\scriptstyle\pm.052}$ & $0.017{\scriptstyle\pm.010}$ & $\mathbf{0.972{\scriptstyle\pm.007}}$ & $0.004{\scriptstyle\pm.002}$ & $0.960{\scriptstyle\pm.041}$ & $0.003{\scriptstyle\pm.002}$ & $0.902{\scriptstyle\pm.056}$ & $0.020{\scriptstyle\pm.023}$ & $0.924{\scriptstyle\pm.061}$ & $0.012{\scriptstyle\pm.021}$ & $0.966{\scriptstyle\pm.026}$ & $0.001{\scriptstyle\pm.000}$ & $\mathbf{0.995{\scriptstyle\pm.003}}$ & $\mathbf{0.000{\scriptstyle\pm.000}}$ & $\mathbf{0.992{\scriptstyle\pm.008}}$ & $0.001{\scriptstyle\pm.001}$\\
HGB & $0.811{\scriptstyle\pm.045}$ & $0.021{\scriptstyle\pm.013}$ & $0.900{\scriptstyle\pm.044}$ & $0.014{\scriptstyle\pm.014}$ & $0.898{\scriptstyle\pm.052}$ & $0.017{\scriptstyle\pm.014}$ & $0.753{\scriptstyle\pm.099}$ & $0.177{\scriptstyle\pm.201}$ & $0.857{\scriptstyle\pm.128}$ & $0.070{\scriptstyle\pm.123}$ & $0.889{\scriptstyle\pm.082}$ & $0.018{\scriptstyle\pm.032}$ & $0.943{\scriptstyle\pm.031}$ & $0.001{\scriptstyle\pm.001}$ & $0.949{\scriptstyle\pm.056}$ & $0.006{\scriptstyle\pm.012}$\\
\midrule
\name{} (Anchored) & $\mathbf{0.977{\scriptstyle\pm.013}}$ & $\mathbf{0.001{\scriptstyle\pm.001}}$ & $\mathbf{0.972{\scriptstyle\pm.032}}$ & $\mathbf{0.000{\scriptstyle\pm.000}}$ & $0.972{\scriptstyle\pm.031}$ & $\mathbf{0.000{\scriptstyle\pm.000}}$ & $0.942{\scriptstyle\pm.028}$ & $\mathbf{0.000{\scriptstyle\pm.000}}$ & $0.969{\scriptstyle\pm.023}$ & $\mathbf{0.000{\scriptstyle\pm.000}}$ & $0.985{\scriptstyle\pm.007}$ & $\mathbf{0.000{\scriptstyle\pm.000}}$ & $0.985{\scriptstyle\pm.007}$ & $\mathbf{0.000{\scriptstyle\pm.000}}$ & $0.988{\scriptstyle\pm.006}$ & $0.002{\scriptstyle\pm.001}$\\
\name{} (Guarded) & $0.924{\scriptstyle\pm.009}$ & $0.008{\scriptstyle\pm.002}$ & $0.906{\scriptstyle\pm.020}$ & $0.004{\scriptstyle\pm.002}$ & $0.930{\scriptstyle\pm.014}$ & $0.001{\scriptstyle\pm.000}$ & $0.940{\scriptstyle\pm.011}$ & $0.002{\scriptstyle\pm.001}$ & $0.936{\scriptstyle\pm.026}$ & $0.001{\scriptstyle\pm.000}$ & $0.948{\scriptstyle\pm.012}$ & $0.001{\scriptstyle\pm.000}$ & $0.916{\scriptstyle\pm.084}$ & $0.001{\scriptstyle\pm.000}$ & $0.946{\scriptstyle\pm.003}$ & $0.001{\scriptstyle\pm.000}$\\
\name{} (Hard Gate) & $0.823{\scriptstyle\pm.023}$ & $0.005{\scriptstyle\pm.002}$ & $0.862{\scriptstyle\pm.031}$ & $0.003{\scriptstyle\pm.001}$ & $0.922{\scriptstyle\pm.025}$ & $0.001{\scriptstyle\pm.000}$ & $0.914{\scriptstyle\pm.020}$ & $0.001{\scriptstyle\pm.000}$ & $0.931{\scriptstyle\pm.012}$ & $0.001{\scriptstyle\pm.000}$ & $0.923{\scriptstyle\pm.016}$ & $\mathbf{0.000{\scriptstyle\pm.000}}$ & $0.927{\scriptstyle\pm.012}$ & $\mathbf{0.000{\scriptstyle\pm.000}}$ & $0.919{\scriptstyle\pm.031}$ & $0.004{\scriptstyle\pm.002}$\\
\name{} (Safety Net) & $0.876{\scriptstyle\pm.018}$ & $0.030{\scriptstyle\pm.008}$ & $0.893{\scriptstyle\pm.025}$ & $0.021{\scriptstyle\pm.007}$ & $0.855{\scriptstyle\pm.037}$ & $0.040{\scriptstyle\pm.015}$ & $0.894{\scriptstyle\pm.025}$ & $0.008{\scriptstyle\pm.006}$ & $0.867{\scriptstyle\pm.045}$ & $0.018{\scriptstyle\pm.008}$ & $0.879{\scriptstyle\pm.038}$ & $0.015{\scriptstyle\pm.007}$ & $0.878{\scriptstyle\pm.046}$ & $0.012{\scriptstyle\pm.006}$ & $0.846{\scriptstyle\pm.055}$ & $0.024{\scriptstyle\pm.010}$\\
\name{} (Same-Class Boost) & $0.644{\scriptstyle\pm.030}$ & $0.071{\scriptstyle\pm.017}$ & $0.697{\scriptstyle\pm.044}$ & $0.061{\scriptstyle\pm.008}$ & $0.771{\scriptstyle\pm.021}$ & $0.037{\scriptstyle\pm.014}$ & $0.758{\scriptstyle\pm.034}$ & $0.035{\scriptstyle\pm.014}$ & $0.838{\scriptstyle\pm.017}$ & $0.021{\scriptstyle\pm.009}$ & $0.858{\scriptstyle\pm.010}$ & $0.010{\scriptstyle\pm.002}$ & $0.822{\scriptstyle\pm.094}$ & $0.016{\scriptstyle\pm.015}$ & $0.870{\scriptstyle\pm.022}$ & $0.013{\scriptstyle\pm.008}$\\
\name{} (Deferred) & $0.951{\scriptstyle\pm.019}$ & $0.007{\scriptstyle\pm.003}$ & $0.970{\scriptstyle\pm.006}$ & $0.003{\scriptstyle\pm.001}$ & $\mathbf{0.983{\scriptstyle\pm.007}}$ & $0.001{\scriptstyle\pm.001}$ & $0.982{\scriptstyle\pm.004}$ & $0.001{\scriptstyle\pm.000}$ & $0.977{\scriptstyle\pm.003}$ & $0.001{\scriptstyle\pm.000}$ & $0.975{\scriptstyle\pm.002}$ & $0.001{\scriptstyle\pm.000}$ & $0.974{\scriptstyle\pm.002}$ & $\mathbf{0.000{\scriptstyle\pm.000}}$ & $0.987{\scriptstyle\pm.001}$ & $0.003{\scriptstyle\pm.001}$\\
\midrule
SPHBI~\cite{elsherif2025sphbi} & $0.896{\scriptstyle\pm.019}$ & $0.012{\scriptstyle\pm.004}$ & $0.943{\scriptstyle\pm.014}$ & $0.007{\scriptstyle\pm.001}$ & $0.966{\scriptstyle\pm.018}$ & $0.003{\scriptstyle\pm.002}$ & $\mathbf{0.985{\scriptstyle\pm.004}}$ & $0.001{\scriptstyle\pm.000}$ & $\mathbf{0.992{\scriptstyle\pm.005}}$ & $0.001{\scriptstyle\pm.000}$ & $\mathbf{0.990{\scriptstyle\pm.010}}$ & $\mathbf{0.000{\scriptstyle\pm.000}}$ & $0.992{\scriptstyle\pm.010}$ & $\mathbf{0.000{\scriptstyle\pm.000}}$ & $0.984{\scriptstyle\pm.023}$ & $\mathbf{0.000{\scriptstyle\pm.000}}$\\
CNN-WDLSTM~\cite{hassan2020hybrid} & $0.742{\scriptstyle\pm.030}$ & $0.034{\scriptstyle\pm.014}$ & $0.773{\scriptstyle\pm.031}$ & $0.026{\scriptstyle\pm.011}$ & $0.817{\scriptstyle\pm.028}$ & $0.013{\scriptstyle\pm.008}$ & $0.818{\scriptstyle\pm.016}$ & $0.008{\scriptstyle\pm.001}$ & $0.845{\scriptstyle\pm.017}$ & $0.008{\scriptstyle\pm.003}$ & $0.853{\scriptstyle\pm.026}$ & $0.012{\scriptstyle\pm.005}$ & $0.884{\scriptstyle\pm.016}$ & $0.007{\scriptstyle\pm.001}$ & $0.858{\scriptstyle\pm.019}$ & $0.011{\scriptstyle\pm.001}$\\
\midrule
Suricata (ET Open)$^\dagger$ & $0.449$ & $0.279$ & $0.449$ & $0.279$ & $0.449$ & $0.279$ & $0.449$ & $0.279$ & $0.449$ & $0.279$ & $0.449$ & $0.279$ & $0.449$ & $0.279$ & $0.449$ & $0.279$\\
Suricata (ET Open + published)$^\dagger$ & $0.624$ & $0.280$ & $0.624$ & $0.280$ & $0.624$ & $0.280$ & $0.624$ & $0.280$ & $0.624$ & $0.280$ & $0.624$ & $0.280$ & $0.624$ & $0.280$ & $0.624$ & $0.280$\\
\bottomrule
\end{tabular}}
\end{table*}

\subsubsection{Flow view}
The flow results in Table~\ref{tab:flow_sota_sweep} strengthen this low-data pattern. At $k=50$, Guarded, Anchored, and Hard Gate obtain MF1 scores of $0.849$, $0.833$, and $0.838$, compared with $0.595$ for the neural baseline and $0.574$ for HGB\@. All three exceed the neural baseline's best sweep result of approximately $0.79$ at $k=10000$, again reducing the equivalent labeled budget from $10{,}000$ to $50$ examples per attack class. Neither component produces this result alone, as the symbolic baseline remains between approximately $0.58$ and $0.66$, so neither branch alone explains the coupled result. At the largest budget, however, HGB reaches $0.936$ MF1 and exceeds the gated variants, which lie between $0.921$ and $0.925$.

The gated policies keep FPR at or below $0.014$ while maintaining DR near $0.90$ to $1.00$. Logistic regression also detects many attacks, but its FPR remains between approximately $0.22$ and $0.28$. Safety Net is the main exception among the coupled models: its permissive rescue path raises FPR to roughly $0.02$ to $0.08$. The flow-view advantage therefore comes from combining high detection with low FPR and correct family assignment.

\begin{table*}[t]
\centering
\scriptsize
\setlength{\tabcolsep}{3pt}
\caption{Flow-view comparison on CIC-IDS2017 across
training sizes $k$. Macro-F1 for multi-class classification models and binary-F1 for anomaly-based baselines of Kitsune and KnowML are reported. Both F1 and FPR are reported as average $\pm$ standard deviation. Best F1 (highest) and FPR (lowest) for each $k$ are shown in bold.}
\label{tab:flow_sota_sweep}
\resizebox{\textwidth}{!}{%
\begin{tabular}{l *{8}{rrr}}
\toprule
& \multicolumn{3}{c}{${k{=}50}$} & \multicolumn{3}{c}{${k{=}100}$} & \multicolumn{3}{c}{${k{=}250}$} & \multicolumn{3}{c}{${k{=}500}$} & \multicolumn{3}{c}{${k{=}1000}$} & \multicolumn{3}{c}{${k{=}2500}$} & \multicolumn{3}{c}{${k{=}5000}$} & \multicolumn{3}{c}{${k{=}10000}$} \\
\cmidrule(lr){2-4}\cmidrule(lr){5-7}\cmidrule(lr){8-10}\cmidrule(lr){11-13}\cmidrule(lr){14-16}\cmidrule(lr){17-19}\cmidrule(lr){20-22}\cmidrule(lr){23-25}
System & MF1 & BF1 & FPR & MF1 & BF1 & FPR & MF1 & BF1 & FPR & MF1 & BF1 & FPR & MF1 & BF1 & FPR & MF1 & BF1 & FPR & MF1 & BF1 & FPR & MF1 & BF1 & FPR \\
\midrule
Neural baseline & $0.595{\scriptstyle\pm.016}$ & $0.902{\scriptstyle\pm.009}$ & $0.071{\scriptstyle\pm.015}$ & $0.617{\scriptstyle\pm.020}$ & $0.906{\scriptstyle\pm.019}$ & $0.071{\scriptstyle\pm.023}$ & $0.670{\scriptstyle\pm.012}$ & $0.946{\scriptstyle\pm.004}$ & $0.039{\scriptstyle\pm.003}$ & $0.709{\scriptstyle\pm.027}$ & $0.957{\scriptstyle\pm.008}$ & $0.033{\scriptstyle\pm.006}$ & $0.762{\scriptstyle\pm.024}$ & $0.972{\scriptstyle\pm.009}$ & $0.021{\scriptstyle\pm.007}$ & $0.774{\scriptstyle\pm.004}$ & $0.976{\scriptstyle\pm.007}$ & $0.018{\scriptstyle\pm.005}$ & $0.789{\scriptstyle\pm.004}$ & $0.983{\scriptstyle\pm.000}$ & $0.013{\scriptstyle\pm.000}$ & $0.786{\scriptstyle\pm.024}$ & $0.970{\scriptstyle\pm.010}$ & $0.022{\scriptstyle\pm.008}$ \\
Symbolic baseline & $0.584{\scriptstyle\pm.060}$ & $0.837{\scriptstyle\pm.000}$ & $0.142{\scriptstyle\pm.000}$ & $0.584{\scriptstyle\pm.060}$ & $0.837{\scriptstyle\pm.000}$ & $0.142{\scriptstyle\pm.000}$ & $0.554{\scriptstyle\pm.073}$ & $0.837{\scriptstyle\pm.000}$ & $0.142{\scriptstyle\pm.000}$ & $0.623{\scriptstyle\pm.005}$ & $0.867{\scriptstyle\pm.015}$ & $0.070{\scriptstyle\pm.037}$ & $0.626{\scriptstyle\pm.002}$ & $0.876{\scriptstyle\pm.000}$ & $0.034{\scriptstyle\pm.000}$ & $0.617{\scriptstyle\pm.000}$ & $0.839{\scriptstyle\pm.000}$ & $0.119{\scriptstyle\pm.000}$ & $0.603{\scriptstyle\pm.055}$ & $0.838{\scriptstyle\pm.001}$ & $0.124{\scriptstyle\pm.009}$ & $0.664{\scriptstyle\pm.060}$ & $0.840{\scriptstyle\pm.000}$ & $0.117{\scriptstyle\pm.000}$ \\
Decision tree & $0.452{\scriptstyle\pm.098}$ & $0.774{\scriptstyle\pm.041}$ & $0.174{\scriptstyle\pm.072}$ & $0.514{\scriptstyle\pm.084}$ & $0.827{\scriptstyle\pm.034}$ & $0.102{\scriptstyle\pm.051}$ & $0.594{\scriptstyle\pm.038}$ & $0.896{\scriptstyle\pm.034}$ & $0.064{\scriptstyle\pm.035}$ & $0.662{\scriptstyle\pm.030}$ & $0.873{\scriptstyle\pm.045}$ & $0.088{\scriptstyle\pm.031}$ & $0.673{\scriptstyle\pm.059}$ & $0.878{\scriptstyle\pm.025}$ & $0.096{\scriptstyle\pm.026}$ & $0.730{\scriptstyle\pm.132}$ & $0.926{\scriptstyle\pm.050}$ & $0.062{\scriptstyle\pm.047}$ & $0.726{\scriptstyle\pm.055}$ & $0.936{\scriptstyle\pm.015}$ & $0.053{\scriptstyle\pm.014}$ & $0.853{\scriptstyle\pm.104}$ & $0.966{\scriptstyle\pm.043}$ & $0.029{\scriptstyle\pm.038}$ \\
Logistic regression & $0.443{\scriptstyle\pm.014}$ & $0.731{\scriptstyle\pm.016}$ & $0.275{\scriptstyle\pm.026}$ & $0.459{\scriptstyle\pm.013}$ & $0.746{\scriptstyle\pm.009}$ & $0.258{\scriptstyle\pm.011}$ & $0.484{\scriptstyle\pm.012}$ & $0.764{\scriptstyle\pm.004}$ & $0.241{\scriptstyle\pm.004}$ & $0.487{\scriptstyle\pm.015}$ & $0.766{\scriptstyle\pm.008}$ & $0.239{\scriptstyle\pm.009}$ & $0.496{\scriptstyle\pm.009}$ & $0.773{\scriptstyle\pm.011}$ & $0.231{\scriptstyle\pm.013}$ & $0.502{\scriptstyle\pm.005}$ & $0.780{\scriptstyle\pm.009}$ & $0.223{\scriptstyle\pm.011}$ & $0.508{\scriptstyle\pm.004}$ & $0.781{\scriptstyle\pm.007}$ & $0.221{\scriptstyle\pm.007}$ & $0.500{\scriptstyle\pm.008}$ & $0.770{\scriptstyle\pm.010}$ & $0.234{\scriptstyle\pm.012}$ \\
Random forest & $0.679{\scriptstyle\pm.044}$ & $0.925{\scriptstyle\pm.020}$ & $0.049{\scriptstyle\pm.021}$ & $0.741{\scriptstyle\pm.028}$ & $0.947{\scriptstyle\pm.017}$ & $0.034{\scriptstyle\pm.014}$ & $0.759{\scriptstyle\pm.009}$ & $0.954{\scriptstyle\pm.010}$ & $0.032{\scriptstyle\pm.008}$ & $0.768{\scriptstyle\pm.009}$ & $0.958{\scriptstyle\pm.009}$ & $0.031{\scriptstyle\pm.008}$ & $0.782{\scriptstyle\pm.033}$ & $0.959{\scriptstyle\pm.014}$ & $0.032{\scriptstyle\pm.011}$ & $0.800{\scriptstyle\pm.012}$ & $0.975{\scriptstyle\pm.013}$ & $0.019{\scriptstyle\pm.011}$ & $0.771{\scriptstyle\pm.011}$ & $0.963{\scriptstyle\pm.003}$ & $0.029{\scriptstyle\pm.003}$ & $0.891{\scriptstyle\pm.055}$ & $0.993{\scriptstyle\pm.004}$ & $0.005{\scriptstyle\pm.003}$ \\
HGB & $0.574{\scriptstyle\pm.077}$ & $0.853{\scriptstyle\pm.055}$ & $0.115{\scriptstyle\pm.074}$ & $0.636{\scriptstyle\pm.130}$ & $0.921{\scriptstyle\pm.065}$ & $0.051{\scriptstyle\pm.066}$ & $0.709{\scriptstyle\pm.059}$ & $0.950{\scriptstyle\pm.025}$ & $0.028{\scriptstyle\pm.020}$ & $0.727{\scriptstyle\pm.043}$ & $0.941{\scriptstyle\pm.028}$ & $0.040{\scriptstyle\pm.017}$ & $0.778{\scriptstyle\pm.058}$ & $0.954{\scriptstyle\pm.022}$ & $0.037{\scriptstyle\pm.019}$ & $0.887{\scriptstyle\pm.050}$ & $\mathbf{0.987{\scriptstyle\pm.014}}$ & $0.010{\scriptstyle\pm.011}$ & $0.825{\scriptstyle\pm.059}$ & $0.967{\scriptstyle\pm.021}$ & $0.027{\scriptstyle\pm.018}$ & $\mathbf{0.936{\scriptstyle\pm.010}}$ & $\mathbf{0.998{\scriptstyle\pm.000}}$ & $\mathbf{0.002{\scriptstyle\pm.000}}$ \\
\midrule
\name{} (Anchored) & $0.833{\scriptstyle\pm.016}$ & $0.955{\scriptstyle\pm.014}$ & $0.008{\scriptstyle\pm.001}$ & $0.842{\scriptstyle\pm.011}$ & $0.965{\scriptstyle\pm.003}$ & $0.009{\scriptstyle\pm.003}$ & $0.841{\scriptstyle\pm.024}$ & $\mathbf{0.969{\scriptstyle\pm.005}}$ & $0.011{\scriptstyle\pm.002}$ & $0.899{\scriptstyle\pm.018}$ & $\mathbf{0.977{\scriptstyle\pm.002}}$ & $0.010{\scriptstyle\pm.002}$ & $0.913{\scriptstyle\pm.005}$ & $0.980{\scriptstyle\pm.001}$ & $0.009{\scriptstyle\pm.001}$ & $0.915{\scriptstyle\pm.003}$ & $0.982{\scriptstyle\pm.000}$ & $0.009{\scriptstyle\pm.000}$ & $0.919{\scriptstyle\pm.006}$ & $0.983{\scriptstyle\pm.001}$ & $0.008{\scriptstyle\pm.001}$ & $0.921{\scriptstyle\pm.006}$ & $0.981{\scriptstyle\pm.002}$ & $0.009{\scriptstyle\pm.001}$ \\
\name{} (Same-Class Boost) & $0.672{\scriptstyle\pm.013}$ & $0.894{\scriptstyle\pm.022}$ & $0.045{\scriptstyle\pm.010}$ & $0.677{\scriptstyle\pm.014}$ & $0.912{\scriptstyle\pm.011}$ & $0.039{\scriptstyle\pm.011}$ & $0.673{\scriptstyle\pm.030}$ & $0.913{\scriptstyle\pm.028}$ & $0.043{\scriptstyle\pm.007}$ & $0.690{\scriptstyle\pm.027}$ & $0.947{\scriptstyle\pm.009}$ & $0.038{\scriptstyle\pm.007}$ & $0.765{\scriptstyle\pm.021}$ & $0.964{\scriptstyle\pm.009}$ & $0.026{\scriptstyle\pm.007}$ & $0.788{\scriptstyle\pm.012}$ & $0.968{\scriptstyle\pm.008}$ & $0.024{\scriptstyle\pm.006}$ & $0.810{\scriptstyle\pm.013}$ & $0.976{\scriptstyle\pm.009}$ & $0.018{\scriptstyle\pm.007}$ & $0.818{\scriptstyle\pm.021}$ & $0.966{\scriptstyle\pm.011}$ & $0.027{\scriptstyle\pm.009}$ \\
\name{} (Guarded) & $0.849{\scriptstyle\pm.016}$ & $0.960{\scriptstyle\pm.005}$ & $0.011{\scriptstyle\pm.005}$ & $0.864{\scriptstyle\pm.013}$ & $0.965{\scriptstyle\pm.008}$ & $\mathbf{0.007{\scriptstyle\pm.002}}$ & $0.847{\scriptstyle\pm.008}$ & $0.967{\scriptstyle\pm.004}$ & $0.014{\scriptstyle\pm.006}$ & $0.860{\scriptstyle\pm.033}$ & $0.975{\scriptstyle\pm.008}$ & $0.013{\scriptstyle\pm.005}$ & $0.904{\scriptstyle\pm.009}$ & $\mathbf{0.982{\scriptstyle\pm.001}}$ & $0.008{\scriptstyle\pm.000}$ & $0.908{\scriptstyle\pm.004}$ & $0.982{\scriptstyle\pm.001}$ & $\mathbf{0.008{\scriptstyle\pm.000}}$ & $\mathbf{0.921{\scriptstyle\pm.002}}$ & $\mathbf{0.985{\scriptstyle\pm.000}}$ & $\mathbf{0.007{\scriptstyle\pm.000}}$ & $0.925{\scriptstyle\pm.003}$ & $0.984{\scriptstyle\pm.002}$ & $0.008{\scriptstyle\pm.002}$ \\
\name{} (Hard Gate) & $0.838{\scriptstyle\pm.018}$ & $0.944{\scriptstyle\pm.018}$ & $0.010{\scriptstyle\pm.004}$ & $0.838{\scriptstyle\pm.022}$ & $0.950{\scriptstyle\pm.011}$ & $0.014{\scriptstyle\pm.004}$ & $0.841{\scriptstyle\pm.017}$ & $0.951{\scriptstyle\pm.023}$ & $0.010{\scriptstyle\pm.005}$ & $0.878{\scriptstyle\pm.030}$ & $0.974{\scriptstyle\pm.006}$ & $0.014{\scriptstyle\pm.005}$ & $0.902{\scriptstyle\pm.022}$ & $0.978{\scriptstyle\pm.006}$ & $0.010{\scriptstyle\pm.004}$ & $0.912{\scriptstyle\pm.006}$ & $0.981{\scriptstyle\pm.004}$ & $0.009{\scriptstyle\pm.003}$ & $0.916{\scriptstyle\pm.006}$ & $0.982{\scriptstyle\pm.005}$ & $0.009{\scriptstyle\pm.004}$ & $0.925{\scriptstyle\pm.005}$ & $0.981{\scriptstyle\pm.003}$ & $0.010{\scriptstyle\pm.003}$ \\
\name{} (Safety Net) & $0.703{\scriptstyle\pm.029}$ & $0.901{\scriptstyle\pm.006}$ & $0.016{\scriptstyle\pm.007}$ & $0.666{\scriptstyle\pm.046}$ & $0.900{\scriptstyle\pm.003}$ & $0.019{\scriptstyle\pm.005}$ & $0.636{\scriptstyle\pm.032}$ & $0.871{\scriptstyle\pm.016}$ & $0.071{\scriptstyle\pm.037}$ & $0.670{\scriptstyle\pm.026}$ & $0.866{\scriptstyle\pm.011}$ & $0.080{\scriptstyle\pm.028}$ & $0.673{\scriptstyle\pm.022}$ & $0.877{\scriptstyle\pm.008}$ & $0.069{\scriptstyle\pm.010}$ & $0.698{\scriptstyle\pm.029}$ & $0.886{\scriptstyle\pm.003}$ & $0.073{\scriptstyle\pm.011}$ & $0.715{\scriptstyle\pm.042}$ & $0.890{\scriptstyle\pm.005}$ & $0.061{\scriptstyle\pm.018}$ & $0.713{\scriptstyle\pm.040}$ & $0.895{\scriptstyle\pm.011}$ & $0.046{\scriptstyle\pm.015}$ \\
\name{} (Deferred) & $0.593{\scriptstyle\pm.020}$ & $0.902{\scriptstyle\pm.018}$ & $0.074{\scriptstyle\pm.020}$ & $0.606{\scriptstyle\pm.020}$ & $0.917{\scriptstyle\pm.004}$ & $0.062{\scriptstyle\pm.007}$ & $0.634{\scriptstyle\pm.024}$ & $0.934{\scriptstyle\pm.008}$ & $0.050{\scriptstyle\pm.006}$ & $0.697{\scriptstyle\pm.021}$ & $0.949{\scriptstyle\pm.009}$ & $0.039{\scriptstyle\pm.007}$ & $0.728{\scriptstyle\pm.008}$ & $0.958{\scriptstyle\pm.001}$ & $0.032{\scriptstyle\pm.001}$ & $0.759{\scriptstyle\pm.027}$ & $0.966{\scriptstyle\pm.008}$ & $0.026{\scriptstyle\pm.006}$ & $0.789{\scriptstyle\pm.008}$ & $0.975{\scriptstyle\pm.009}$ & $0.020{\scriptstyle\pm.007}$ & $0.783{\scriptstyle\pm.014}$ & $0.966{\scriptstyle\pm.009}$ & $0.027{\scriptstyle\pm.007}$ \\
\midrule
Kitsune~\cite{mirsky2018kitsune} + Flow features & - & $0.754{\scriptstyle\pm.022}$ & $0.066{\scriptstyle\pm.017}$ & - & $0.747{\scriptstyle\pm.018}$ & $0.057{\scriptstyle\pm.028}$ & - & $0.741{\scriptstyle\pm.005}$ & $0.066{\scriptstyle\pm.040}$ & - & $0.746{\scriptstyle\pm.004}$ & $0.025{\scriptstyle\pm.005}$ & - & $0.740{\scriptstyle\pm.005}$ & $0.028{\scriptstyle\pm.004}$ & - & $0.741{\scriptstyle\pm.010}$ & $0.023{\scriptstyle\pm.010}$ & - & $0.752{\scriptstyle\pm.004}$ & $0.018{\scriptstyle\pm.006}$ & - & $0.755{\scriptstyle\pm.003}$ & $0.032{\scriptstyle\pm.031}$ \\
Kitsune~\cite{mirsky2018kitsune} + AfterImage (faithful)$^\dagger$ & - & $0.734$ & $0.102$ & - & $0.734$ & $0.102$ & - & $0.734$ & $0.102$ & - & $0.734$ & $0.102$ & - & $0.734$ & $0.102$ & - & $0.734$ & $0.102$ & - & $0.734$ & $0.102$ & - & $0.734$ & $0.102$ \\
KnowML~\cite{guo2026knowml} + KitNET (unsupervised) & - & $0.883{\scriptstyle\pm.019}$ & $0.132{\scriptstyle\pm.025}$ & - & $0.890{\scriptstyle\pm.010}$ & $0.122{\scriptstyle\pm.014}$ & - & $0.889{\scriptstyle\pm.017}$ & $0.124{\scriptstyle\pm.022}$ & - & $0.874{\scriptstyle\pm.015}$ & $0.143{\scriptstyle\pm.020}$ & - & $0.851{\scriptstyle\pm.036}$ & $0.175{\scriptstyle\pm.051}$ & - & $0.793{\scriptstyle\pm.053}$ & $0.260{\scriptstyle\pm.076}$ & - & $0.792{\scriptstyle\pm.024}$ & $0.260{\scriptstyle\pm.035}$ & - & $0.829{\scriptstyle\pm.032}$ & $0.207{\scriptstyle\pm.046}$ \\
\bottomrule
\end{tabular}}
\end{table*}

\subsubsection{Hybrid view}
Table~\ref{tab:hybrid_couplings} combines packet and flow neural evidence with predicates from both views. A \name{} policy leads at every training size, although the best policy varies with $k$. Deferred leads at $k=50$ and $k=500$, Product of Experts (PoE) at $k=100$, $250$, $2500$, and $5000$, Hard Gate at $k=1000$, and Safety Net at $k=10000$. At $k=50$, Deferred reaches $0.870$ MF1, compared with $0.782$ for the neural product and $0.827$ for calibrated DST-IDS fusion. PoE reaches $0.978$ at $k=250$ and $0.985$ at $k=5000$. By $k=10000$, Safety Net reaches $0.983$, only $0.006$ above the neural product and $0.010$ above DST-IDS.

Policy choice has a larger effect on hybrid FPR. Anchored reaches $0.003$ at $k=50$, but rises to $0.048$ to $0.096$ at several larger budgets when evidence from either view can trigger an alert. PoE, Guarded, Same-Class Boost, and Hard Gate are generally better controlled. At $k=10000$, Safety Net combines $0.983$ MF1 and $0.998$ DR with an FPR of $0.007$, compared with $0.001$ for the neural product. Its small MF1 gain therefore carries a small false-alert cost.

\begin{table*}[t]
\centering
\scriptsize
\setlength{\tabcolsep}{3pt}
\caption{Hybrid-view comparison on CIC-IDS2018 across
per-class training sizes $k$. Results report F1 and FPR as average $\pm$ standard deviation. Best F1 (highest) and FPR (lowest) for each $k$ are shown in bold.}
\label{tab:hybrid_couplings}
\resizebox{\textwidth}{!}{%
\begin{tabular}{l *{8}{rr}}
\toprule
& \multicolumn{2}{c}{${k{=}50}$} & \multicolumn{2}{c}{${k{=}100}$} & \multicolumn{2}{c}{${k{=}250}$} & \multicolumn{2}{c}{${k{=}500}$} & \multicolumn{2}{c}{${k{=}1000}$} & \multicolumn{2}{c}{${k{=}2500}$} & \multicolumn{2}{c}{${k{=}5000}$} & \multicolumn{2}{c}{${k{=}10000}$} \\
\cmidrule(lr){2-3} \cmidrule(lr){4-5} \cmidrule(lr){6-7} \cmidrule(lr){8-9} \cmidrule(lr){10-11} \cmidrule(lr){12-13} \cmidrule(lr){14-15} \cmidrule(lr){16-17}
System & F1 & FPR & F1 & FPR & F1 & FPR & F1 & FPR & F1 & FPR & F1 & FPR & F1 & FPR & F1 & FPR \\
\midrule
Symbolic-only baseline & $0.841{\scriptstyle\pm.023}$ & $0.006{\scriptstyle\pm.000}$ & $0.739{\scriptstyle\pm.021}$ & $0.099{\scriptstyle\pm.024}$ & $0.801{\scriptstyle\pm.072}$ & $0.077{\scriptstyle\pm.063}$ & $0.818{\scriptstyle\pm.082}$ & $0.060{\scriptstyle\pm.067}$ & $0.857{\scriptstyle\pm.014}$ & $0.086{\scriptstyle\pm.018}$ & $0.835{\scriptstyle\pm.024}$ & $0.093{\scriptstyle\pm.021}$ & $0.849{\scriptstyle\pm.001}$ & $0.077{\scriptstyle\pm.007}$ & $0.849{\scriptstyle\pm.001}$ & $0.077{\scriptstyle\pm.007}$ \\
Neural product (baseline) & $0.782{\scriptstyle\pm.064}$ & $0.029{\scriptstyle\pm.015}$ & $0.888{\scriptstyle\pm.019}$ & $0.014{\scriptstyle\pm.004}$ & $0.936{\scriptstyle\pm.030}$ & $0.006{\scriptstyle\pm.003}$ & $0.955{\scriptstyle\pm.018}$ & $0.003{\scriptstyle\pm.002}$ & $0.950{\scriptstyle\pm.022}$ & $\mathbf{0.003}{\scriptstyle\pm.001}$ & $\mathbf{0.967{\scriptstyle\pm.011}}$ & $0.003{\scriptstyle\pm.000}$ & $0.970{\scriptstyle\pm.012}$ & $\mathbf{0.002}{\scriptstyle\pm.001}$ & $0.977{\scriptstyle\pm.010}$ & $\mathbf{0.001}{\scriptstyle\pm.001}$ \\
DST-IDS fusion~\cite{QIU2022} & $0.827{\scriptstyle\pm.065}$ & $0.024{\scriptstyle\pm.014}$ & $0.879{\scriptstyle\pm.040}$ & $0.012{\scriptstyle\pm.004}$ & $0.920{\scriptstyle\pm.027}$ & $0.010{\scriptstyle\pm.005}$ & $0.943{\scriptstyle\pm.018}$ & $0.006{\scriptstyle\pm.002}$ & $0.962{\scriptstyle\pm.014}$ & $\mathbf{0.003}{\scriptstyle\pm.001}$ & $0.966{\scriptstyle\pm.005}$ & $0.004{\scriptstyle\pm.001}$ & $0.970{\scriptstyle\pm.008}$ & $\mathbf{0.002}{\scriptstyle\pm.001}$ & $0.973{\scriptstyle\pm.010}$ & $0.002{\scriptstyle\pm.001}$ \\
\midrule
\name{} (PoE) & $0.867{\scriptstyle\pm.035}$ & $0.020{\scriptstyle\pm.016}$ & $\mathbf{0.925{\scriptstyle\pm.031}}$ & $\mathbf{0.009}{\scriptstyle\pm.004}$ & $\mathbf{0.978{\scriptstyle\pm.006}}$ & $0.004{\scriptstyle\pm.003}$ & $\mathbf{0.974{\scriptstyle\pm.008}}$ & $\mathbf{0.001}{\scriptstyle\pm.001}$ & $0.961{\scriptstyle\pm.023}$ & $0.006{\scriptstyle\pm.005}$ & $\mathbf{0.967{\scriptstyle\pm.029}}$ & $0.036{\scriptstyle\pm.048}$ & $\mathbf{0.985{\scriptstyle\pm.002}}$ & $0.005{\scriptstyle\pm.004}$ & $0.952{\scriptstyle\pm.028}$ & $0.012{\scriptstyle\pm.003}$ \\
\name{} (Guarded) & $0.759{\scriptstyle\pm.036}$ & $0.008{\scriptstyle\pm.004}$ & $0.823{\scriptstyle\pm.077}$ & $0.023{\scriptstyle\pm.029}$ & $0.865{\scriptstyle\pm.024}$ & $0.008{\scriptstyle\pm.005}$ & $0.899{\scriptstyle\pm.024}$ & $0.005{\scriptstyle\pm.003}$ & $0.952{\scriptstyle\pm.015}$ & $0.004{\scriptstyle\pm.004}$ & $0.962{\scriptstyle\pm.019}$ & $\mathbf{0.002}{\scriptstyle\pm.001}$ & $0.971{\scriptstyle\pm.015}$ & $\mathbf{0.002}{\scriptstyle\pm.001}$ & $0.970{\scriptstyle\pm.013}$ & $0.003{\scriptstyle\pm.001}$ \\
\name{} (Anchored) & $0.805{\scriptstyle\pm.019}$ & $\mathbf{0.003}{\scriptstyle\pm.002}$ & $0.753{\scriptstyle\pm.029}$ & $0.092{\scriptstyle\pm.029}$ & $0.843{\scriptstyle\pm.047}$ & $0.067{\scriptstyle\pm.040}$ & $0.815{\scriptstyle\pm.055}$ & $0.048{\scriptstyle\pm.058}$ & $0.818{\scriptstyle\pm.032}$ & $0.091{\scriptstyle\pm.024}$ & $0.867{\scriptstyle\pm.019}$ & $0.071{\scriptstyle\pm.000}$ & $0.828{\scriptstyle\pm.044}$ & $0.096{\scriptstyle\pm.035}$ & $0.840{\scriptstyle\pm.022}$ & $0.079{\scriptstyle\pm.010}$ \\
\name{} (Same-Class Boost) & $0.709{\scriptstyle\pm.040}$ & $0.049{\scriptstyle\pm.016}$ & $0.820{\scriptstyle\pm.084}$ & $0.051{\scriptstyle\pm.030}$ & $0.913{\scriptstyle\pm.039}$ & $0.012{\scriptstyle\pm.006}$ & $0.934{\scriptstyle\pm.004}$ & $0.006{\scriptstyle\pm.001}$ & $0.954{\scriptstyle\pm.007}$ & $0.005{\scriptstyle\pm.004}$ & $0.957{\scriptstyle\pm.003}$ & $0.004{\scriptstyle\pm.001}$ & $0.960{\scriptstyle\pm.011}$ & $0.003{\scriptstyle\pm.000}$ & $0.968{\scriptstyle\pm.007}$ & $0.003{\scriptstyle\pm.000}$ \\
\name{} (Deferred) & $\mathbf{0.870{\scriptstyle\pm.037}}$ & $0.018{\scriptstyle\pm.016}$ & $0.760{\scriptstyle\pm.176}$ & $0.026{\scriptstyle\pm.035}$ & $0.894{\scriptstyle\pm.102}$ & $\mathbf{0.003}{\scriptstyle\pm.003}$ & $\mathbf{0.974{\scriptstyle\pm.008}}$ & $\mathbf{0.001}{\scriptstyle\pm.001}$ & $0.961{\scriptstyle\pm.023}$ & $0.006{\scriptstyle\pm.005}$ & $\mathbf{0.967{\scriptstyle\pm.028}}$ & $0.036{\scriptstyle\pm.048}$ & $\mathbf{0.985{\scriptstyle\pm.002}}$ & $0.005{\scriptstyle\pm.004}$ & $0.932{\scriptstyle\pm.039}$ & $0.031{\scriptstyle\pm.031}$ \\
\name{} (Hard Gate) & $0.749{\scriptstyle\pm.041}$ & $0.016{\scriptstyle\pm.014}$ & $0.818{\scriptstyle\pm.068}$ & $0.027{\scriptstyle\pm.027}$ & $0.897{\scriptstyle\pm.023}$ & $0.008{\scriptstyle\pm.005}$ & $0.927{\scriptstyle\pm.045}$ & $0.011{\scriptstyle\pm.014}$ & $\mathbf{0.963{\scriptstyle\pm.014}}$ & $0.004{\scriptstyle\pm.001}$ & $0.966{\scriptstyle\pm.012}$ & $0.004{\scriptstyle\pm.000}$ & $0.969{\scriptstyle\pm.010}$ & $0.003{\scriptstyle\pm.000}$ & $0.974{\scriptstyle\pm.009}$ & $0.002{\scriptstyle\pm.001}$ \\
\name{} (Safety Net) & $0.862{\scriptstyle\pm.043}$ & $0.017{\scriptstyle\pm.011}$ & $0.922{\scriptstyle\pm.019}$ & $0.011{\scriptstyle\pm.006}$ & $0.956{\scriptstyle\pm.011}$ & $0.008{\scriptstyle\pm.005}$ & $0.971{\scriptstyle\pm.008}$ & $0.004{\scriptstyle\pm.002}$ & $0.956{\scriptstyle\pm.038}$ & $0.020{\scriptstyle\pm.020}$ & $0.944{\scriptstyle\pm.048}$ & $0.045{\scriptstyle\pm.055}$ & $0.982{\scriptstyle\pm.006}$ & $0.008{\scriptstyle\pm.002}$ & $\mathbf{0.983{\scriptstyle\pm.004}}$ & $0.007{\scriptstyle\pm.003}$ \\
\bottomrule
\end{tabular}}
\end{table*}

\subsubsection{Cross-view assessment}
The packet and flow sweeps show that symbolic predicates provide useful class structure before the neural models have seen enough examples to learn the same distinctions. The $200\times$ comparison concerns performance-equivalent label budgets, not training time or uniform improvement at every operating point. The effect is smaller in the hybrid view because its neural baseline already combines complementary packet and flow representations.

The benefit also depends on the policy and the quality of the predicates. Strict policies control FPR but may suppress correct neural predictions when symbolic support is incomplete, whereas permissive policies preserve detection at the cost of more false alarms. At large budgets, HGB leads in the flow view, the packet random forest matches or exceeds the strongest couplings, and the hybrid baselines close most of the gap. The results therefore support a bounded conclusion: \name{} is most useful when labels are scarce, while its high-data benefit depends on how the coupling handles incomplete or inaccurate symbolic evidence.

\subsection{Threshold-free Detection Performance}
\label{secapp:per_class_performance}

The preceding results use a fixed operating threshold. To separate ranking quality from threshold selection, we compare \name{} with two unsupervised detectors in a one-versus-benign setting. ROC-AUC measures rank separation independently of prevalence, whereas the PR-AUC baseline depends on attack-family prevalence and is more demanding for rare families. \name{} uses $P(\text{family})$ as its score; the unsupervised systems use their native anomaly scores.

The systems retain their native test bases: \name{} and KnowML+KitNET use a held-out $20\%$ split of $565{,}525$ flows, while faithful Kitsune uses the streamed Tuesday--Friday original captures. The comparison therefore concerns family-level separation rather than identical test instances.

\begin{figure}[h]
\centering
\begin{subfigure}[t]{0.49\columnwidth}
    \centering
    \includegraphics[width=0.65\linewidth]{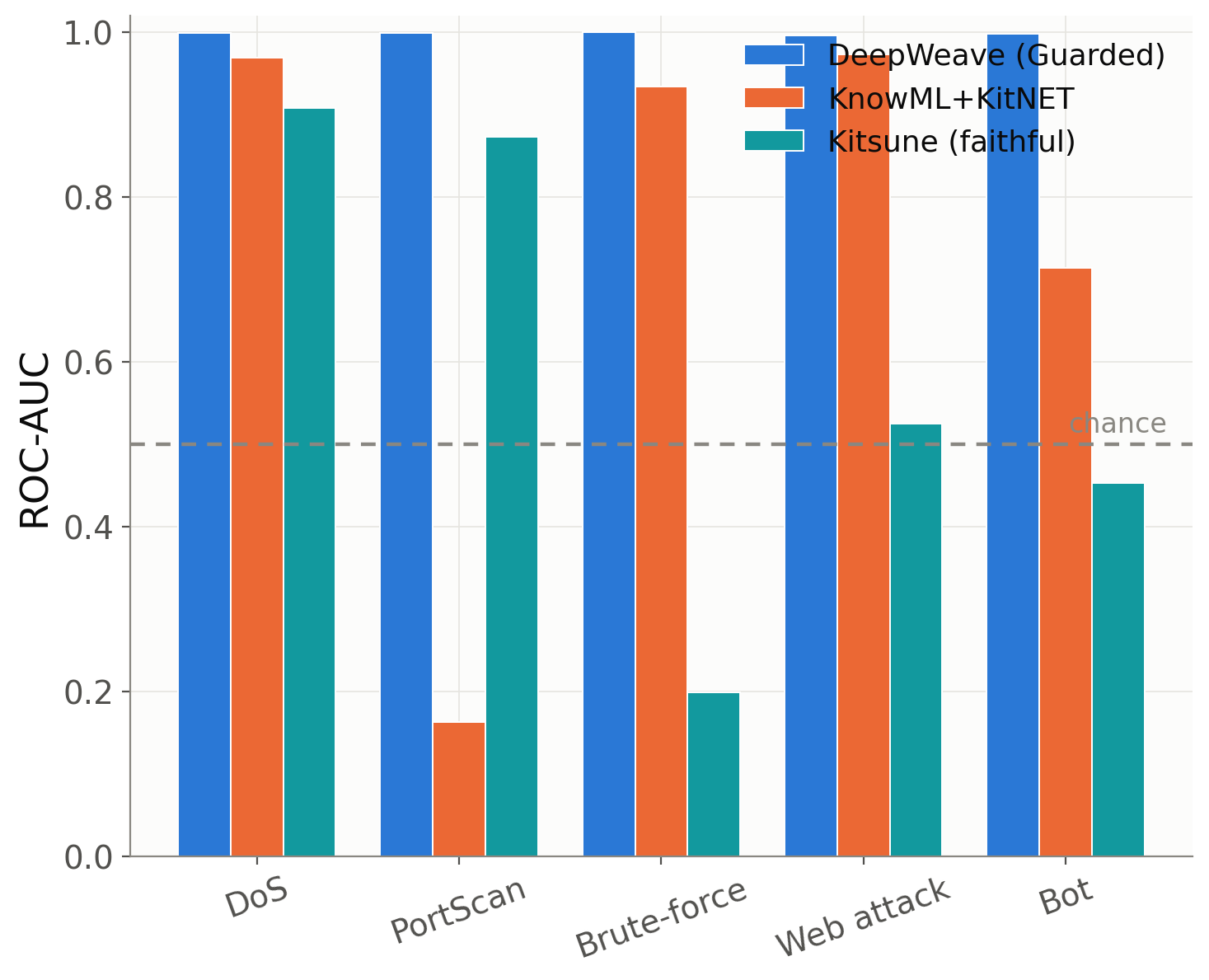}
    \caption{Per family ROC-AUC}
    \label{fig:per_family_roc_auc}
\end{subfigure}
\hfill
\begin{subfigure}[t]{0.49\columnwidth}
    \centering
    \includegraphics[width=0.65\linewidth]{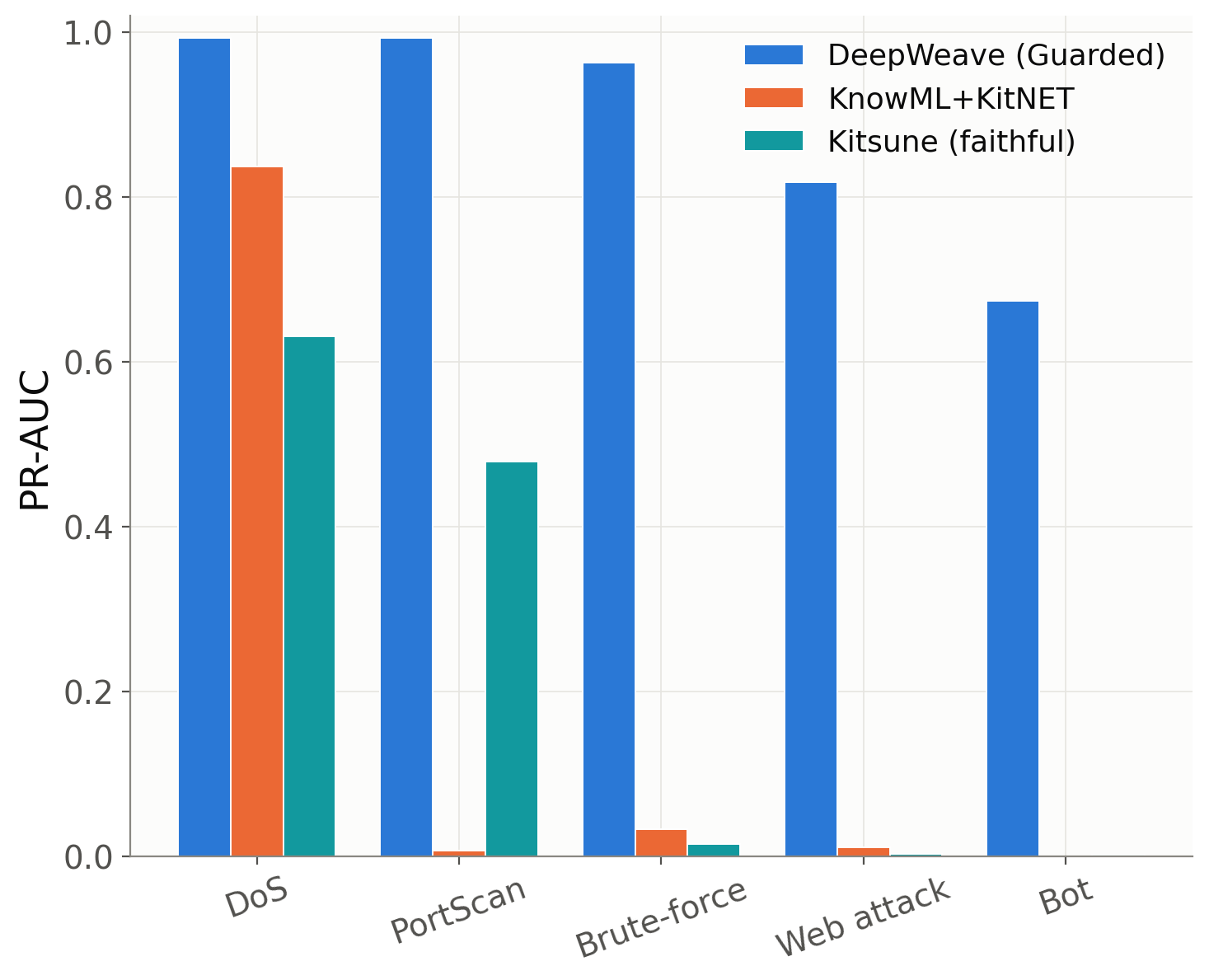}
    \caption{Per family PR-AUC}
    \label{fig:per_family_pr_auc}
\end{subfigure}
\caption{Per-family threshold-free detection quality on CIC-IDS2017 flow: ROC-AUC (left) and PR-AUC (right), each attack family scored one-versus-benign. The dashed line marks chance (ROC-AUC $=0.5$); bars below it are anti-correlated with the attack.}
\label{fig:perfamily_auc_bars}
\end{figure}

\ourparagraph{Per-class Detection Performance.} Figure~\ref{fig:perfamily_auc_bars} shows that \name{} separates all five families, with ROC-AUC between $0.997$ and $1.000$. PR-AUC remains $0.818$ for web attacks and $0.674$ for bot traffic despite their low prevalence, indicating that the lower precision-recall scores mainly reflect class imbalance.

The unsupervised detectors have complementary blind spots. Faithful Kitsune performs well on DoS and PortScan but is near or below chance on web attacks, bot traffic, and brute force. FTP-Patator is its clearest failure (ROC-AUC $0.045$), as the anomaly score ranks attacks below benign traffic. KnowML+KitNET separates DoS, web attacks, and brute force, but fails on PortScan ($0.210$) and is only slightly above chance on bot traffic ($0.594$). Since both baselines use KitNET, their different failures point to the feature representation. Neither covers all five families.

\begin{figure}[h]
\centering
\begin{subfigure}[t]{0.49\linewidth}
    \centering
    \includegraphics[width=0.65\linewidth]{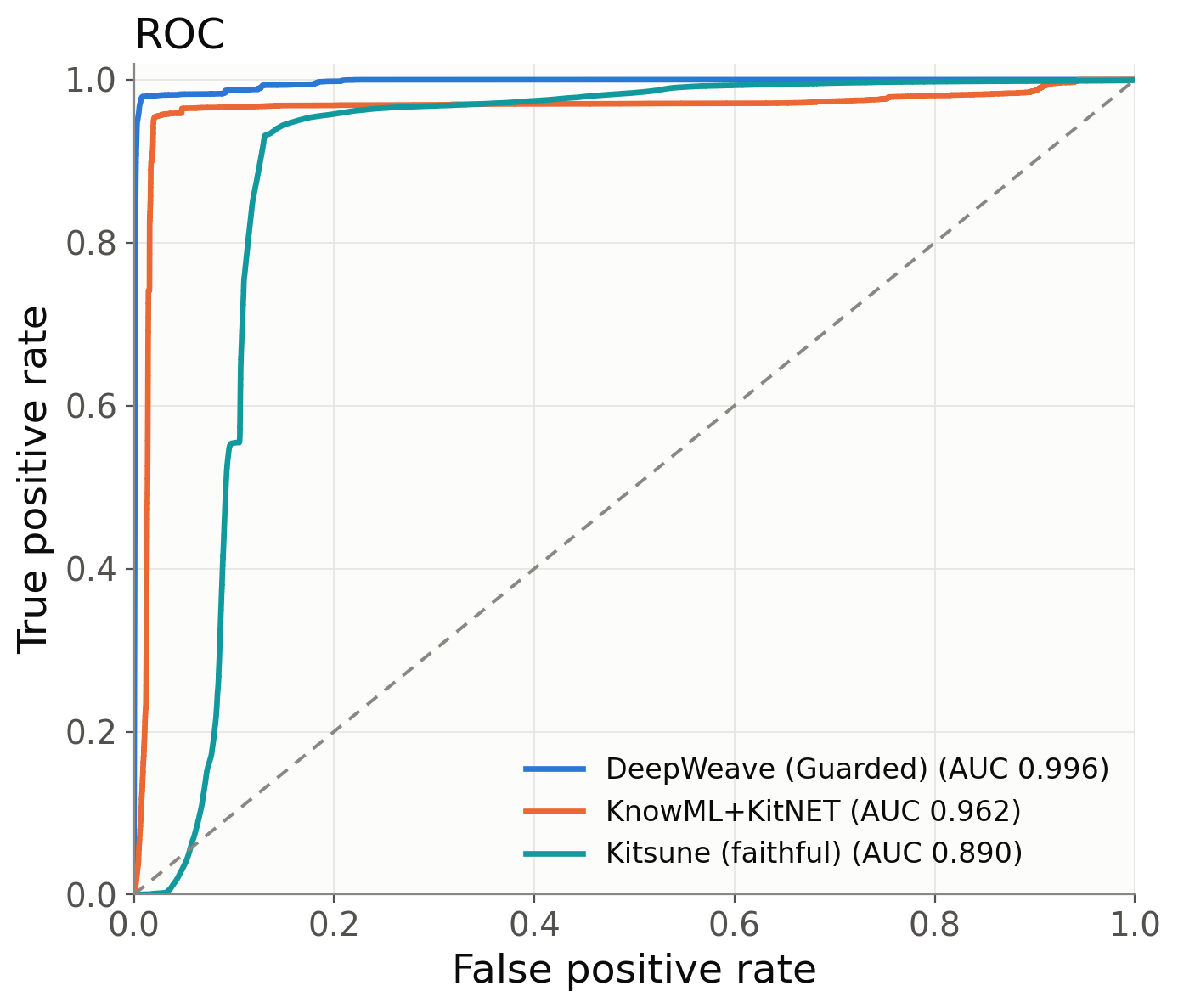}
    \caption{Overall ROC-AUC curve}
    \label{fig:overall_roc_auc}
\end{subfigure}
\begin{subfigure}[t]{0.49\linewidth}
    \centering
    \includegraphics[width=0.65\linewidth]{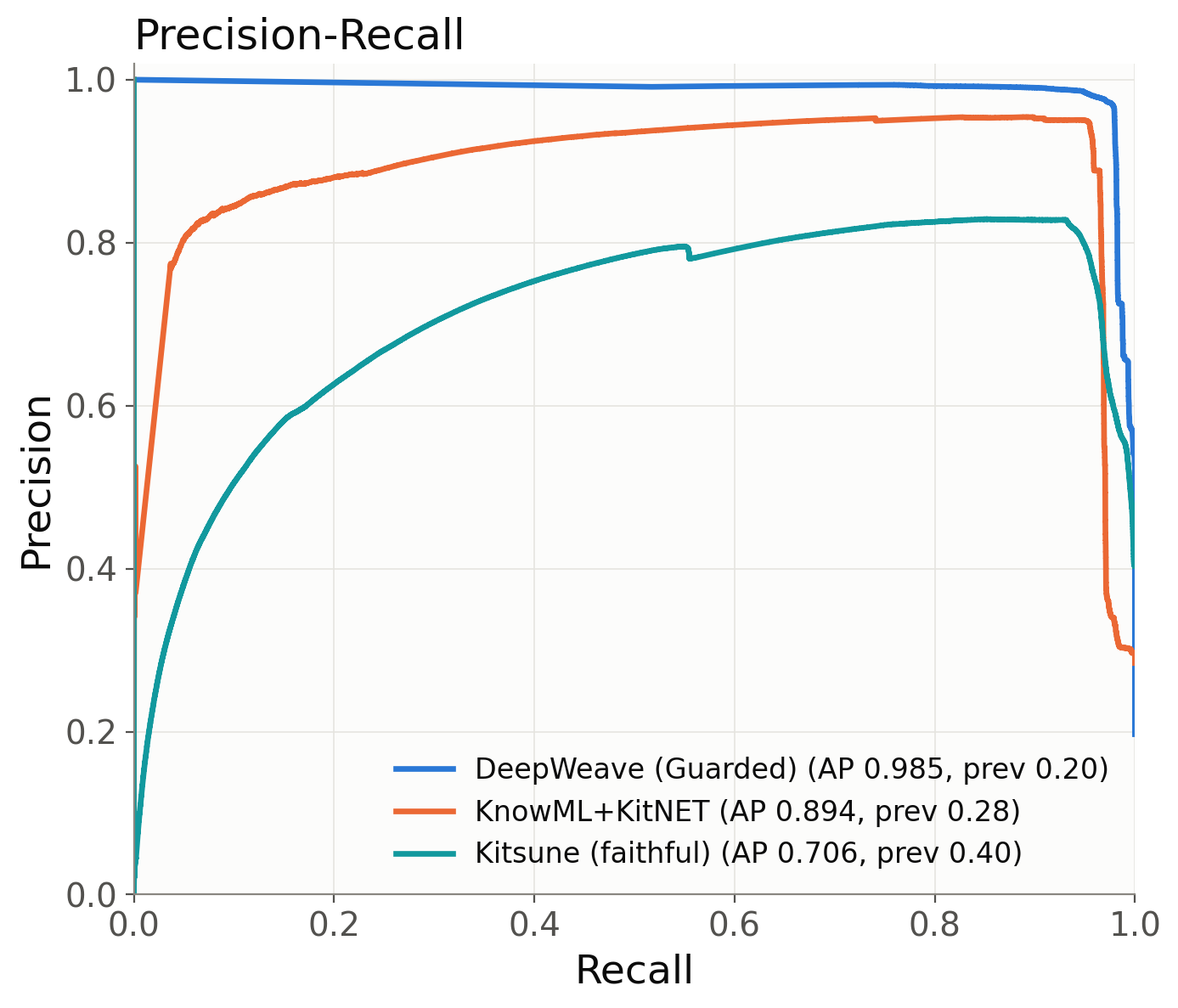}
    \caption{Overall PR-AUC curve}
    \label{fig:overll_pr_auc}
\end{subfigure}
\caption{Overall pooled attack-versus-benign ROC (left) and precision--recall (right) curves, one per system. Prevalence differs per system and is annotated on the precision--recall panel.}
\label{fig:overall_auc_curves}
\end{figure}

\ourparagraph{Overall Discrimination.} Figure~\ref{fig:overall_auc_curves} pools all attacks against benign traffic. \name{} reaches $0.996$ ROC-AUC, compared with $0.962$ for KnowML+KitNET and $0.890$ for faithful Kitsune. Abundant volumetric attacks narrow the pooled gap and conceal some family-level failures. ROC-AUC provides the safer cross-system comparison because attack prevalence differs across the evaluations ($0.20$, $0.28$, and $0.40$); PR-AUC should be interpreted within each system.

\subsection{Additional Adversarial Robustness Results}\label{secapp:robustness_extra}

Table~\ref{tab:pants_relative_per_class} provides a detailed view of the Attack success rate (ASR) per attack class.

\begin{table*}[t]
\centering
\scriptsize
\setlength{\tabcolsep}{4pt}
\caption{Attack success rate (ASR) by attack class under the PANTS~\cite{jin2025robustifying}-inspired proportional per-flow feature-space attack. Parentheses in the column headings give each class's share of the 111,301 attack flows in the natural-prior test split. Weighted ASR is flow-weighted by these frequencies; class mean is the unweighted mean across attack classes. Lower is better.}
\label{tab:pants_relative_per_class}
\resizebox{\textwidth}{!}{%
\begin{tabular}{lrrrrrrr}
\toprule
System & DoS (68.2\%) & PortScan (28.5\%) & BruteForce (2.5\%) & WebAttack (0.4\%) & Bot (0.4\%) & Weighted ASR & Class mean \\
\midrule
Symbolic baseline & $32.2\%$ & $3.6\%$ & $0.0\%$ & $9.9\%$ & $4.1\%$ & $23.1\%$ & $10.0\%$ \\
Neural baseline & $13.3\%$ & $72.5\%$ & $44.7\%$ & $33.8\%$ & $72.4\%$ & $31.3\%$ & $47.3\%$ \\
\midrule
\name{} (Safety Net) & $26.2\%$ & $48.8\%$ & $49.6\%$ & $13.0\%$ & $59.7\%$ & $33.3\%$ & $39.5\%$ \\
\name{} (Guarded) & $31.2\%$ & $50.3\%$ & $38.0\%$ & $2.5\%$ & $4.4\%$ & $36.6\%$ & $25.3\%$ \\
\name{} (Anchored) & $28.5\%$ & $61.5\%$ & $55.3\%$ & $2.0\%$ & $44.0\%$ & $38.5\%$ & $38.3\%$ \\
\name{} (Hard Gate) & $31.2\%$ & $75.8\%$ & $36.5\%$ & $7.5\%$ & $32.0\%$ & $44.0\%$ & $36.6\%$ \\
\bottomrule
\end{tabular}}
\end{table*}

\subsection{Additional Explainability Results}\label{secapp:explainability}

\ourparagraph{Additional Explanation Visualization.}\label{sec:additional_explanations}

\begin{figure*}[h]
\centering 
\begin{subfigure}[t]{0.49\columnwidth} 
\centering 
\includegraphics[ width=0.65\linewidth ]{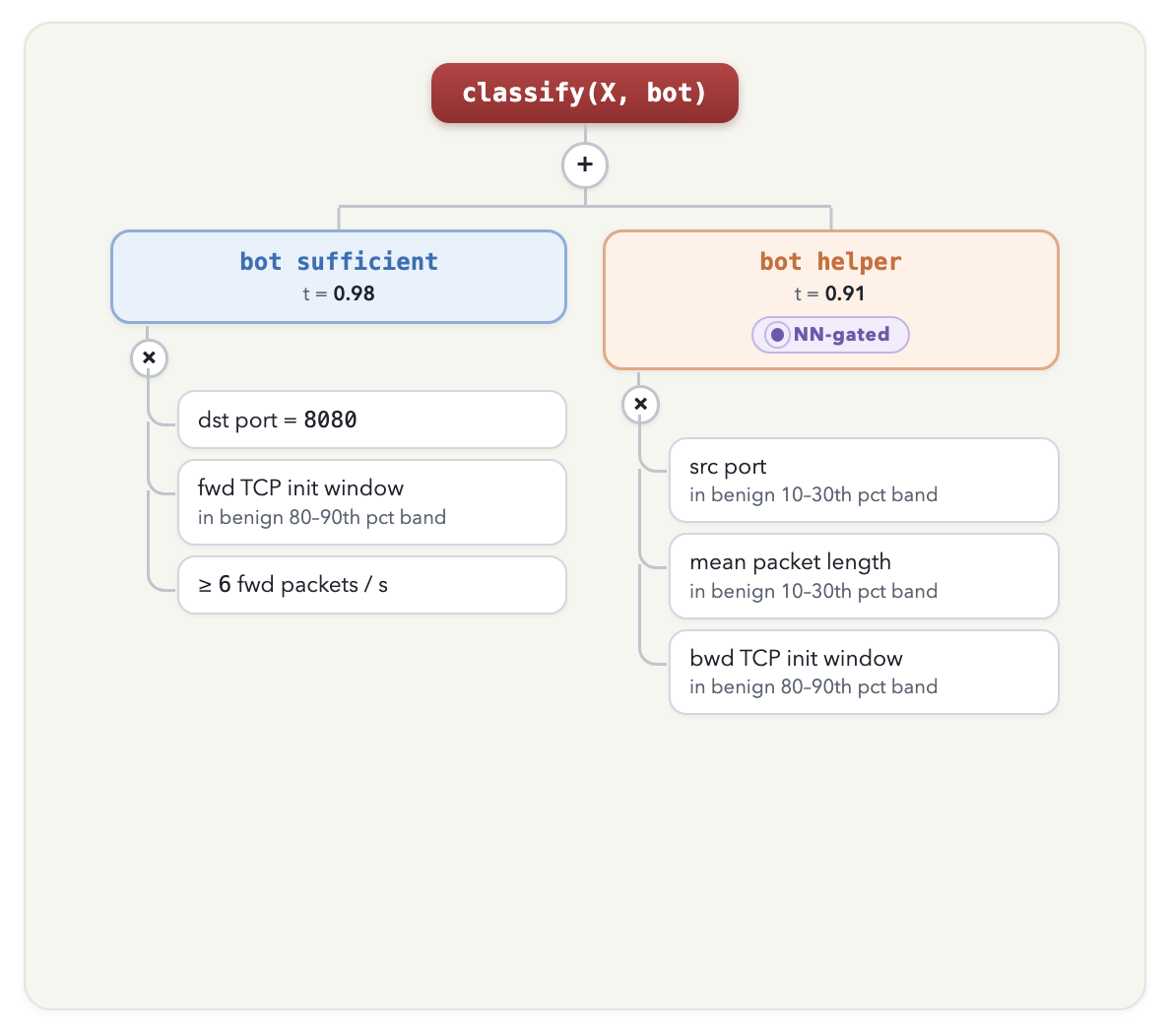} \caption{\name{} global explanation for detection of bot attacks.} \label{fig:global_explanation_bot} \end{subfigure} 
\hfill 
\begin{subfigure}[t]{0.49\columnwidth} 
\centering \includegraphics[ width=0.65\linewidth ]{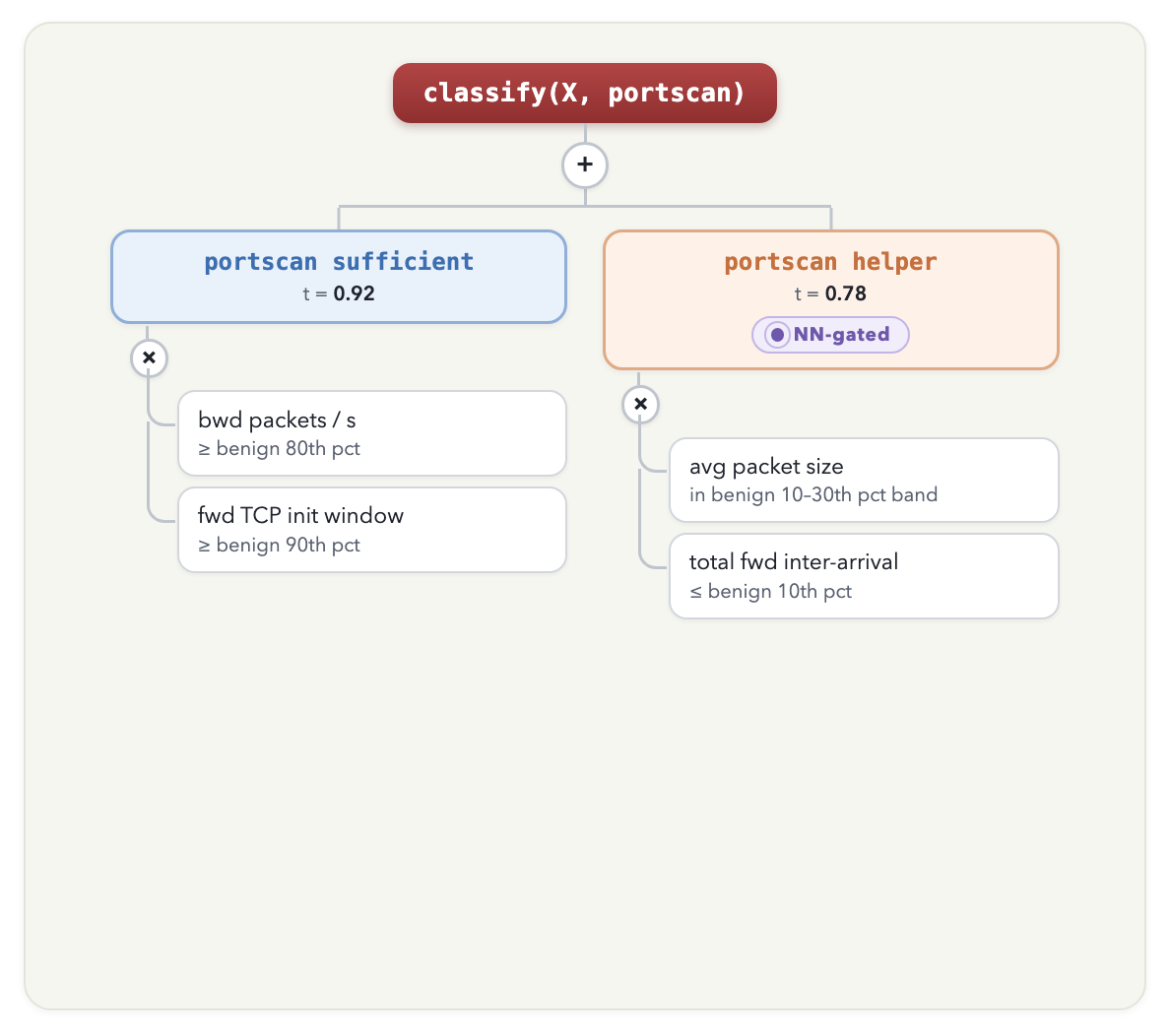} \caption{\name{} global explanation for detection of port scan attacks.} \label{fig:global_explanation_portscan} 
\end{subfigure} 
\hfill
\begin{subfigure}[t]{0.49\columnwidth} 
\centering 
\includegraphics[ width=0.65\linewidth ]{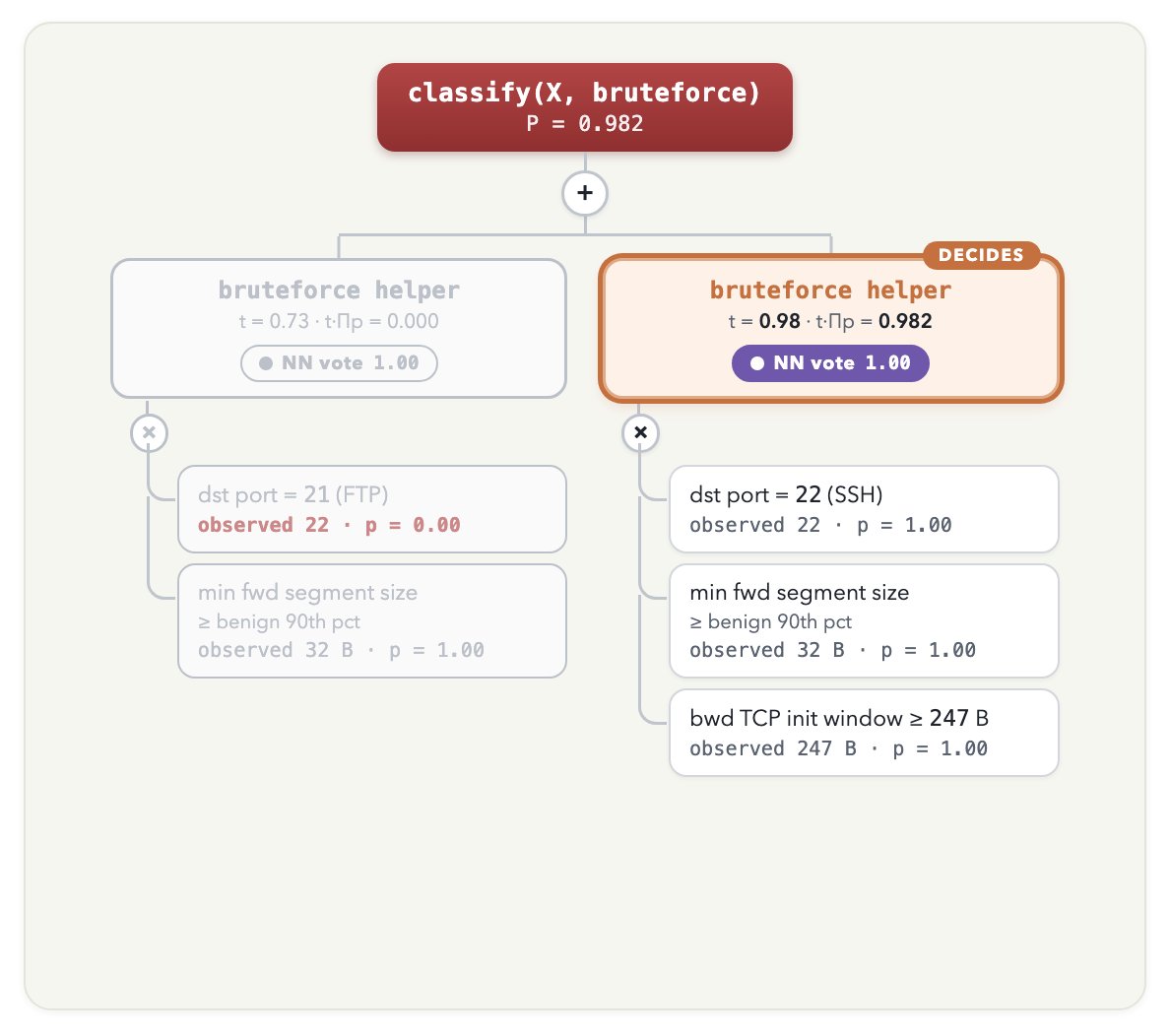} \caption{\name{} local explanation given for the detection of a brute force attack.} \label{fig:local_explanation_bruteforce} \end{subfigure} 
\hfill 
\begin{subfigure}[t]{0.49\columnwidth} 
\centering \includegraphics[ width=0.65\linewidth ]{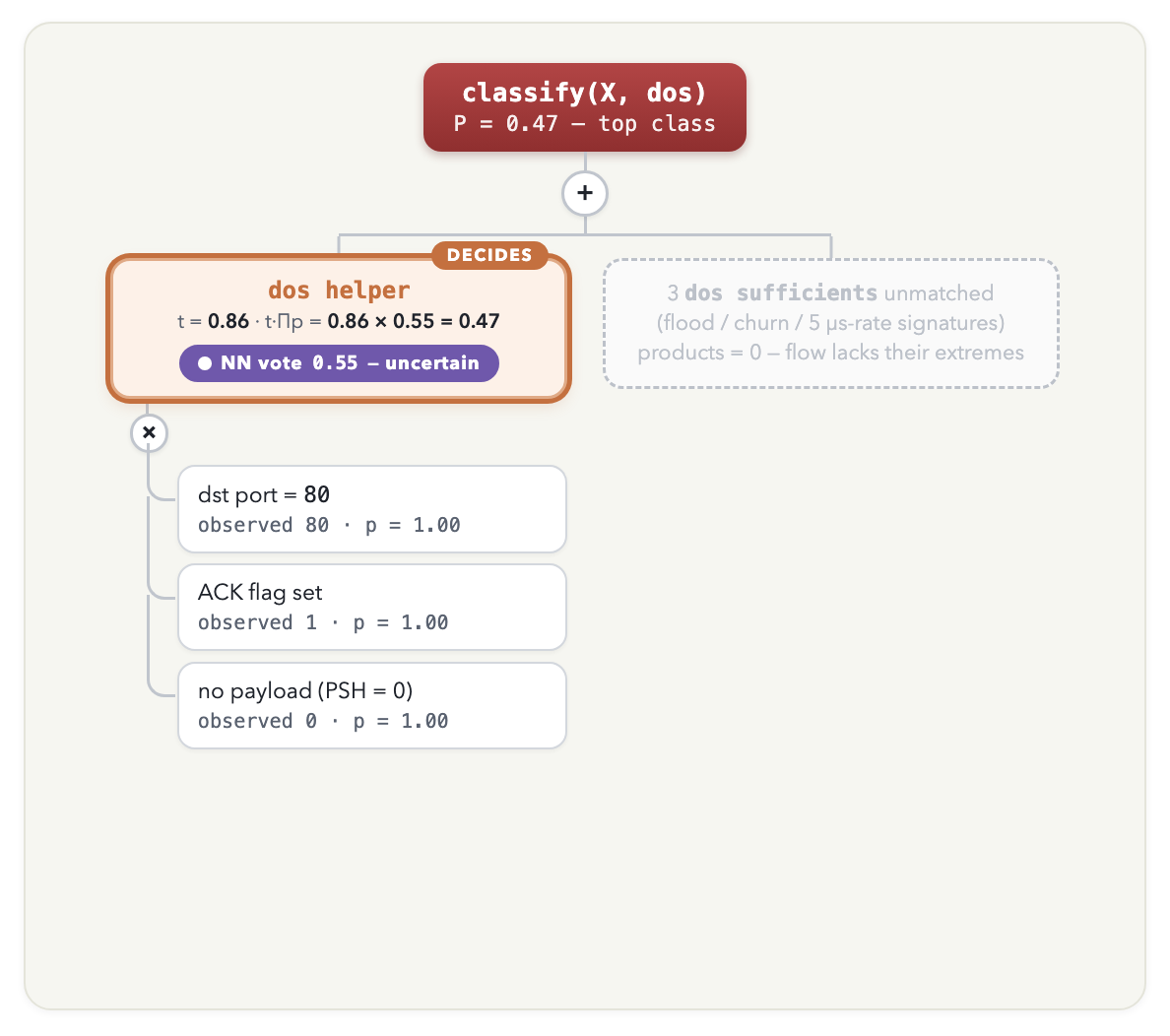} \caption{\name{} local explanation given for the detection of a DoS attack.} \label{fig:local_explanation_dos} 
\end{subfigure} 
\caption{\name{}'s explanations visualization.} 
%\caption{Motivation for neuro-symbolic intrusion detection.}
\label{fig:additional_explanation_examples} 
\end{figure*}

\begin{figure}[h]
\centering
\begin{subfigure}[t]{0.49\columnwidth}
    \centering
    \includegraphics[width=0.65\linewidth]{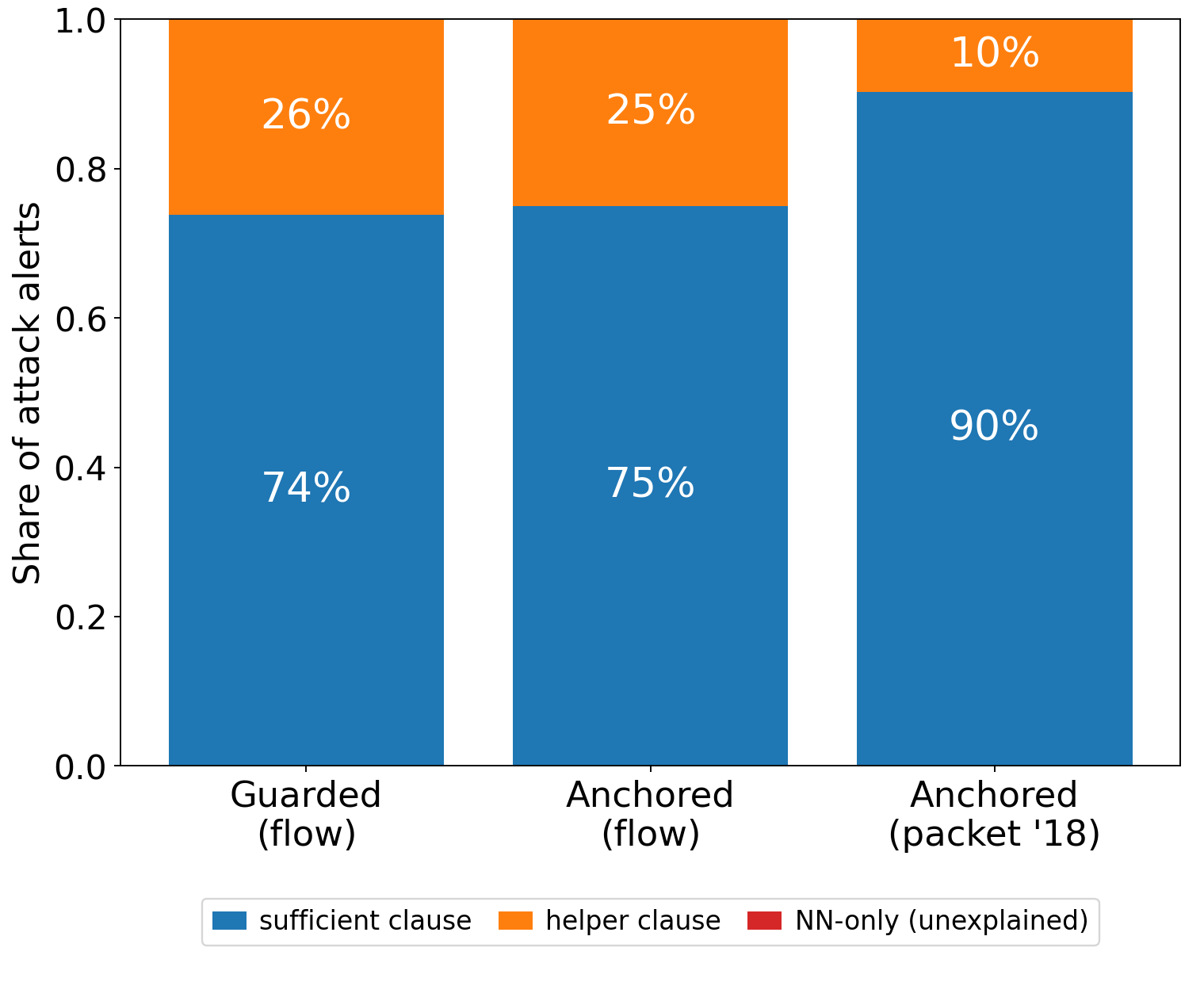}
    \caption{Composition of attack-alert proof paths across coupling
    architectures and views.}
    \label{fig:explain_coverage}
\end{subfigure}
\hfill
\begin{subfigure}[t]{0.49\columnwidth}
    \centering
    \includegraphics[width=0.65\linewidth]{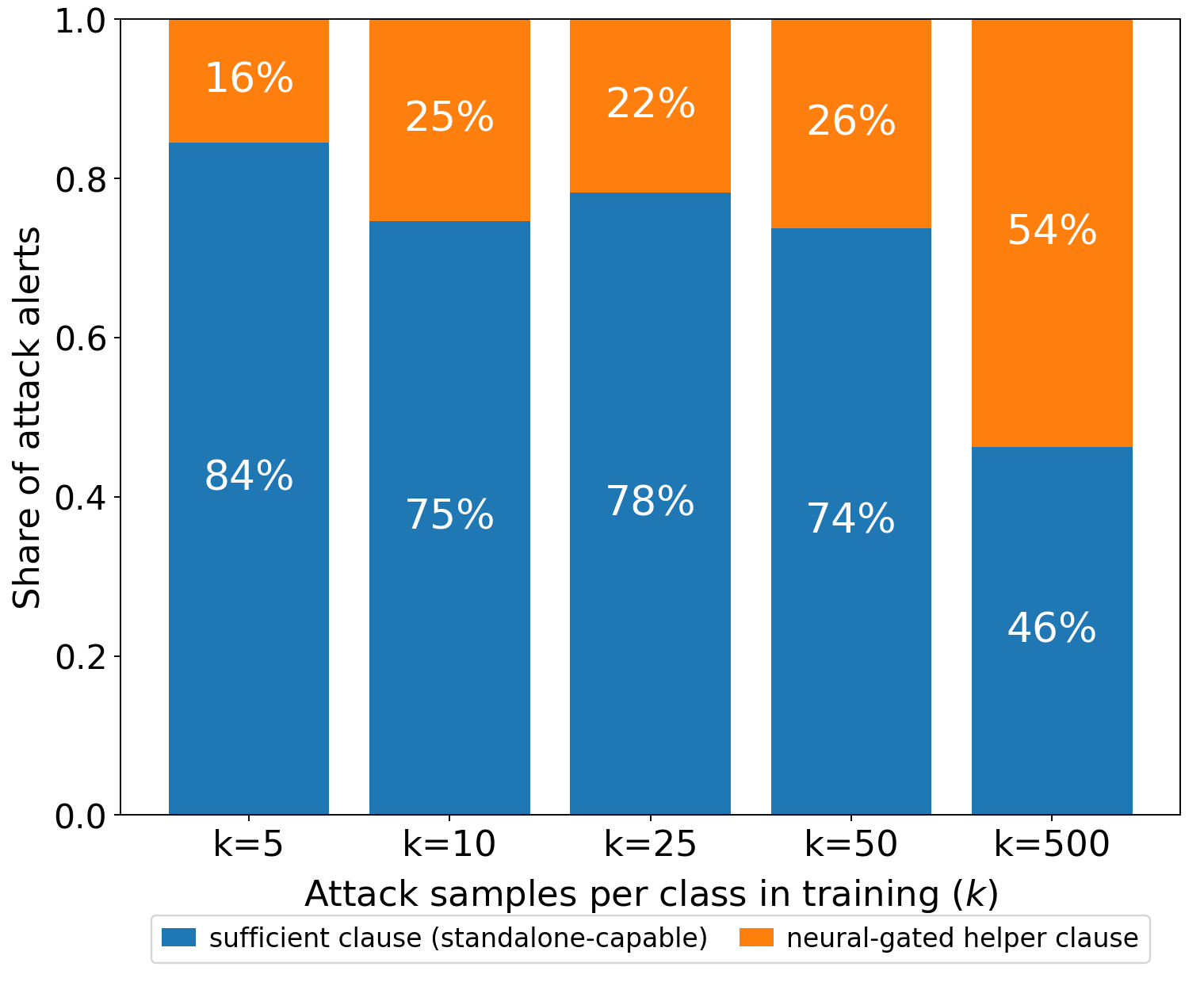}
    \caption{Attack-alert proof composition across training sizes for
    CIC-IDS2017 flow traffic.}
    \label{fig:explain_coverage_k}
\end{subfigure}
\caption{\name{}'s Attack-alert proof paths.}
\label{fig:attack_alert_proof_paths}
\end{figure}

\ourparagraph{Proof-path coverage.}
\label{sec:rq3_coverage}
Coverage measures the fraction of attack alerts whose decision can be
attributed to at least one named symbolic clause.
We evaluate coverage on CIC-IDS2017 flow traffic and
NF-CSE-CIC-IDS2018-v3 packet traffic.
The packet setting is particularly informative because its coupling retains an
open neural path, allowing the model in principle to raise an attack alert
without using a symbolic rule.
Figure~\ref{fig:explain_coverage} separates alerts decided by standalone
sufficient rules, neural helper rules, and the neural path alone.
Guarded coupling attributes every attack alert to a rule by construction.
Under Anchored coupling, approximately three quarters of alerts are supported by standalone sufficient rules, while the remainder use neural-supported clauses.
More importantly, the packet model produces no attack alerts through its open
neural path alone, even though the architecture permits that outcome.
Every observed alert is therefore associated with a named rule across the evaluated datasets and traffic views.

The composition of the explanations changes with the labelled-data budget.
As shown in Figure~\ref{fig:explain_coverage_k}, sufficient rules account for \(84.5\%\) of attack alerts at \(k=5\), compared with \(46.3\%\) at \(k=500\).
As the neural model receives more labelled examples, the model increasingly
uses helper clauses that combine neural evidence with broader symbolic conditions.
High-precision sufficient rules therefore dominate in the low-data setting, while neural-supported proofs become more prominent as the classifier improves.

\ourparagraph{Clause fidelity.}
\label{sec:rq3_fidelity}
Coverage alone does not establish whether the deciding clause is reliable.
We therefore measure each clause's precision on held-out test traffic and the
recall of the attack class it supports.
Clause precision measures how often an alert attributed to the clause is
correct, while class recall indicates whether high precision is obtained only
by abstaining on most attacks.
\begin{table}[t]
\centering
\small
\setlength{\tabcolsep}{3pt}
\caption{Clause fidelity for Guarded coupling on CIC-IDS2017 flow traffic at
\(k=50\).}
\label{tab:clause_fidelity}
\begin{tabular}{llrrr}
\toprule
\textbf{Clause} &
\textbf{Role} &
\textbf{Alerts} &
\textbf{Precision} &
\textbf{Recall} \\
\midrule
Benign gate & Disqualifier & 15,625 & 0.990 & 0.987 \\
\midrule
Bot sufficient & Sufficient & 82 & 0.902 & \multirow{2}{*}{0.796} \\
Bot helper & Helper & 27 & 0.593 & \\
\midrule
Brute-force helper & Helper & 245 & 0.784 & 0.985 \\
\midrule
DoS sufficient & Sufficient & 2,440 & 0.993 & \multirow{2}{*}{0.964} \\
DoS helper & Helper & 287 & 0.693 & \\
\midrule
PortScan sufficient & Sufficient & 605 & 0.997 & \multirow{2}{*}{0.981} \\
PortScan helper & Helper & 588 & 0.968 & \\
\midrule
Web Attack sufficient & Sufficient & 101 & 0.990 & 0.877 \\
\bottomrule
\end{tabular}
\end{table}
Table~\ref{tab:clause_fidelity} shows that sufficient rules are highly selective, with test precision between \(0.90\) and \(1.00\).
Helper clauses are intentionally broader and have precision between
\(0.59\) and \(0.97\), which explains why their authority is conditioned on
compatible neural evidence.
Class recall remains between \(0.80\) and \(0.99\), indicating that the rule
base does not obtain high precision simply by explaining a negligible fraction
of attacks.
The observed results therefore support both the role assignment and the
decision to prevent lower-precision helper clauses from raising alerts
independently.

\ourparagraph{Analyst-facing interpretation.}
\label{sec:rq3_interpretability}
We next examine whether the generated explanations can be expressed using
traffic properties and attack behaviors familiar to an analyst.
Table~\ref{tab:mitre} translates representative clauses into plain-language
descriptions and relates them to MITRE ATT\&CK techniques.
The mapping is an analyst interpretation of the recovered traffic conditions,
not an automatically inferred ATT\&CK classification.

\begin{table*}[t]
\centering
\small
\setlength{\tabcolsep}{4pt}
\caption{Representative CIC-IDS2017 clauses and their analyst-assigned
MITRE ATT\&CK interpretations.}
\label{tab:mitre}
\begin{tabular}{p{2.5cm}p{6.5cm}p{3.8cm}c}
\toprule
\textbf{Clause} &
\textbf{Analyst reading} &
\textbf{MITRE ATT\&CK interpretation} &
\textbf{Precision} \\
\midrule
Bot sufficient &
Destination port \(8080\) and forward-window tier &
T1071.001 Web Protocols &
0.90 \\
Bot helper &
Low source port and small-payload tiers &
T1071 Application Layer Protocol / T1573 Encrypted Channel &
0.59 \\
Brute-force helper &
Destination port \(21\) or \(22\) and minimum-segment tier &
T1110.001 Password Guessing &
0.78 \\
DoS sufficient &
Destination port \(80\) and backward-length variance &
T1499 Endpoint Denial of Service &
0.99 \\
DoS sufficient &
Destination port \(80\) and FIN flag &
T1499.002 Service Exhaustion Flood &
0.99 \\
DoS sufficient &
Destination port \(80\), ACK-only traffic, and very short inter-arrival time &
T1499 Endpoint Denial of Service &
0.99 \\
PortScan sufficient &
High backward-packet rate and default forward-window tier &
T1046 Network Service Discovery &
1.00 \\
PortScan helper &
Small packets and near-zero flow duration &
T1046 Network Service Discovery &
0.97 \\
Web Attack sufficient &
Regular forward inter-arrival time and minimal payload &
T1190 Exploit Public-Facing Application &
0.99 \\
\bottomrule
\end{tabular}
\end{table*}
The recovered conditions align with recognizable behaviors, including
web-based command and control, credential guessing, denial-of-service floods,
service discovery, and exploitation of public-facing applications.
These clauses are generated from feature statistics and validation behavior
rather than manually authored attack descriptions.
Their correspondence with documented behavior suggests that the construction
pipeline can recover analyst-meaningful conditions.

This interpretation has two limitations.
First, helper clauses are broader and less reliable than sufficient clauses by design.
Second, some conditions, particularly port values, reflect the evaluated
captures and may not generalize to different deployments.
Timing, flag, size, and directional relationships often describe the attack
shape more generally, but the mining procedure does not guarantee that every
retained rule will correspond cleanly to a documented technique.
The results should therefore be interpreted as evidence of structural
interpretability rather than a substitute for analyst validation.

\ourparagraph{Stability across training replicates.}
\label{sec:rq3_stability}
A useful explanation should remain stable when the model is retrained on
different samples from the same distribution.
We train five replicates using a fixed test set and measure how often they
produce the same explanation for each observation.
For \name{}, agreement means selecting the same winning clause.
For the neural baseline, agreement means selecting the same highest-ranked
feature using Integrated Gradients or SHAP.
%

\begin{figure}[ht]
\centering
\begin{subfigure}[t]{0.49\columnwidth}
    \centering
    \includegraphics[width=0.65\linewidth]{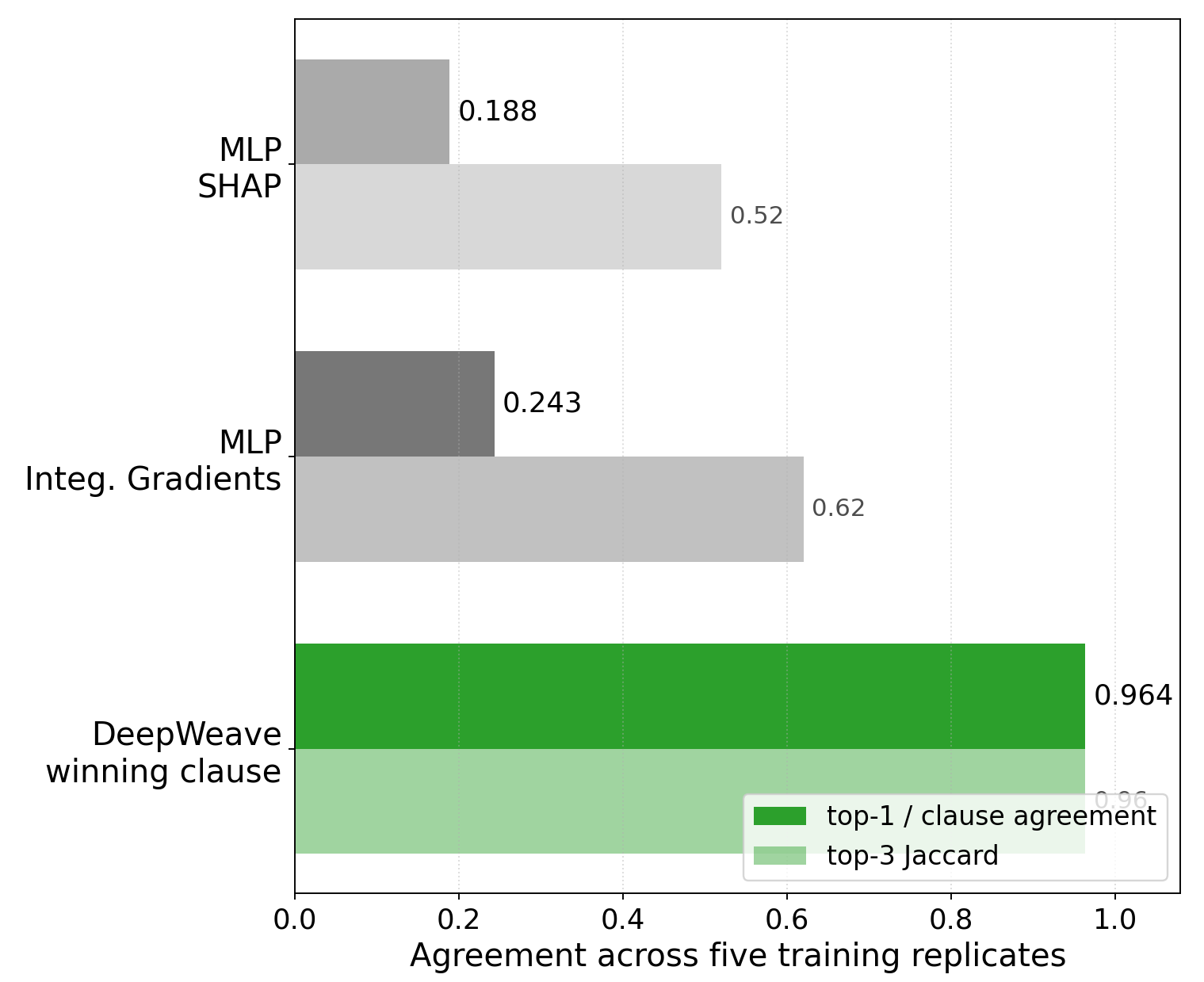}
    \caption{Explanation agreement of explainability methods across five CIC-IDS2017 training replicates}
    \label{fig:explain_stability}
\end{subfigure}
\hfill
\begin{subfigure}[t]{0.49\columnwidth}
    \centering
    \includegraphics[width=0.65\linewidth]{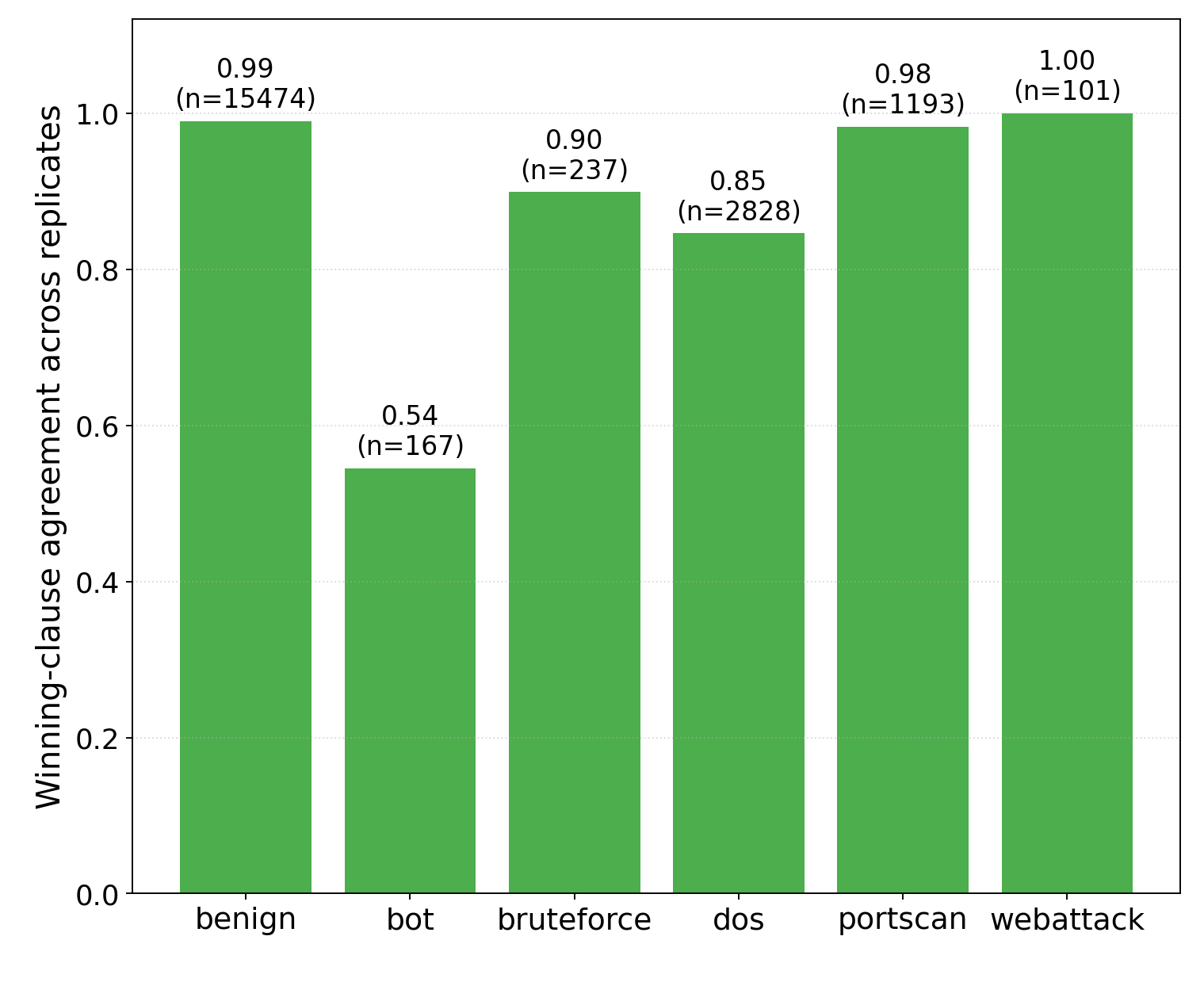}
    \caption{Per-class agreement of symbolic winning clauses across replicates}
    \label{fig:explain_class_stability}
\end{subfigure}
\caption{Explanation stability of \name{}}
\label{fig:explanation_stability}
\end{figure}

% \begin{figure}[t]
% \centering
% \includegraphics[width=0.72\columnwidth]{img/explain_stability.png}
% \caption{Explanation agreement of explainability methods across five CIC-IDS2017 training replicates}
% \label{fig:explain_stability}
% \end{figure}
%
Figure~\ref{fig:explain_stability} shows that the symbolic winning clause
agrees across replicates for \(96.4\%\) of observations.
When all replicates also agree on the predicted label, clause agreement rises
to approximately \(99.9\%\).
Integrated Gradients selects the same top feature for \(24.3\%\) of
observations, while SHAP does so for \(18.8\%\).
Even top-three agreement reaches only \(62\%\) and \(52\%\), respectively.
Thus, the symbolic explanations are substantially less sensitive to training variation than post-hoc feature rankings.

This stability varies by attack family.
Web Attack, benign traffic, and PortScan obtain agreement of \(1.00\),
\(0.99\), and \(0.98\), respectively.
Bot obtains only \(0.54\), consistent with its smaller test support and the presence of competing sufficient and helper proofs.
Classes with selective, high-precision sufficient rules therefore also produce the most stable explanations.

\ourparagraph{Contrast with neural attribution.}
\label{sec:rq3_contrast}
Table~\ref{tab:explain_contrast} illustrates the aggregate findings using
representative observations.
The comparison places the winning \name{} clause beside Integrated Gradients
and SHAP attributions for the neural baseline on the same traffic sample.

\begin{table}[t]
\centering
\setlength{\tabcolsep}{4pt}
\caption{Intrinsic symbolic explanations and post-hoc neural attributions on
representative CIC-IDS2017 flow observations.}
\label{tab:explain_contrast}
\resizebox{\columnwidth}{!}{%
\begin{tabular}{p{1.6cm}p{6.3cm}p{4.0cm}p{3.2cm}}
\toprule
\textbf{True class} &
\textbf{DeepWeave } &
\textbf{Integrated Gradients} &
\textbf{SHAP} \\
\midrule
Bot &
Destination port \(8080\) and forward initial-window tier &
Destination Port; neural prediction: benign &
Source Port \\
Brute Force &
Destination port \(21\) and minimum-segment-size tier &
Forward PSH Flags; neural prediction: benign &
Forward PSH Flags \\
DoS &
Destination port \(80\) and backward-packet-length variance &
Packet Length Variance; neural prediction: DoS &
Packet Length Variance \\
PortScan &
Backward packets/s and forward initial-window tier &
Destination Port; neural prediction: PortScan &
Total Backward Packets \\
\bottomrule
\end{tabular}}
\end{table}
The symbolic explanation identifies the concrete conditions and thresholds that
participated in the decision.
The attribution methods instead produce feature rankings whose leading feature
often changes across replicates.
For Bot and Brute Force, the neural baseline predicts benign, so its
attribution explains a missed detection rather than the correct attack label.
\name{}'s proof explains the attack decision that the model actually made.

The DoS case demonstrates that the two explanation types can agree when the
neural decision is stable.
Both neural attribution methods emphasize packet-length variance, while the
symbolic proof contains the corresponding packet-length condition.
The largest contrast therefore occurs on uncertain observations, where the
neural baseline and its post-hoc explanations are least stable and where
explicit proof evidence is most valuable.

\ourparagraph{Summary.}
\name{} provides named proof paths for all observed attack alerts in the
evaluated settings.
High-precision sufficient clauses yield reliable and stable explanations, while broader
helper clauses contribute only with neural agreement.
Many recovered clauses correspond to recognizable attack behaviors, although
dataset-specific conditions still require analyst scrutiny.
Compared with neural feature attribution, the symbolic explanations are more
stable across retraining and describe the computation supporting the predicted
class rather than approximating it after the decision.

\subsection{Additional Computational Efficiency Results}
\label{secapp:additional_computational_efficiency}

Table~\ref{tab:appendix_complexity_ladder} reports inference, memory, and training costs under the vectorized evaluator for six programs, from the rule-free L0 classifier to the Anchored-tiered (i.e. multiple quantiles per feature) L5 model. Each configuration is executed five times on CIC-IDS2017 flow data at $k=500$, with the test set, hardware, and evaluation batch size of $1{,}024$ fixed.

\begin{table}[htbp]
    \centering
    \caption{System cost across six symbolic-program configurations on CIC-IDS2017 flow data at $k=500$.
    Peak RSS covers the complete evaluation process.}
    \label{tab:appendix_complexity_ladder}
    \resizebox{\columnwidth}{!}{%
    \begin{tabular}{lrrrrrr}
        \toprule
        \textbf{Program} &
        \textbf{Neural leaves} &
        \textbf{SDD edges} &
        \textbf{Throughput (flows/s)} &
        \textbf{Latency (ms/flow)} &
        \textbf{Eval. RSS (MB)} &
        \textbf{Train (s/epoch)} \\
        \midrule
        L0 neural-only          & 1  & 267 & 53,964 & 0.0194 & 2,283.1 & 1.42 \\
        L1 single Anchored       & 3  & 333 & 20,317 & 0.0493 & 2,200.3 & 2.56 \\
        L2 gated                & 8  & 538 & 6,993  & 0.1431 & 1,408.8 & 5.95 \\
        L3 gated + numeric      & 10 & 484 & 5,948  & 0.1696 & 1,493.9 & 8.09 \\
        L4 tiered               & 10 & 484 & 5,749  & 0.1819 & 583.6   & 13.32 \\
        L5 Anchored-tiered        & 15 & 461 & 3,919  & 0.2707 & 929.2   & 32.85 \\
        \bottomrule
    \end{tabular}%
    }
\end{table}

\begin{figure}[htbp]
    \centering
    \begin{subfigure}[h]{0.49\columnwidth}
        \centering
        \includegraphics[width=0.65\linewidth]{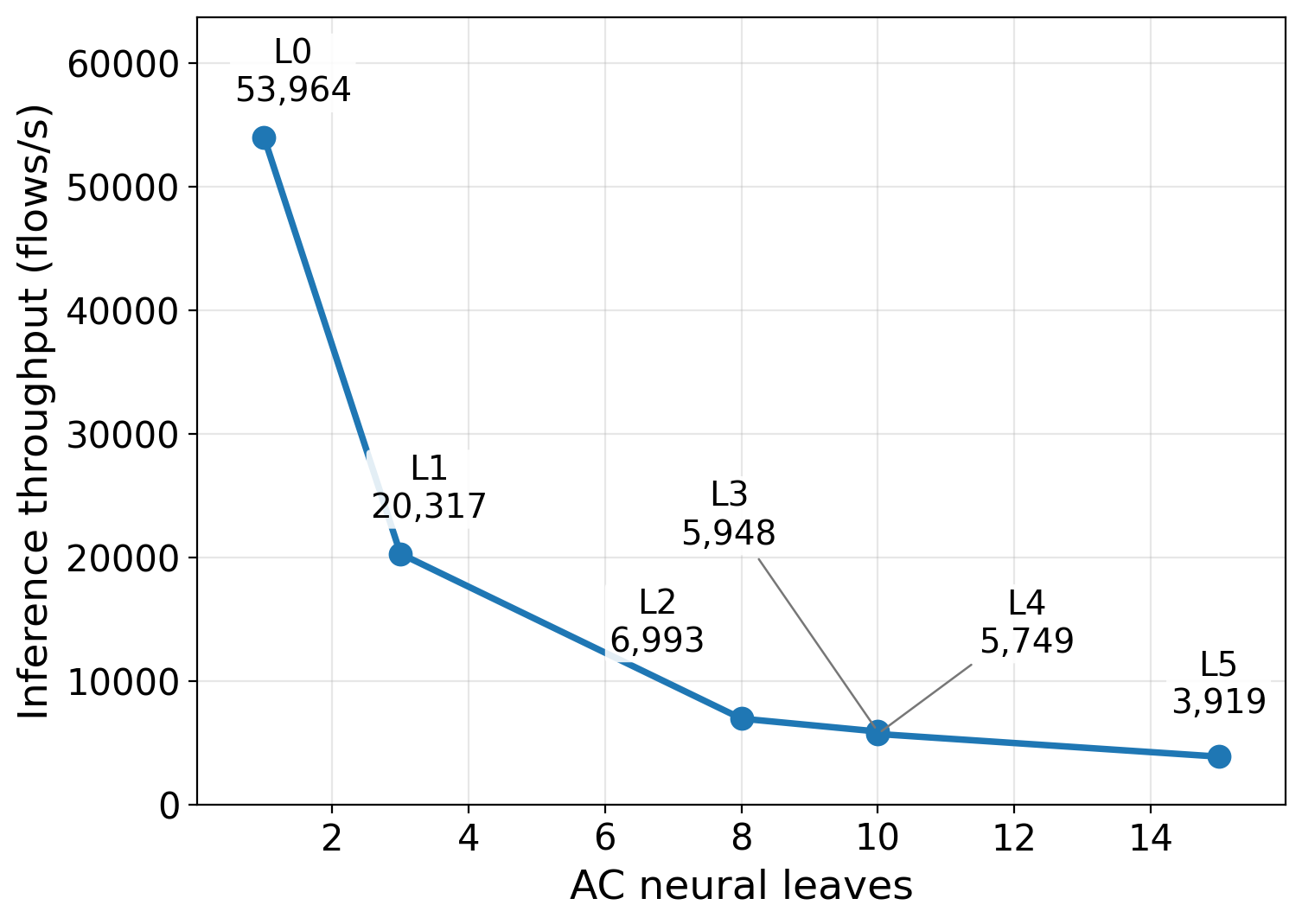}
        \caption{Inference throughput against neural-leaf count.}
        \label{fig:appendix_complexity_throughput}
    \end{subfigure}
    \begin{subfigure}[h]{0.49\columnwidth}
        \centering
        \includegraphics[width=0.65\linewidth]{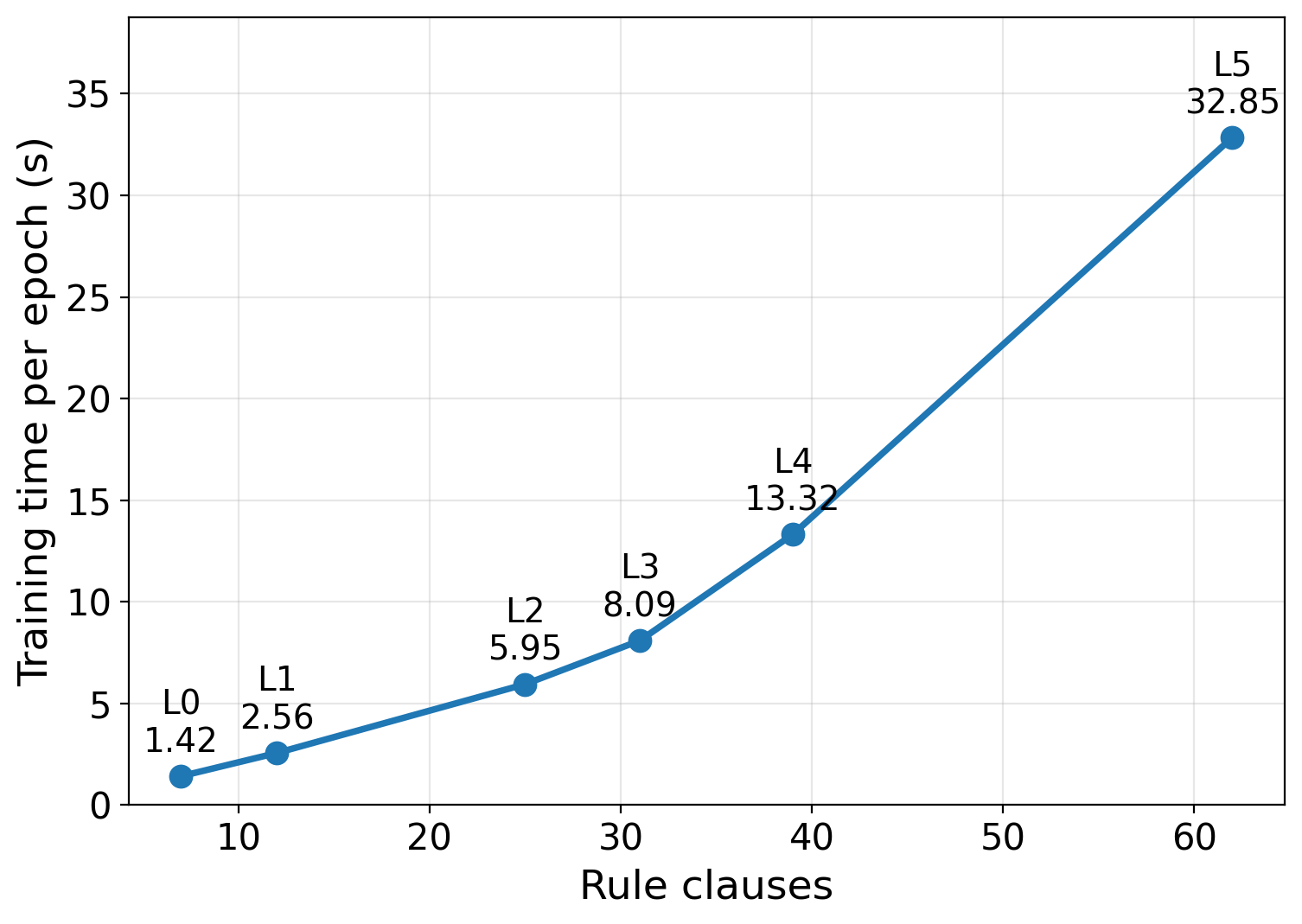}
        \caption{Training time per epoch against rule-clause count.}
        \label{fig:appendix_complexity_training}
    \end{subfigure}
    \caption{\name{} inference throughput (left) and training time (right) with increasing complexity.}
\end{figure}

Figure~\ref{fig:appendix_complexity_throughput} relates inference throughput to neural-leaf count: from L0 to L5, throughput falls from $53{,}964$ to $3{,}919$ flows per second and latency rises from $0.0194$ to $0.2707$ ms per flow. 
In contrast, Figure~\ref{fig:appendix_complexity_training} shows the increase in training cost. L3 and L4 share ten neural leaves and $484$ SDD edges, so their throughput differs by only $3.5\%$, yet training rises from $8.09$ to $13.32 $seconds per epoch. Tiered predicates (i.e. multiple quantiles per numerical feature) therefore affect training more than inference.

\begin{figure}[htbp]
    \centering
    \includegraphics[width=0.38\columnwidth]{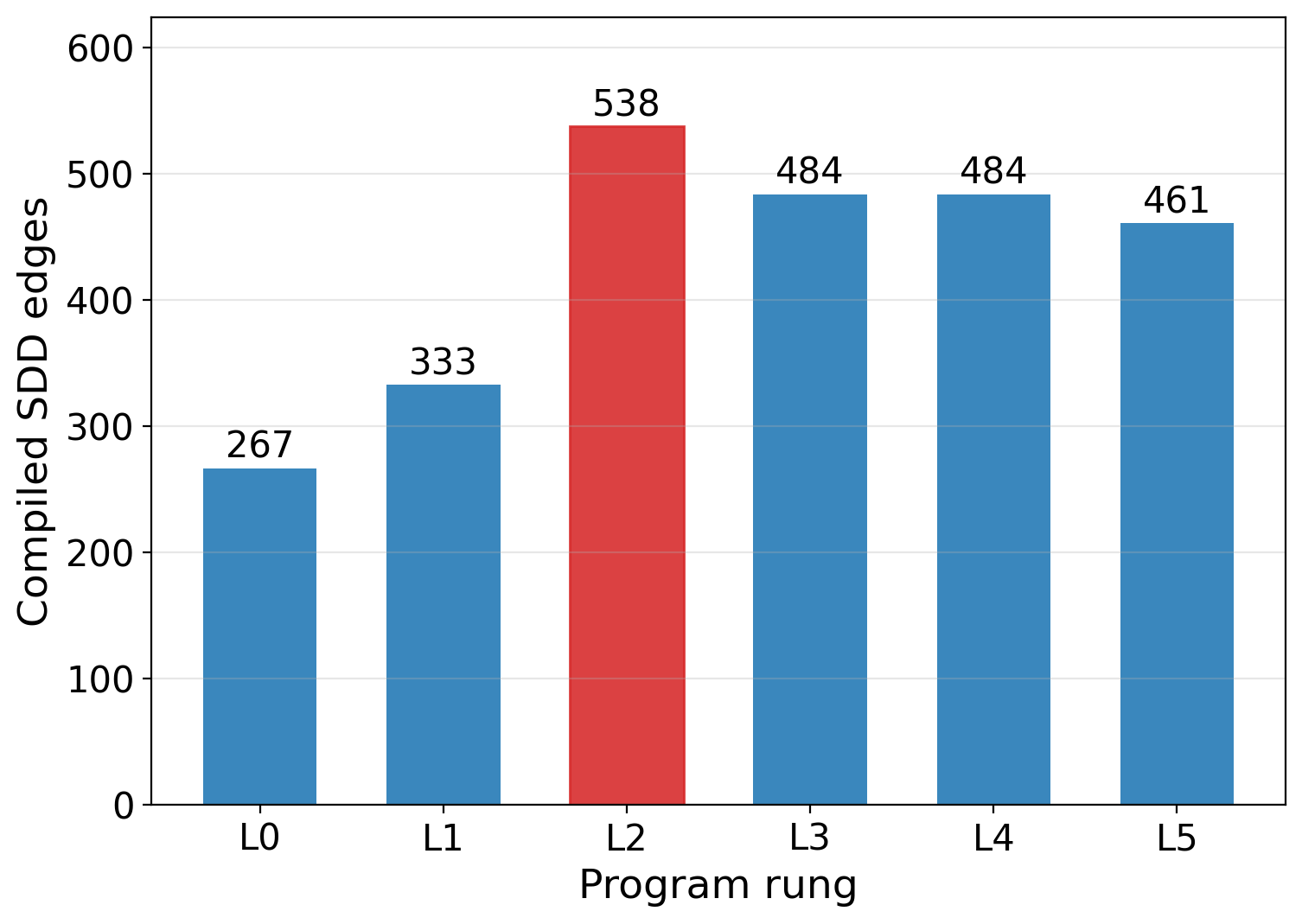}
    \caption{Compiled SDD size across the six program rungs.}
    \label{fig:appendix_complexity_circuit_size}
\end{figure}

Figure~\ref{fig:appendix_complexity_circuit_size} shows that compiled size peaks at $538$ SDD edges at L2 before falling to $461$ at L5. RSS is not a circuit-size proxy because it covers the complete process. Leaf count, circuit size, and tiered predicates therefore remain distinct costs; their absolute values are specific to the Apple M3 system used here.

\section{Automatic Rule Generation}\label{secapp:automatic_rule_generation}

\subsection{Rule Generation Approaches}
\label{secapp:rule_generation_approaches}
The pipeline supports three sources of candidate rules.

Scenario~A derives rules from labelled traffic. For each attack class, it ranks features by their separation from benign or non-target traffic, instantiates numeric and categorical predicates, and forms short conjunctions. Identifiers, timestamp-derived attributes, constant features, and other disallowed variables are removed before candidates enter the common Rule IR.

Scenario~B converts expert knowledge into dataset-compatible rules. MITRE ATT\&CK techniques are expressed as network-observable behaviours and mapped to available features. Thresholds and categorical values are instantiated from the construction reference, and each mapping records whether the behaviour is fully, partially, or not observable. Claims that require unavailable temporal, payload, or cross-flow information are retained as documentation but not converted into predicates.

Scenario~C asks an LLM to propose candidates from a fixed catalogue. The provider-neutral request supplies feature and rule schemas, class descriptions, complete feature rankings with directionality, construction and reference statistics, class-local literal catalogues, and the MITRE ATT\&CK evidence used in Scenario~B. The model may combine listed predicates but cannot introduce features, operators, or thresholds. It returns Rule IR JSON, and the pipeline rejects malformed or unsupported proposals before evaluation.

All three scenarios use the same validation and selection procedure. Candidates are rejected or assigned sufficient, helper, or disqualifier roles according to their precision, recall, FPR, and construction-to-validation stability. Pareto and firing-set filters remove dominated or near-duplicate rules, while per-class and per-role limits bound the shortlist. The coupling screen selects the final Rule IR, which is compiled deterministically into Prolog.

\ourparagraph{Generation pipeline hyperparameters. } Shared parameters define the role gates, permitted stability gaps, redundancy threshold, shortlist limits, validation FAR budgets, screening depth, and final rule budget. These settings govern validation and selection rather than candidate generation.

Scenario-specific parameters only bound how each candidate pool is constructed. Scenario~A configures feature ranking, quantile grids, categorical-value filters, conjunction length, support requirements, and search limits. Scenario~B selects the expert profile, ATT\&CK-to-feature mappings, template expansion rules, and proposal limits. Scenario~C records the provider, model, structured request, supplied evidence, literal catalogue, output limits, and supported decoding settings. In every case, candidates must satisfy the same role and stability gates before they can enter the final program.

\subsection{Scenario Performance}

\ourparagraph{Experiment hyperparameters. } All scenarios emit the same Rule IR and use shared validation gates. Each arm starts from the same neural checkpoint and independently optimizes its neural and differentiable rule parameters under Guarded coupling with soft predicates. Shared gates require sufficient-rule precision/FPR of $0.90/0.015$, helper recall/FPR of $0.35/0.05$, and disqualifier precision/FPR of $0.90/0.02$. Construction-to-validation precision and FPR gaps are limited to $0.20$ and $0.05$. Rules contain at most three literals. The shortlist retains 32 candidates per class and role, removes firing sets with Jaccard similarity above $0.98$, and passes at most 12 rules to the coupling screen. Validation macro-F1 selects checkpoints under an active $1\%$ FAR budget.

Scenario~A retains the 12 highest-ranked stable features per attack-versus-benign contrast. Its ranking combines the Kolmogorov-Smirnov statistic, Cliff's delta, Hellinger distance, AUC, Cram\'{e}r's~$V$, and total variation distance with weights $0.30$, $0.20$, $0.20$, $0.20$, $0.05$, and $0.05$. Numeric predicates use lower quantiles $\{0.001,0.003,0.005,0.03,0.10,0.30\}$ and upper quantiles $\{0.80,0.90,0.97,0.995,0.997,0.999\}$. Conjunctions draw from the 40 highest-ranked predicates and are limited to $5{,}000$ candidates, 120 seconds, and 10 validation true positives. The process generated 19,326 candidates and retained 12.

Scenario~B uses the semantic-v4 expert profile and numeric quantiles $\{0.05,0.25,0.50,0.75,0.95\}$. Its 15 templates contain at most three literals and emit at most $4{,}096$ candidates, with categorical search limited to 31 subsets.

Scenario~C uses a structured request containing complete feature rankings, construction and reference statistics, class-local literal catalogues, and Scenario~B's MITRE ATT\&CK evidence. Claude proposed 20 candidates and validation retained 10 helpers; Codex proposed seven and retained all seven. The corrected matcher equates integral integer and floating-point categorical values, allowing numeric port literals to match the dataset.

\ourparagraph{Automated Generation Results. } Table~XVIII reports the results after independent end-to-end optimization from the common checkpoint. Codex leads the automatic arms with $0.7671$ macro-F1 and $0.7691$ detection at $0.7820\%$ FAR. Claude reaches $0.7033$ macro-F1 and $0.6177$ detection at $0.6032\%$ FAR. Scenarios~A and~B reach $0.4300$ and $0.4399$ macro-F1, while the historical rules remain strongest at $0.8408$.

\begin{table}[tp]
    \centering
    \fontsize{8}{9.5}\selectfont   % not needed if you use the updated template.tex
    \caption{CIC-IDS2017 fully automated rule generation results at $k=500$, one replicate, using Guarded coupling.}
    \label{tab:rule-generation-k500}
    \sbox0{%
    \begin{tabular}{lcccc}
        \toprule
        Rule source & Macro-F1 & Accuracy & Detection & FAR (\%) \\
        \midrule
        Scenario~A: data-driven & 0.4300 & 0.9045 & 0.5171 & 0.0627 \\
        Scenario~B: expert/MITRE & 0.4399 & 0.9396 & 0.7016 & 0.2080 \\
        Scenario~C: Codex & \underline{0.7671} & 0.9481 & 0.7691 & 0.7820 \\
        Scenario~C: Claude & 0.7033 & 0.9198 & 0.6177 & 0.6032 \\
        \midrule
        Existing Rules (Guarded)$^{\dagger}$ & \textbf{0.8408} & \textbf{0.9836} & \textbf{0.9648} & 1.1640 \\
        \bottomrule
    \end{tabular}}%
    \usebox0

    \smallskip
    \begin{minipage}{\wd0}   % note is exactly as wide as the table
        \raggedright
        \textit{Note:} Macro-F1 covers benign and five attack classes. $^{\dagger}$The existing Guarded system is a historical baseline, not a fourth generation arm. Generated-arm results use one replicate.
    \end{minipage}
\end{table}

\end{document}